\documentclass[12pt,a4paper,dvips]{article}
\usepackage{a4p}
\usepackage{cite,mcite}
\usepackage{graphicx}
\usepackage{rotating}
\usepackage{l3_title,ifthen}
\journalname{The European Physical Journal C}
\date{February 04, 2000}
\preprint{2000-022}
\newlength{\fepsy}                
\newcommand{\pltdir}{}

\newlength{\capindent}
\newlength{\capwidth}
\newlength{\figwidth}
\newcommand{\icaption}[2][!*!,!]{\hspace*{\capindent}%
  \begin{minipage}{\capwidth}
    \ifthenelse{\equal{#1}{!*!,!}}%
      {\caption{#2}}%
      {\caption[#1]{#2}}
  \end{minipage}}
\def\Zo{\ensuremath{\mathrm {Z}}}
\def\Photon{\ensuremath{\mathrm {\gamma}}}

\def\TeV{\ensuremath{\mathrm{Te\kern -0.1em V}}}
\def\GeV{\ensuremath{\mathrm{Ge\kern -0.1em V}}}
\def\MeV{\ensuremath{\mathrm{Me\kern -0.1em V}}}
\def\keV{\ensuremath{\mathrm{ke\kern -0.1em V}}}
\def\eV{\ensuremath{\mathrm{e\kern -0.1em V}}}
\def\pb{\ensuremath{\mathrm{pb}}}
\def\nb{\ensuremath{\mathrm{nb}}}
\def\pbinv{\ensuremath{\mathrm{pb^{-1}}}}%
\def\MZbar{\ensuremath{\overline{m}_{\mathrm{Z}}}}%
\def\MZ{\ensuremath{m_{\mathrm{Z}}}}%
\def\MW{\ensuremath{m_{\mathrm{W}}}}%
\def\Mt{\ensuremath{m_{\mathrm{t}}}}%
\def\MH{\ensuremath{m_{\mathrm{H}}}}%
\def\GZbar{\ensuremath{\overline{\Gamma}_{\mathrm{Z}}}}
\def\GZ{\ensuremath{\Gamma_{\mathrm{Z}}}}
\def\GE{\ensuremath{\Gamma_{\mathrm{e}}}}
\def\GM{\ensuremath{\Gamma_{\mu}}}
\def\GT{\ensuremath{\Gamma_{\tau}}}
\def\Gh{\ensuremath{\Gamma_{\mathrm{had}}}}
\def\GL{\ensuremath{\Gamma_{\ell}}}
\def\GN{\ensuremath{\Gamma_{\nu}}}
\def\GB{\ensuremath{\Gamma_{\mathrm{b}}}}
\def\Gf{\ensuremath{\Gamma_{\mathrm{f}}}}
\def\Ginv{\ensuremath{\Gamma_{\mathrm{inv}}}} 
\def\ee{\ensuremath{\mathrm{e^+ e^-}}}%
\def\mm{\ensuremath{\mathrm{\mu^+ \mu^-}}}%
\def\tautau{\ensuremath{\mathrm{\tau^+ \tau^-}}}%
\def\ellell{\ensuremath{\mathrm{\ell^+ \ell^-}}}%
\def\antibar#1{\ensuremath{\mathrm{#1\bar{#1}}}}%
\def\nnbar{\antibar{\nu}}%
\def\bbar{\antibar{b}}
\def\EEHADG{\ensuremath{\ee \rightarrow \mathrm{hadrons}(\Photon)}}
\def\EEEEG{\ensuremath{\ee \rightarrow \ee(\Photon)}}
\def\EEMMG{\ensuremath{\ee \rightarrow \mm(\Photon)}}
\def\EETTG{\ensuremath{\ee \rightarrow \tautau(\Photon)}}
\def\EELLG{\ensuremath{\ee \rightarrow \ellell(\Photon)}}
\def\ffbar{\ensuremath{\mathrm{f\bar{f}}}}

\def\EENNG{\ensuremath{\ee \rightarrow \nnbar\Photon(\Photon)}}
\def\EEGGG{\ensuremath{\ee \rightarrow \Photon\Photon(\Photon)}}
\def\gVb{\ensuremath{\bar{g}_{\mathrm{V}}}}
\def\gAb{\ensuremath{\bar{g}_{\mathrm{A}}}}
\def\IL{\ensuremath{{\cal L}}}
\def\bb{\ensuremath{\mathrm{b\bar{b}}}}
\def\QFB{\ensuremath{Q_{\mathrm{FB}}}}
\def\AFB{\ensuremath{A_{\mathrm{FB}}}}
\def\AFBZ{\ensuremath{A^0_{\mathrm{FB}}}}
\def\Ebeam{\ensuremath{E_{\mathrm{b}}}}
\def\aqed{\ensuremath{\alpha}}
\newcommand{\aqedMZ}{\aqed(\MZ)}
\def\Da5had{\ensuremath{\Delta\alpha^{(5)}_{\mathrm{had}} }}
\def\as{\hbox{$\alpha_{\mathrm{s}}$}}
\def\s0had{\ensuremath{\sigma^0_{\mathrm{had}}}}
\def\RE{\ensuremath{R_{\mathrm{e}}}}
\def\RM{\ensuremath{R_{\mathrm{\mu}}}}
\def\RT{\ensuremath{R_{\mathrm{\tau}}}}
\def\RL{\ensuremath{R_{\mathrm{\ell}}}}
\def\AFBZe{\ensuremath{\AFB^{\mathrm{0,e}}}}
\def\AFBZm{\ensuremath{\AFB^{\mathrm{0,\mu}}}}
\def\AFBZt{\ensuremath{\AFB^{\mathrm{0,\tau}}}}
\def\AFBZl{\ensuremath{\AFB^{\mathrm{0,\ell}}}}
\def\AFBZb{\ensuremath{\AFB^{0,\mathrm{b}}}}
\def\gVbe{\ensuremath{\gVb^{\mathrm{e}}}}
\def\gAbe{\ensuremath{\gAb^{\mathrm{e}}}}
\def\gVbm{\ensuremath{\gVb^{\mathrm{\mu}}}}
\def\gAbm{\ensuremath{\gAb^{\mathrm{\mu}}}}
\def\gVbt{\ensuremath{\gVb^{\mathrm{\tau}}}}
\def\gAbt{\ensuremath{\gAb^{\mathrm{\tau}}}}
\def\gVbl{\ensuremath{\gVb^{\mathrm{\ell}}}}
\def\gAbl{\ensuremath{\gAb^{\mathrm{\ell}}}}

\def\swsq{\ensuremath{\sin^2\!\theta_{\mathrm{W}}}}%
\def\cwsq{\ensuremath{\cos^2\!\theta_{\mathrm{W}}}}%
\def\ctwsq{\ensuremath{\cot^2\!\theta_{\mathrm{W}}}}%
\def\swsqb{\ensuremath{\sin^2\!\overline{\theta}_{\mathrm{W}}}}%
\def\rtxsha{\ensuremath{r^{\mathrm{tot}}_{\mathrm{had}}}}
\def\rtxsel{\ensuremath{r^{\mathrm{tot}}_{\mathrm{e}}}}   
\def\rtxsmu{\ensuremath{r^{\mathrm{tot}}_{\mathrm{\mu}}}} 
\def\rtxsta{\ensuremath{r^{\mathrm{tot}}_{\mathrm{\tau}}}}
\def\rtxsle{\ensuremath{r^{\mathrm{tot}}_{\mathrm{\ell}}}}
\def\jtxsha{\ensuremath{j^{\mathrm{tot}}_{\mathrm{had}}}} 
\def\jtxsel{\ensuremath{j^{\mathrm{tot}}_{\mathrm{e}}}}   
\def\jtxsmu{\ensuremath{j^{\mathrm{tot}}_{\mathrm{\mu}}}} 
\def\jtxsta{\ensuremath{j^{\mathrm{tot}}_{\mathrm{\tau}}}}
\def\jtxsle{\ensuremath{j^{\mathrm{tot}}_{\mathrm{\ell}}}}
\def\rafbel{\ensuremath{r^{\mathrm{FB}}_{\mathrm{e}}}}      
\def\rafbmu{\ensuremath{r^{\mathrm{FB}}_{\mathrm{\mu}}}}    
\def\rafbta{\ensuremath{r^{\mathrm{FB}}_{\mathrm{\tau}}}}   
\def\rafble{\ensuremath{r^{\mathrm{FB}}_{\mathrm{\ell}}}}   
\def\jafbel{\ensuremath{j^{\mathrm{FB}}_{\mathrm{e}}}}      
\def\jafbmu{\ensuremath{j^{\mathrm{FB}}_{\mathrm{\mu}}}}    
\def\jafbta{\ensuremath{j^{\mathrm{FB}}_{\mathrm{\tau}}}}   
\def\jafble{\ensuremath{j^{\mathrm{FB}}_{\mathrm{\ell}}}}   
\def\RS{\ensuremath{\sqrt{s}}}
\def\RSprime{\ensuremath{\sqrt{s^\prime}}}
\def\Evis{\ensuremath{E_{\mathrm{vis}}}}
\def\Elong{\ensuremath{E_\parallel}}
\def\Etrans{\ensuremath{E_\perp}}
\def\GF{\ensuremath{G_{\mathrm{F}}}}
\newcommand{\GZint}{$\Photon/\Zo$-inter\-fer\-ence}
\newcommand{\stint}{$s/t$-inter\-fer\-ence}
\newcommand{\cms}{centre-of-mass}
\newcommand{\MR}{\mathrm}
\newcommand{\MC}{\multicolumn}
\newcommand{\SM}{SM}

\newcommand{\SMD}{silicon microvertex detector}
\newcommand{\TEC}{central tracking chamber}
\font\TI=ptmr
\newcommand {\permille} {{\TI\char189}}       
\def\etal{{\it et~al.}}%
\def\coll{{Collab.}}%

\begin{document}
%
%
%
\def\mZqval{\ensuremath{91\,189.8}}   
\def\mZqerr{\ensuremath{3.1}}
\def\GZqval{\ensuremath{2\,502.4}}    
\def\GZqerr{\ensuremath{4.2}}
\def\Ghqval{\ensuremath{1\,751.1}}    
\def\Ghqerr{\ensuremath{3.8}}
\def\Glqval{\ensuremath{84.14}}       
\def\Glqerr{\ensuremath{0.17}}
\def\Ginvqval{\ensuremath{499.1}}     
\def\Ginvqerr{\ensuremath{2.9}}       
\def\nnqval{\ensuremath{2.978}}       
\def\nnqerr{\ensuremath{0.014}}
\def\Alepval{\ensuremath{0.1575}}     
\def\Aleperr{\ensuremath{0.0067}}
%
\def\gVqval{\ensuremath{-0.0397}}     
\def\gVqerr{\ensuremath{0.0017}}      
\def\gAqval{\ensuremath{-0.50153}}    
\def\gAqerr{\ensuremath{ 0.00053}}    
\def\swsqval{\ensuremath{0.23093}}    
\def\swsqerr{\ensuremath{0.00066}}
%
\def\mZSmatrixqval{\ensuremath{91\,185.2}}
\def\mZSmatrixqerr{\ensuremath{10.3}}
\def\mZjhadcor{\ensuremath{-0.95}}
%
\def\mtlval{\ensuremath{197}}         
\def\mtlerrup{\ensuremath{30}}        
\def\mtlerrlo{\ensuremath{16}}        
%
\def\mhlimit{\ensuremath{133}}        

%
%
%
\begin{titlepage}
\title{Measurements of Cross Sections and Forward-Backward Asymmetries 
       at the \protect\boldmath\Zo\ Resonance and \\
       Determination of Electroweak Parameters}
\vspace*{-15mm}
\author{The L3 Collaboration}
%
%
\vspace*{-15mm}
\footnotesize
\begin{abstract}
We report on measurements of hadronic and leptonic cross sections and 
leptonic forward-backward asymmetries performed with the L3 detector 
in the years $1993-95$.
A total luminosity of $103\ \pbinv$ was collected at 
\cms\  energies $\RS \approx \MZ$ and $\RS \approx \MZ \pm 1.8\ \GeV$
which corresponds to 2.5 million hadronic and 245 thousand leptonic 
events selected.
These data lead to a significantly improved determination
of \Zo\ parameters.
From the total cross sections, combined with our measurements in $1990-92$,
we obtain the final results:
\begin{eqnarray*}
   \MZ = \mZqval \pm  \mZqerr\ \MeV \, ,   & &
   \GZ = \GZqval \pm  \GZqerr\ \MeV \, ,   \\
   \Gh = \Ghqval \pm  \Ghqerr\ \MeV \, ,   & &
   \GL = \Glqval \pm  \Glqerr\ \MeV \,.
 \label{eq:Zpara_abstract}
\end{eqnarray*}
An invisible width of $\Ginv = \Ginvqval \pm \Ginvqerr\ \MeV$ 
is derived which in the Standard Model yields for the number
of light neutrino species $N_{\MR{\nu}} = \nnqval \pm \nnqerr$.

Adding our results on the leptonic forward-backward asymmetries and the
tau polarisation, the effective vector and axial-vector 
coupling constants of the neutral weak current to charged leptons 
are determined to be
   $\gVbl = \gVqval \pm \gVqerr$ and
   $\gAbl = \gAqval \pm \gAqerr$.
Including our measurements of the $\Zo\rightarrow\bb$ forward-backward 
and quark charge asymmetries a value for the effective electroweak
mixing angle of $\swsqb =  \swsqval \pm \swsqerr$ is derived. 

All these measurements are in good agreement with the Standard Model
of electroweak interactions.
Using all our measurements of electroweak observables an upper limit 
on the mass of the Standard Model Higgs boson of 
$\MH < \mhlimit\ \GeV$ is set at 95\% confidence level.
\end{abstract}
\normalsize

\centerline{\it{Dedicated to the memory of Prof. Dr. Klaus Schultze}}
%
%
\vspace*{-10mm}
\submitted
\end{titlepage}

\section{Introduction}\label{sec:introduction}

The Standard Model (SM) of electroweak 
interactions~\cite{standard_model,Veltman_SM}
is tested with great precision by the experiments performed at the 
LEP and SLC \ee\ colliders running at \cms\ energies, \RS, close 
to the \Zo\ mass.
From measurements of the total cross sections and forward-backward
asymmetries in the reactions 
\begin{eqnarray}
 \EEHADG\, , && \EEEEG \, ,\nonumber \\
 \EEMMG \, , && \EETTG \, ,
\label{eq:reactions}
\end{eqnarray}
the mass, total and partial widths of the \Zo\ and other electroweak
parameters are obtained by L3~\cite{l3-28,l3-69}
and other experiments~%
\cite{
aleph-29, 
delphi-21,
opal-33,  
sld-05    
}.
The $(\Photon)$ indicates the presence of radiative photons.

The large luminosity collected in the years $1993 - 95$ enables a significant
improvement on our previous measurements of \Zo\ parameters.
An integrated luminosity of $103\ \pbinv$ was collected,
corresponding to the selection of $2.5\cdot 10^6$ hadronic and 
$2.5\cdot 10^5$ leptonic events.
Most of the data were collected at a \cms\ energy corresponding to the
maximum annihilation cross section.

In 1993 and 1995 scans, of the \Zo\ resonance were performed where runs at the 
\Zo\ pole alternated with runs at about $1.8\ \GeV$ on either side of the 
peak.
Compared to previous measurements, our event samples on the wings of the
\Zo\ resonance are increased by more than a factor of five.

The LEP beam energies were precisely calibrated at the three energy points in 
$1993-95$ using the method of resonant depolarisation~\cite{lep-09}.
As a result, the contributions to the errors on the \Zo\ mass and total width
from the uncertainty on the \cms\ energy are reduced by factors of about
five and three, respectively, as compared to the data collected before.

The installation of silicon strip detectors in front of the small angle
electromagnetic calorimeters allows a much more precise 
determination of the fiducial volume used for the luminosity 
measurement~\cite{l3-SLUM}.
This improvement, together with the reduced theoretical uncertainty on the
small angle Bhabha cross section~\cite{BHLUMI404,Ward_lumi}, allows more precise
measurements of the cross sections, in particular that for 
\EEHADG.
This results in a better determination of the invisible \Zo\ width, 
from which the number of light neutrino generations is deduced.

In this article measurements of hadronic and leptonic cross sections
and leptonic forward-backward asymmetries, obtained from the data
collected between 1993 and 1995, are presented.
These measurements are combined with our published results from the
data collected in $1990-92$~\cite{l3-69}.            
The complete integrated luminosity collected by L3 at the \Zo\ resonance
is $143\ \pbinv$, consisting of about $3.5\cdot10^6$ hadronic and $3.5\cdot10^5$
leptonic events.
The results on the properties of the \Zo\ boson and on other electroweak
observables presented here are based on the final analyses of the complete data
set collected at the \Zo\ resonance.

This article is organised as follows:
After a brief description of the L3 detector in Section~\ref{sec:L3detector},
we summarise in Section~\ref{sec:data_analysis} features of the $1993-95$
data analysis common to all final states investigated.
Section~\ref{sec:LEP_Ecal} addresses issues related to the LEP \cms\ energy.
The measurement of luminosity is described in Section~\ref{sec:luminosity}.
The event selection and the analysis of the reactions in (\ref{eq:reactions})
are discussed in Sections~\ref{sec:hadrons} to~\ref{sec:electrons}
and the results on the measurements of total cross sections and 
forward-backward asymmetries are presented in Section~\ref{sec:xsafbresults}.
A general description of the fits performed to our data is given in
Section~\ref{sec:fit_intro}.
Various fits for \Zo\ parameters are performed in Section~\ref{sec:fit_Zpara}
and the results of the fits in the framework of the \SM\ are given
in Section~\ref{sec:fit_SMpara}.
We summarise and conclude in Section~\ref{sec:conclusion}.
The Appendices~\ref{app:tchannel} and~\ref{app:covariance} give details
on the treatment of the $t$-channel contributions in \EEEEG\ and on
technicalities of the fit procedures, respectively.

%
%
\section{The L3 Detector}
        \label{sec:L3detector}

The L3 detector~\cite{l3-00} consists of 
a \SMD~\cite{l3-SMD}, 
a \TEC,
a high resolution electromagnetic calorimeter composed of BGO crystals,
a lead-scintillator ring calorimeter at low polar angles~\cite{l3-ALR},
a scintillation counter system,
a uranium hadron calorimeter with proportional wire chamber readout
and an accurate muon spectrometer.
Forward-backward muon chambers, completed for the 1995 data taking,
extend the polar angle coverage of the muon system down to 
24 degrees~\cite{l3-FBmuon} with respect to the beam line.
All detectors are installed in a 12~m diameter magnet which provides 
a solenoidal field of $0.5\ \MR{T}$ in the central region and 
a toroidal field of $1.2\ \MR{T}$ in the forward-backward region.
The luminosity is measured using BGO calorimeters preceded by silicon
trackers~\cite{l3-SLUM} situated on each side of the detector.

In the L3 coordinate system the direction of the 
$\mathrm{e^{-}}$ beam defines the $z$ direction. The $xy$, or $r\phi$ plane,
is the bending plane of the magnetic field, with the $x$ direction pointing 
to the centre of the LEP ring. The coordinates $\phi$ and $\theta$ denote 
the azimuthal and polar angles.

%
%
\section{Data Analysis}
         \label{sec:data_analysis}

The data collected between 1993 and 1995 are split into nine samples according 
to the year and the \cms\ energy.
Data samples at $\RS \approx \MZ$ are referred to as peak,
those at off-peak energies are referred to as peak$-2$ and peak$+2$.
The peak samples in 1993 and 1995 are further split into data taken early 
in the year (pre-scan) and those peak runs interspersed with off-peak data 
taking (scan) which coincide with the precise LEP energy calibration
(see Section \ref{sec:LEP_Ecal}).
Cross sections and leptonic forward-backward asymmetries are determined
for each data sample.

Acceptances, background contaminations and trigger efficiencies are studied 
for all nine data samples separately to take into account their possible
dependence on the \cms\ energy and the time dependence of the detector 
status.
Systematic errors are determined for the data samples individually.
Average values for uncertainties are used if no dependence on
the \cms\ energy or the data taking period is observed.
Correlations of the systematic errors among the data sets are estimated and
are taken into account in the analyses to determine electroweak parameters.

Acceptances and background contaminations from \ee-interactions
are determined by Monte Carlo simulations.
The following event generator programs are used for the various signal
and background processes:
JETSET~\cite{JETSET_PYTHIA} and HERWIG~\cite{HERWIG57} for \EEHADG;
KORALZ~\cite{KORALZ40} for \EEMMG\ and \EETTG;
BHAGENE~\cite{BHAGENE3}, BHWIDE~\cite{BHWIDE101} 
and BABAMC~\cite{BABAMC} for large angle \EEEEG;
BHLUMI~\cite{BHLUMI404} for small angle \EEEEG;
GGG~\cite{GGG} for \EEGGG;
DIAG36~\cite{DIAG36} for $\ee\rightarrow\ee\,\ellell$;
DIAG36, PHOJET~\cite{PHOJET105} and PYTHIA~\cite{JETSET_PYTHIA}
for $\ee\rightarrow\ee\,\mathrm{hadrons}$.
For the simulation of hadronic final states the fragmentation parameters of 
JETSET and HERWIG are tuned to describe our data as discussed
in Reference~\cite{l3-38}.
%

The generated events are passed through a complete detector simulation.
The response of the L3 detector is modelled with the GEANT~\cite{geant}
detector simulation program which includes the effects of energy loss,
multiple scattering and showering in the detector materials.
Hadronic showers are simulated with the GHEISHA~\cite{GHEISHA} program.
The performance of the detector, including inefficiencies and their time
dependence as observed during data taking, is taken into account 
in the simulation.
With this procedure, experimental systematic errors on cross sections and 
forward-backward asymmetries are minimized.

%
%
\section{LEP Energy Calibration}
         \label{sec:LEP_Ecal}

The average \cms\ energy of the colliding particles at the L3 interaction
point is calculated using the results provided by the Working Group on LEP 
Energy~\cite{lep-09}.
Every 15 minutes the average \cms\ energy is determined from measured LEP 
machine parameters, applying the energy model which is based on 
calibration by resonant depolarisation~\cite{lep-07}.
This model traces the time variation of the \cms\ energy of typically 
$1\ \MeV$ per hour.
The average \cms\ energies are calculated for each data sample individually 
as luminosity weighted averages.
Slightly different values are obtained for different reactions because of
small differences in the usable luminosity.

The errors on the \cms\ energies and their correlations for the 1994 data
and for the two scans performed in 1993 and 1995 are given in form of a 
$7\times 7$ covariance matrix in Table~\ref{tab:leperror1}.
The uncertainties on the \cms\ energy for the data samples not included in 
this matrix, i.e. the 1993 and 1995 pre-scans, 
are $ 18\ \MeV$ and $ 10\ \MeV$, respectively.
Details of the treatment of these errors in the fits can be found in
Appendix~\ref{app:covariance}.

The energy distribution of the particles circulating in an \ee-storage
ring has a finite width due to synchrotron oscillations.
An experimentally observed cross section is therefore a convolution of 
cross sections at energies which are distributed around the average 
value in a gaussian form.
The spread of the \cms\ energy for the L3 interaction point as obtained from 
the observed longitudinal length of the particle bunches in LEP is listed in 
Table~\ref{tab:cms_spread}~\cite{lep-09}.
The time variation of the average energy causes a similar, but smaller, effect 
which is included in these numbers.

All cross sections and forward-backward asymmetries quoted below are 
corrected for the energy spread to the average value of the \cms\ energy.
The relative corrections on the measured hadronic cross sections amount to
$+1.7$ per mill (\permille) at the \Zo\ pole and to $-1.1$\permille\ and 
$-0.6$\permille\ at the peak$-2$ and peak$+2$ energy, 
respectively.
The absolute corrections on the forward-backward asymmetries are very small.
The largest correction is $-0.0002$ for the muon and tau peak$-2$ data sets.
The error on the energy spread is propagated into the fits, resulting
in very small contributions to the errors of the fitted parameters
(see Appendix~\ref{app:covariance}).
The largest effect is on the total width of the \Zo, contributing 
approximately $0.3\ \MeV$ to its error.

During the operation of LEP, no evidence for an average longitudinal
polarisation of the electrons or positrons has been observed.
Stringent limits on residual polarisation during luminosity runs are set 
such that the uncertainties on the determination of electroweak observables
are negligible compared to their experimental errors~\cite{LEP_plong}. 

The determination of the LEP \cms\ energy in $1990-92$ is described in
References~\cite{lep-02}.
From these results the LEP energy error matrix given in 
Table~\ref{tab:leperror9092} is derived.

%
%

\section{Luminosity Measurement}
         \label{sec:luminosity}

The integrated luminosity \IL\ is determined by measuring the number of
small-angle Bhabha interactions \EEEEG.
For this purpose two cylindrical calorimeters consisting of arrays of BGO 
crystals are located on either side of the interaction point.
Both detectors are divided into two half-rings in the vertical plane 
to allow the opening of the detectors during filling of LEP.
A silicon strip detector, consisting of two layers measuring the polar angle, 
$\theta$, and one layer measuring the azimuthal angle, $\phi$, 
is situated in front of each calorimeter to precisely define the fiducial 
volume.
A detailed description of the luminosity monitor and the luminosity 
determination can be found in Reference~\cite{l3-SLUM}.

The selection of small-angle Bhabha events is based on the energy depositions
in adjacent crystals of the BGO calorimeters which are grouped to form 
clusters. 
The highest-energy cluster on each side is considered for the luminosity 
analysis. 
For about 98\% of the cases a hit in the silicon detectors is matched with
a cluster and its coordinate is used;
otherwise the BGO coordinate is retained.

\begin{figure}[htbp]
 \begin{center}
 \vspace*{-10mm}%
 \includegraphics[width=\figwidth]{\pltdir 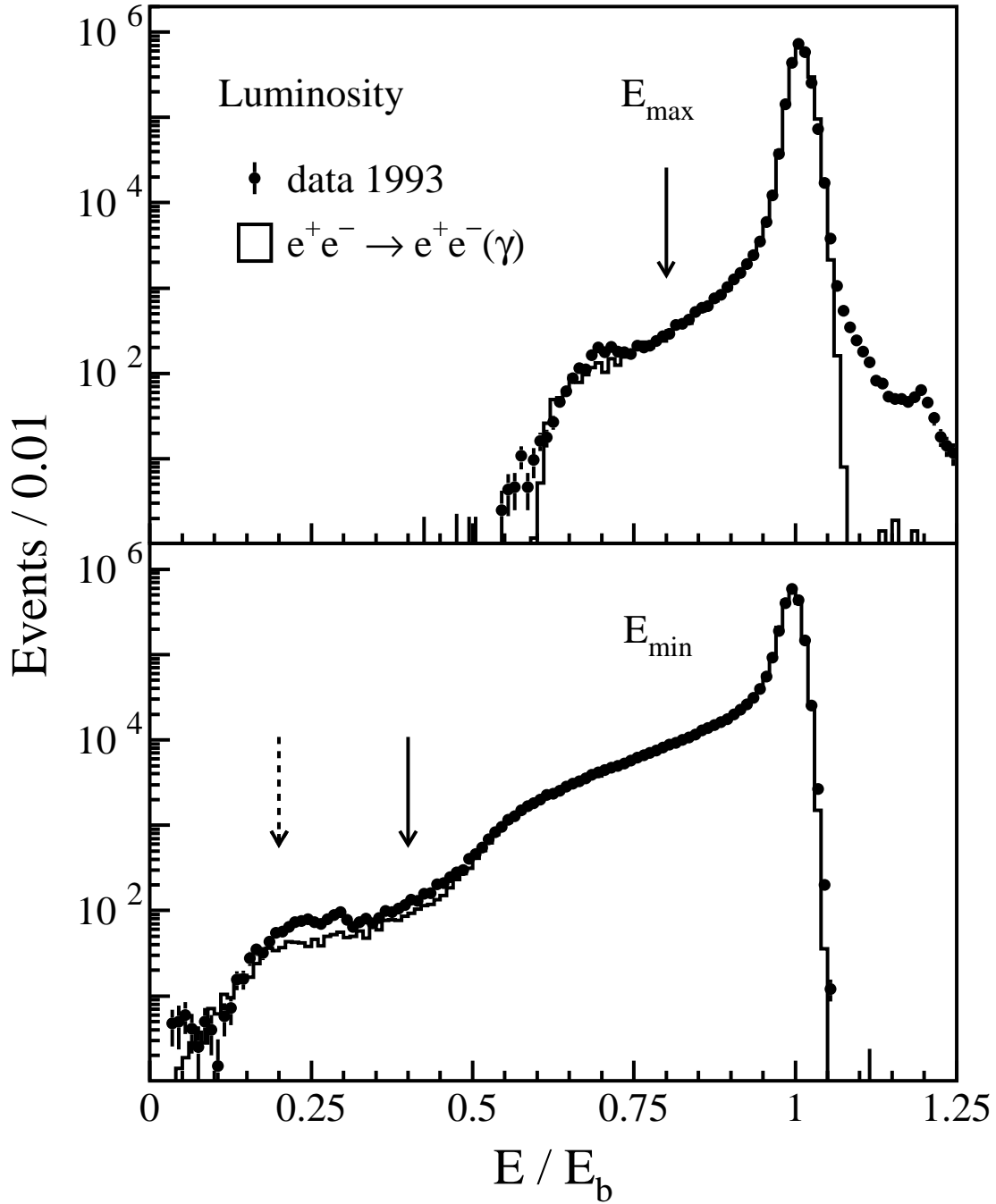}
 \end{center}
 \caption{The distributions of the energies measured in the 
          luminosity detectors for small angle Bhabha candidates in 1993.
          The top plot contains the most energetic cluster, $E_{\MR{max}}$,
          and the lower plot shows the energy of the cluster on the
          opposite side. 
          All selection cuts except the one under study are applied.
          In this and the following figures, the dots are the data and 
          the histograms represent Monte Carlo simulations.
          The vertical arrows indicate the positions of the selection
          cuts (see text).
         }
 \label{fig:lumi_eminmax}
\end{figure}

\begin{figure}[htbp]
 \begin{center}
 \includegraphics[width=\figwidth]{\pltdir 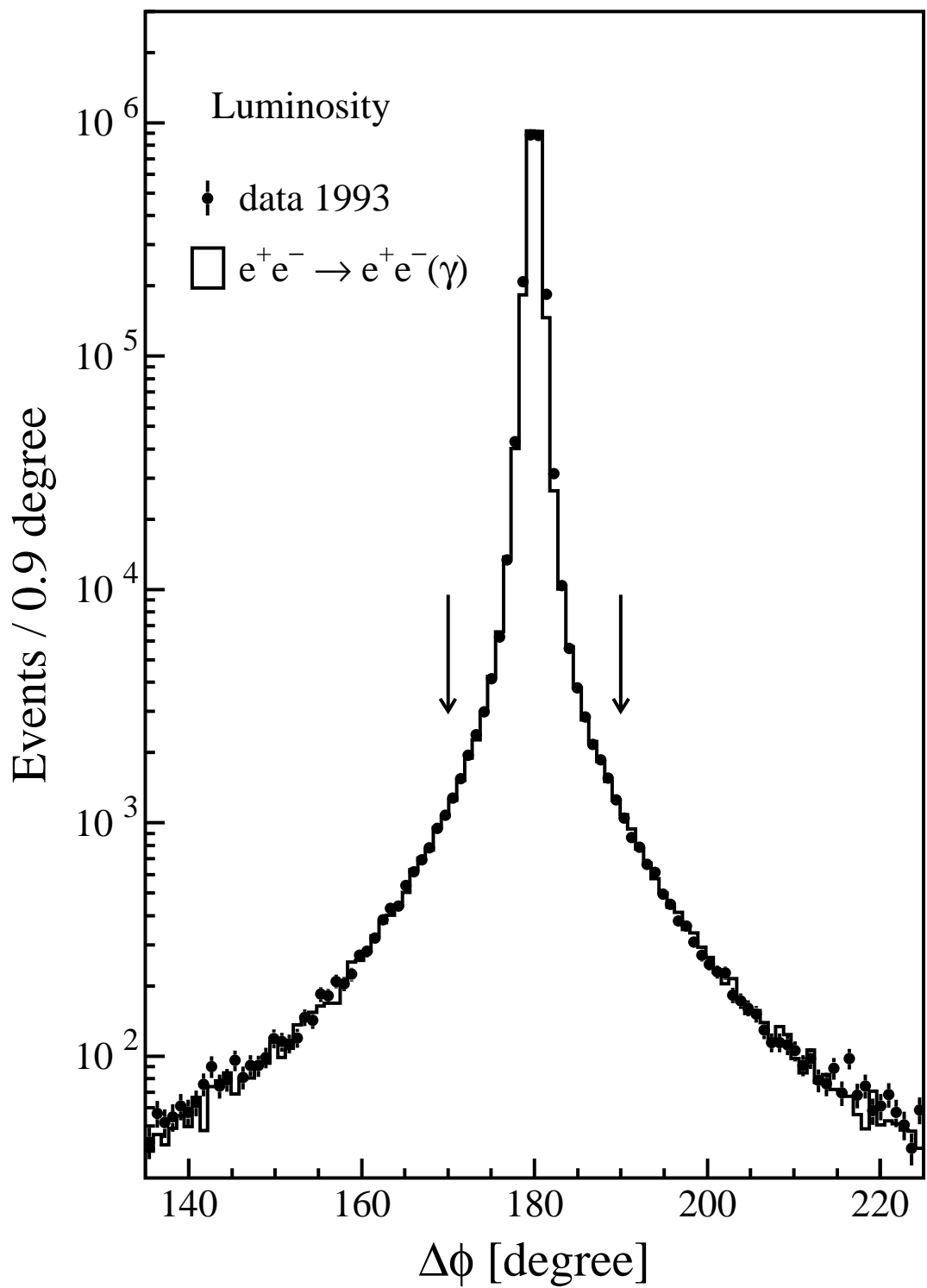}
 \end{center}
 \caption{The distribution of the coplanarity angle $\Delta\phi$
          for small angle Bhabha candidates.
         }
 \label{fig:lumi_acop}
\end{figure}

The event selection criteria are:
\begin{enumerate}
 \item The energy of the most energetic cluster is required to exceed 
       $0.8\Ebeam$ and the energy on the opposite side must 
       be greater than
       $0.4\Ebeam$, where \Ebeam\ is the beam energy.
       If the energy of the most energetic cluster is within $\pm 5\%$
       of \Ebeam\ the minimum energy requirement on the opposite side is 
       reduced to $0.2\Ebeam$ in order to recover events with energy lost 
       in the gaps between crystals.
       The distributions of the energy of the most energetic cluster and 
       the cluster on the opposite side as measured in the luminosity
       monitors are shown in Figure~\ref{fig:lumi_eminmax} for the 1993 data.
       All selection cuts except the one under study are applied.
\item The cluster on one side must be confined to a tight fiducial volume:
       \begin{itemize}
        \item 32 mrad $< \theta <$ 54 mrad;
              $|\phi - 90^{\circ}| > 11.25^{\circ}$ and 
              $|\phi - 270^{\circ}| > 11.25^{\circ}$.
       \end{itemize}
       The requirements on the azimuthal angle remove the regions where the
       half-rings of the detector meet.
       The cluster on the opposite side is required to be within a larger
       fiducial volume:
       \begin{itemize}
        \item 27 mrad $< \pi - \theta <$ 65 mrad;
              $|\phi - 90^{\circ}| > 3.75^{\circ}$ and 
              $|\phi - 270^{\circ}| > 3.75^{\circ}$.
       \end{itemize}
       This ensures that the event is fully contained in the detectors
       and edge effects in the reconstruction are avoided.
 \item The coplanarity angle $\Delta\phi = \phi(z<0) - \phi(z>0)$
       between the two clusters must 
       satisfy $|\Delta\phi - 180^{\circ}| < 10^{\circ}$.
\end{enumerate}
The distribution of the coplanarity angle is shown in Figure~\ref{fig:lumi_acop}.
Very good agreement with the Monte Carlo simulation is observed.

Four samples of Bhabha events are defined by applying the tight fiducial 
volume cut to one of the $\theta$-measuring silicon layers.
Taking the average of the luminosities obtained from these samples minimizes
the effects of relative offsets between the interaction point and the detectors.
The energy and coplanarity cuts reduce the background from random beam-gas
coincidences. 
The remaining contamination is very small: $(3.4 \pm 2.2) \cdot 10^{-5}$.
This number is estimated using the sidebands of the coplanarity distribution,
$10^{\circ} < |\Delta\phi -180^{\circ}| < 30^{\circ}$,
after requiring that neither of the two clusters have an energy within 
$\pm5\%$ of \Ebeam.

The accepted cross section is determined from Monte Carlo 
\EEEEG\ samples generated with the  BHLUMI event
generator at a fixed \cms\ energy of $\RS = 91.25\ \GeV$.
The dependence on the \cms\ energy, as well as the contributions of 
\Zo-exchange and  $\Photon\Zo$ interference, are calculated with the 
BHLUMI program.
At $\RS = 91.25\ \GeV$ the accepted cross section is determined to be
$69.62\ \nb$.
The statistical error on the Monte Carlo sample contributes $0.35$\permille\ to
the uncertainty of the luminosity measurement.
The theoretical uncertainty on the Bhabha cross section in our fiducial volume
is estimated to be $0.61$\permille~\cite{Ward_lumi}.

The experimental errors of the luminosity measurement are small.
Important sources of systematic errors are:
geometrical uncertainties due to the internal alignment of the silicon 
detectors ($0.15$\permille\ to $0.27$\permille),   
temperature expansion effects ($0.14$\permille) 
and the knowledge on the longitudinal position of the silicon detectors
($0.16$\permille\ to $0.60$\permille). 
The precision depends on the accuracy of the detector surveys and on the
stability of the detector and wafer positions during the different years.

\begin{figure}[htbp]
 \begin{center}
 \includegraphics[width=\figwidth]{\pltdir 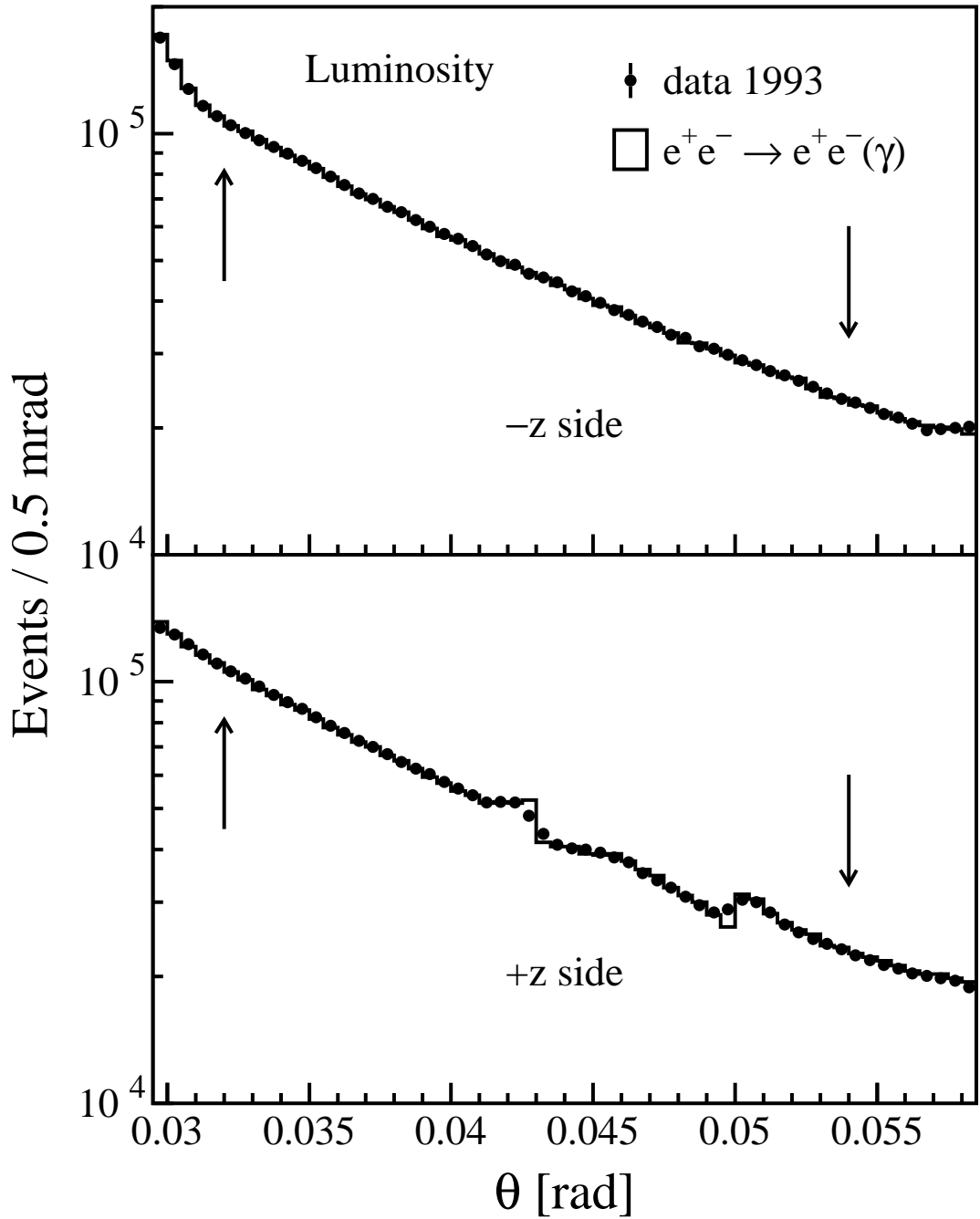}
 \end{center}
 \caption{The polar angle distribution of small angle Bhabha events
          used for luminosity measurement as observed in the two 
          detectors at $-z$ and $+z$.
          The structure seen in the central part of the $+z$ side is due 
          to the flare in the beam pipe on this side.
         }
 \label{fig:lumi_theta}
\end{figure}

The polar angle distribution of Bhabha scattering events used for the luminosity
measurement is shown in Figure~\ref{fig:lumi_theta}.
The structure seen in the central part of the $+z$ side is due to the
flare in the beam pipe on this side.
The imperfect description in the Monte Carlo does not pose any problem as it
is far away from the edges of the fiducial volume.

The overall agreement between the data and Monte Carlo distributions of the 
selection quantities is good. 
Small discrepancies in the energy distributions at high energies are due 
to contamination of Bhabha events with beam-gas interactions and, 
at low energies, due to an imperfect description of the cracks between 
crystals.
The selection uncertainty is estimated by varying the selection criteria over
reasonable ranges and summing in quadrature the resulting contributions. 
This procedure yields errors between $0.42$\permille\ and $0.48$\permille\ for
different years. 
The luminosities determined from the  four samples described above agree
within these errors.
The trigger inefficiency is measured using a sample of events triggered 
by only requiring an energy deposit exceeding $30\ \GeV$ on one side.
It is found to be negligible. 

The various sources of uncertainties are summarized in 
Table~\ref{tab:lumi_sys}. 
Combining them in quadrature yields total experimental errors on the 
luminosity of 
$0.86$\permille, $0.64$\permille\ and $0.68$\permille\
in 1993, 1994 and 1995.
Correlations of the total experimental systematic errors between different 
years are studied and the correlation matrix is given in 
Table~\ref{tab:lumi_correlations}.
The error from the theory is fully correlated.

Because of the $1/s$ dependence of the small angle Bhabha cross section,
the uncertainty on the \cms\ energies causes a small additional uncertainty on 
the luminosity measurement.
For instance, this amounts to $0.1$\permille\ for the high statistics data 
sample of 1994.
This effect is included in the fits performed in Section~\ref{sec:fit_Zpara} 
and~\ref{sec:fit_SMpara}, see Appendix~\ref{app:covariance}.

The statistical error on the luminosity measurement from the number of
observed small angle Bhabha events is also included in those fits.
Table~\ref{tab:lumi_statistics} lists the number of observed  Bhabha events
for the nine data samples and the corresponding errors on cross section 
measurements.
Combining all data sets taken in $1993-95$ at $\RS \approx \MZ$ the 
statistical error on the luminosity contributes 0.45\permille\ to the 
uncertainty on the pole cross section measurements.

\begin{figure}[htbp]
 \begin{center}
 \includegraphics[width=\figwidth]{\pltdir 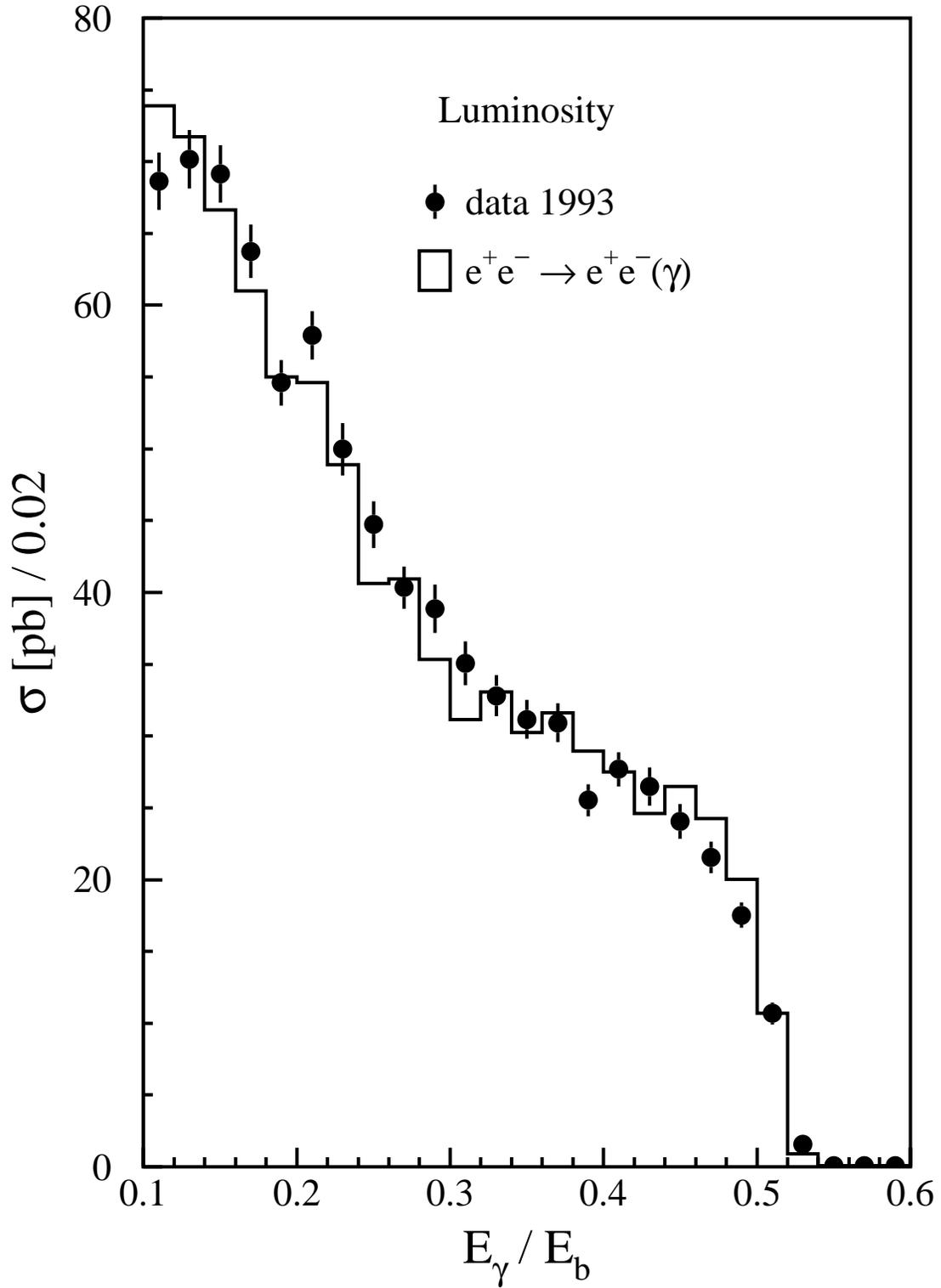}
 \end{center}
 \caption{The distribution of the photon energy, $E_\Photon$, 
          as measured in the luminosity monitors, normalized to the beam energy,
          \Ebeam.
         }
 \label{fig:lumi_photon}
\end{figure}

Higher order corrections from photon radiation to the small angle Bhabha cross 
section are studied with the photon spectrum of luminosity events.
For this analysis events with two distinct energy clusters exceeding
$0.1\Ebeam$ in one of the calorimeters are selected.
The photon is identified as the lower energy cluster.
The fraction of radiative events with $E_\gamma >0.1\Ebeam$ 
in the total low-angle Bhabha sample is 2\% and the measured cross section, 
normalised to the expectation, is found to be $0.993 \pm 0.16$.
The observed spectrum from 1993 is shown in Figure~\ref{fig:lumi_photon} and good
agreement is found with the Monte Carlo expectation.

%
%
\section{\protect\boldmath\EEHADG}
         \label{sec:hadrons}

\subsubsection*{Event Selection}

Hadronic \Zo\ decays are identified by their large energy deposition and 
high multiplicity in the electromagnetic and hadron calorimeters.
The selection criteria are similar to those applied in our previous 
analysis~\cite{l3-69}:
\begin{enumerate}
  \item The total energy observed in the detector, \Evis,
        normalised to the \cms\ energy must satisfy
        $0.5 < \Evis/\RS < 2.0$;
  \item The energy imbalance along the beam direction, \Elong,
        must satisfy
        $|\Elong|/\Evis < 0.6$;
  \item The transverse energy imbalance, \Etrans,
        must satisfy
        $\Etrans/\Evis < 0.6$;
  \item The number of clusters, $N_{\mathrm{cl}}$, formed from energy 
        depositions in the calorimeters is required to be: \\
        a) $N_{\mathrm{cl}} \geq 13$ for $|\cos\theta_t| \leq 0.74$
           (barrel region), \\
        b) $N_{\mathrm{cl}} \geq 17$ for $|\cos\theta_t| > 0.74$
           (end-cap region), \\
           where $\theta_t$ is the polar angle of the event thrust axis.
\end{enumerate}

\begin{figure}[htbp]
 \begin{center}
 \includegraphics[width=\figwidth]{\pltdir 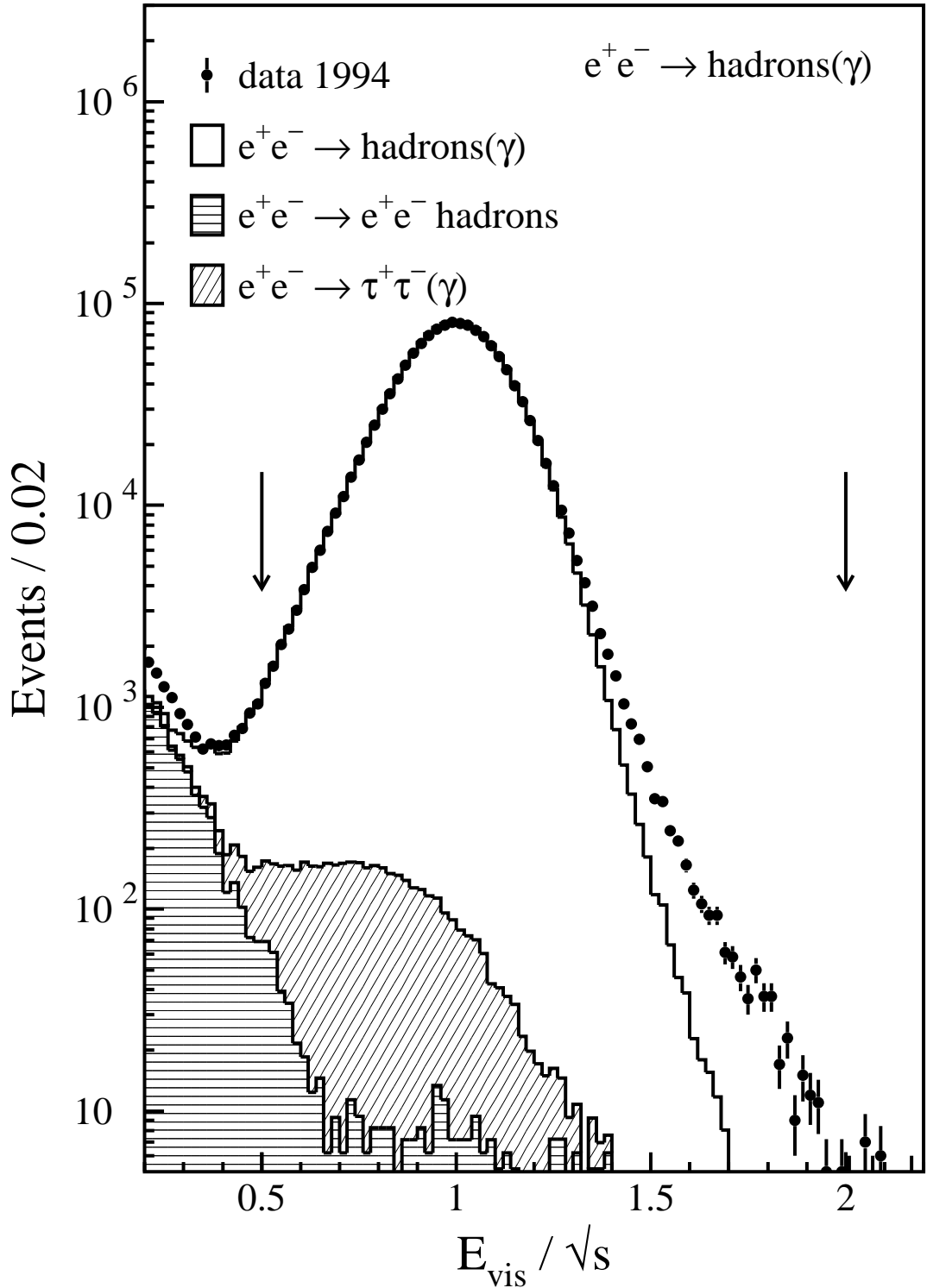}
 \end{center}
 \caption{The distribution of the visible energy normalized to the
          \protect\cms\ energy for \protect\EEHADG\ candidates
          collected in 1994.
          In this and the following figures the data are presented
          as dots, the Monte Carlo simulations of the signal as open
          and of the different background sources as shaded histograms.
         }
 \label{fig:hadron_evis}
\end{figure}

\begin{figure}[htbp]
 \begin{center}
 \includegraphics[width=\figwidth]{\pltdir 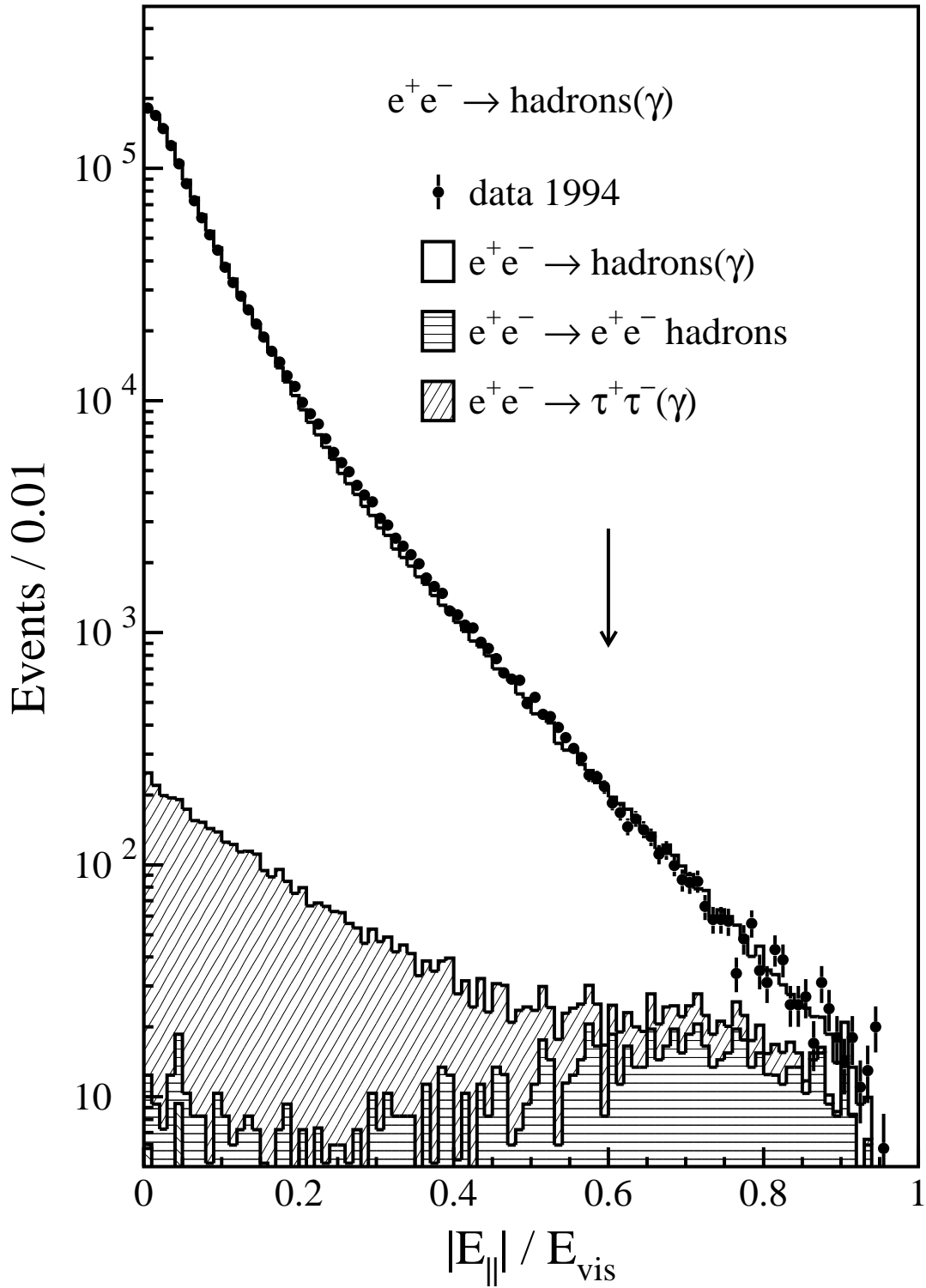}
 \end{center}
 \caption{The distribution of the longitudinal energy imbalance
          for \protect\EEHADG\ candidates.
         }
 \label{fig:hadron_long}
\end{figure}

\begin{figure}[htbp]
 \begin{center}
 \includegraphics[width=\figwidth]{\pltdir 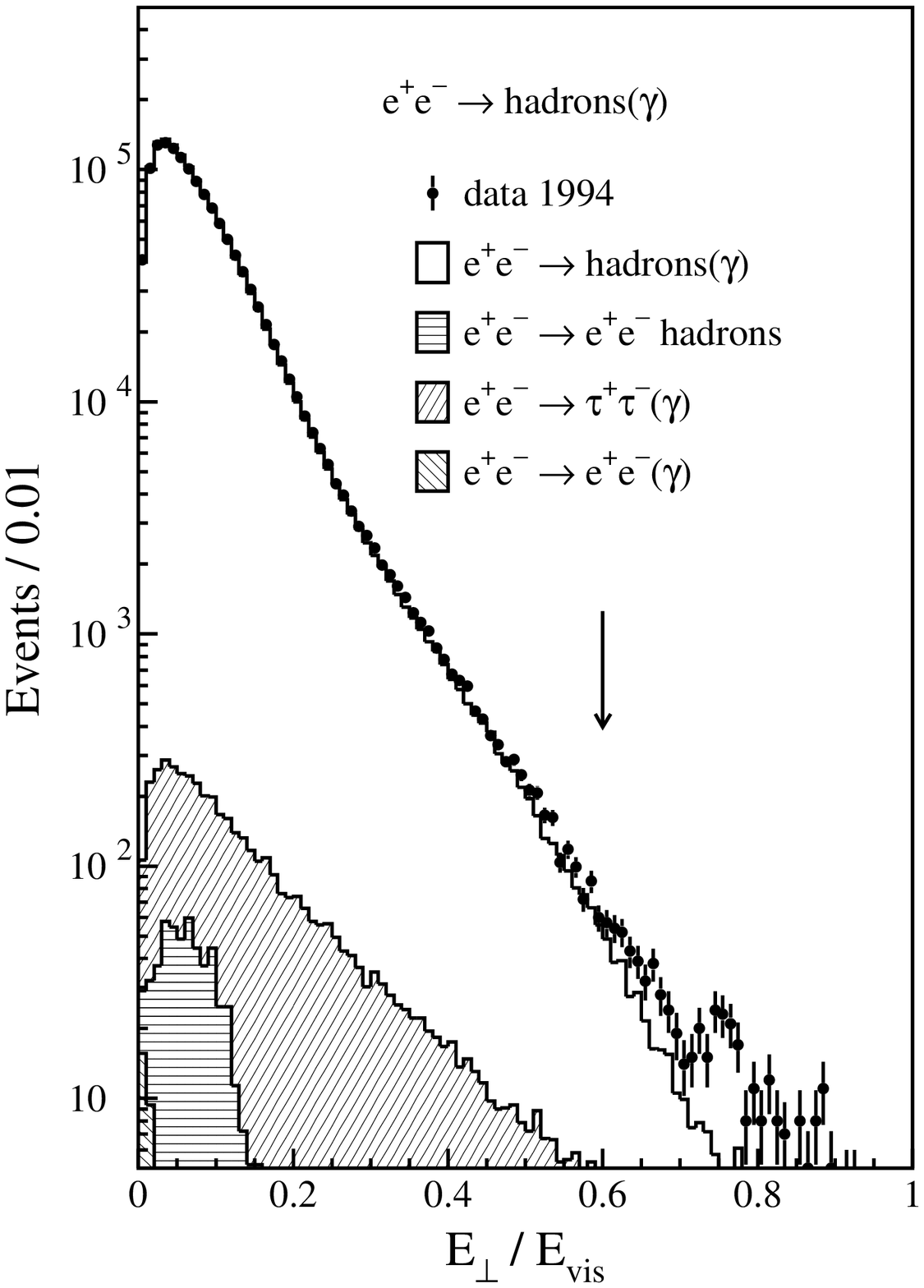}
 \end{center}
 \caption{The distribution of the transverse energy imbalance
          for \protect\EEHADG\ candidates.
         }
 \label{fig:hadron_trans}
\end{figure}

\begin{figure}[htbp]
 \begin{center}
 \includegraphics[width=\figwidth]{\pltdir 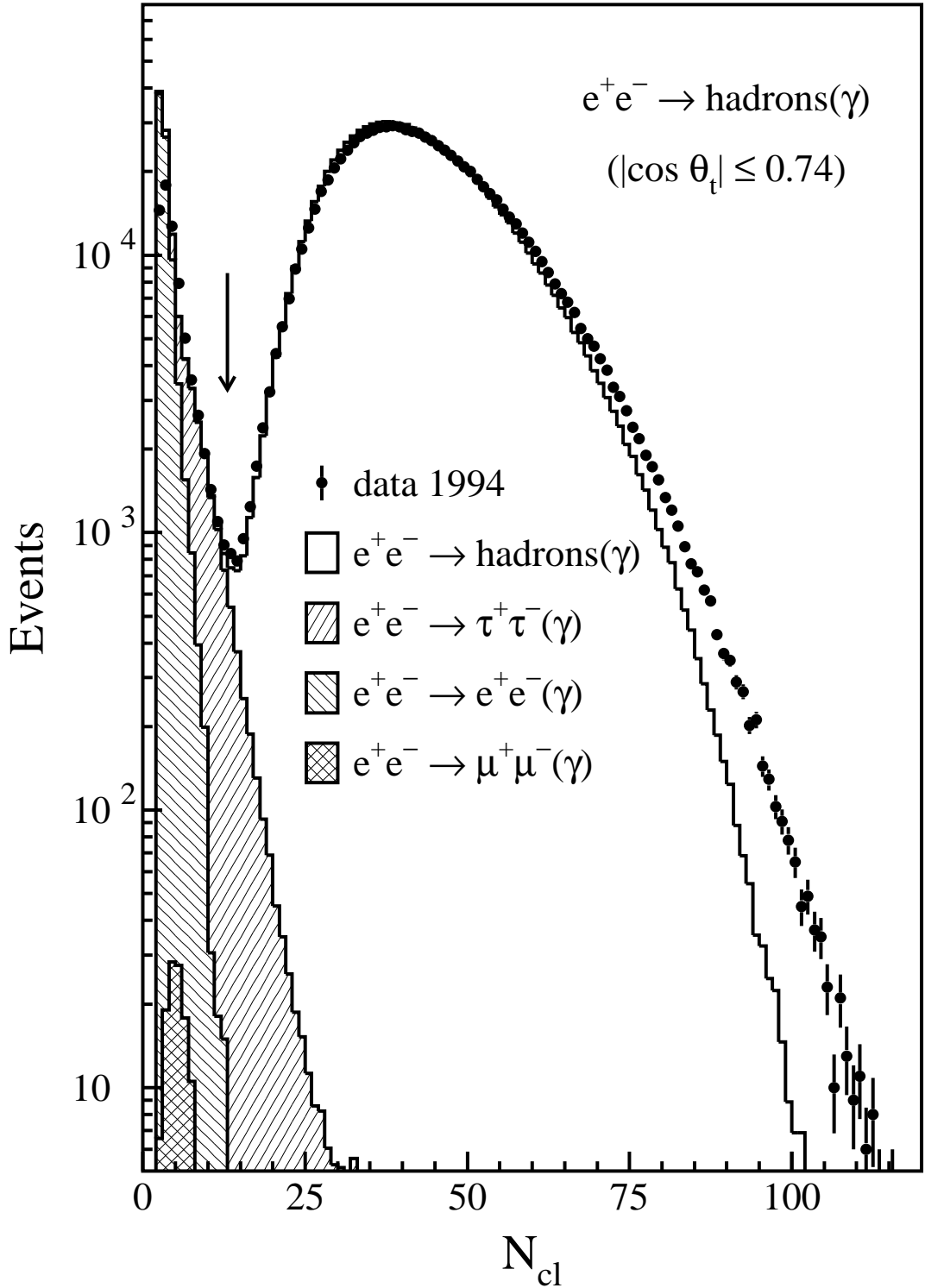}
 \end{center}
 \caption{The distribution of the number of energy clusters in the
          calorimeters for \protect\EEHADG\ candidates in the
          barrel region ($|\cos\theta_t| \leq 0.74$).
         }
 \label{fig:hadron_ncba}
\end{figure}

\begin{figure}[htbp]
 \begin{center}
 \includegraphics[width=\figwidth]{\pltdir 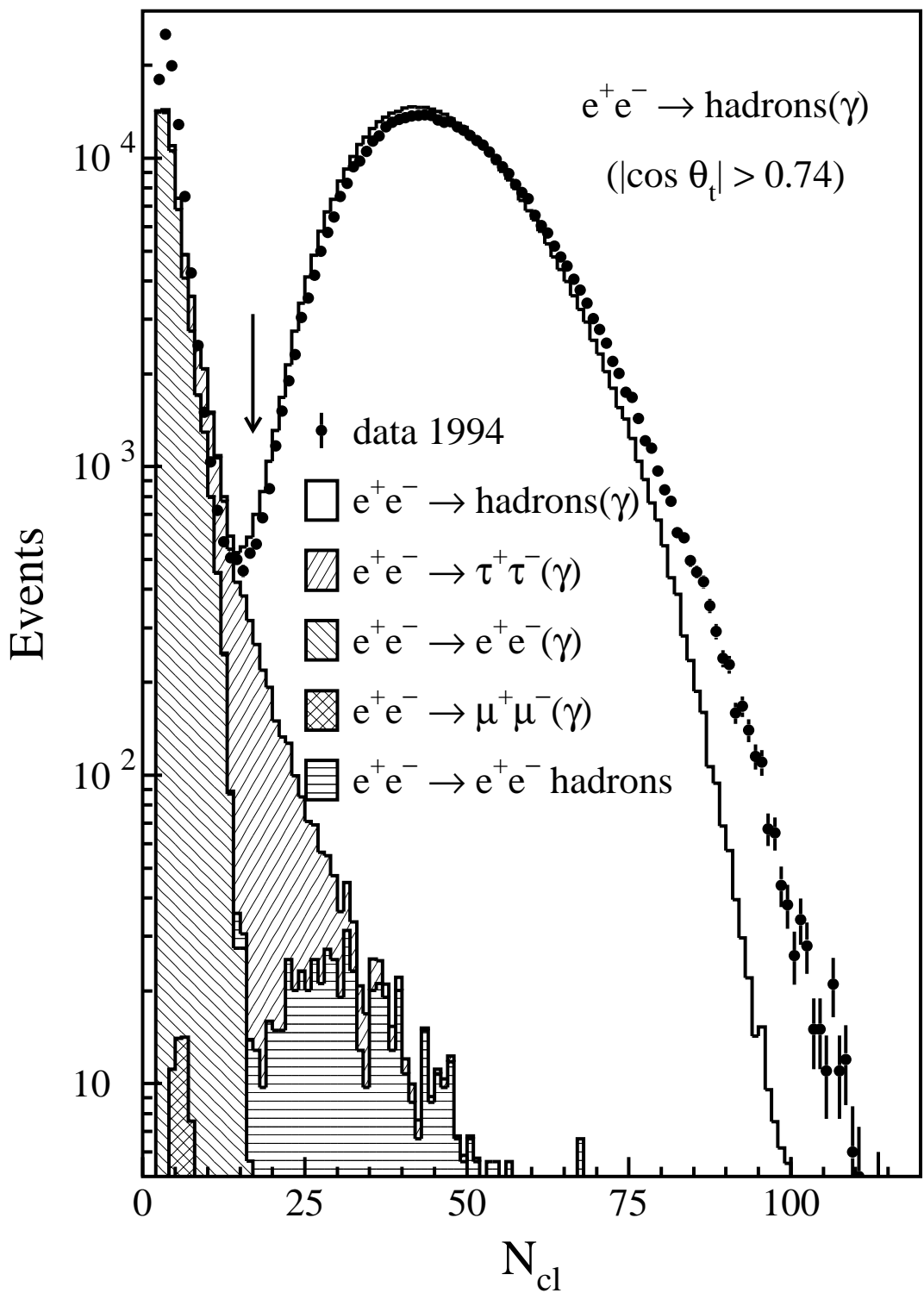}
 \end{center}
 \caption{Same as Figure~\protect\ref{fig:hadron_ncba} for
          events in the endcap region ($|\cos\theta_t| > 0.74$).
         }
 \label{fig:hadron_ncec}
\end{figure}

Detailed analyses of the large data samples collected have been used to improve 
the Monte Carlo simulation of the detector response.
Figures \ref{fig:hadron_evis} to~\ref{fig:hadron_ncec}
show the distributions of the quantities used to 
select hadronic \Zo\ decays and the comparisons to the Monte Carlo
predictions.
In these plots all selection cuts are applied, except the one under study.
Good agreement is observed between our data and the Monte Carlo 
simulations.

\subsubsection*{Total Cross Section}

The acceptance for \EEHADG\ events is determined from large samples of 
Monte Carlo events generated with the JETSET program.
Applying the selection cuts, between $99.30\%$ and $99.42\%$
of the events are accepted depending on the year of the data taking
and on differences in initial-state photon radiation at the various 
\cms\ energies.
Monte Carlo events are generated with $\RSprime > 0.1 \RS$
where \RSprime\ is the effective \cms\ energy after initial state photon 
radiation.
The acceptance for events in the data with $\RSprime \leq 0.1 \RS$ is 
estimated to be negligible.
They are not considered as part of the signal and hence not corrected for.

The interference between initial and final state photon radiation is not
accounted for in the event generator.
This effect modifies the angular distribution of the events in 
particular at very low polar angles where the detector inefficiencies are 
largest.
However, the error from the imperfect simulation on the measured cross section,
which includes initial-final state interference as part of the signal,
is estimated to be very small ($\ll 0.1\ \pb$) in the \cms\ energy range
considered here.
Quark pairs originating from pair production from initial state radiation
are considered as part of the signal if their invariant mass exceeds 
50\% of \RS.

To estimate the uncertainty on the acceptance on the modelling of
the quark fragmentation, the determination of the acceptance is repeated
using the HERWIG program.
The detector simulations of both Monte Carlo programs are tuned in the same
way to describe as closely as possible our data, e.g. in terms of energy 
resolution and cluster multiplicity.
The remaining difference in acceptance is $0.42$\permille\ and we assign half 
of it as an estimate of the uncertainty on the acceptance of \EEHADG\ events
due to the modelling of quark fragmentation.
Differences of the implementation of QED effects in both programs are 
studied and found to have negligible impact on the acceptance.

Hadronic \Zo\ decays are triggered by the energy, central track, muon
or scintillation counter multiplicity triggers.
The combined trigger efficiency is obtained from the fraction of events with 
one of these triggers missing as a function of the polar angle of the 
event thrust axis.
This takes into account most of the correlations among 
triggers.
A sizeable inefficiency is only observed for events in the very forward region 
of the detector, where hadrons can escape through the beam pipe.
Trigger efficiencies, including all steps of the trigger system,
between $99.829\%$ and $99.918\%$ are obtained for the various data 
sets.
Trigger inefficiencies determined for data sets taken in the same year 
are statistically compatible.
Combining those data sets results in statistical errors of at most
$0.12$\permille\ which is assigned as systematic error to all data 
sets.

The background from other \Zo\ decays is found to be small: 
$2.9$\permille\ essentially only from \EETTG.
The uncertainty on this number is negligible compared to the total 
systematic error.

The determination of the non-resonant background, mainly 
$\ee\rightarrow \ee\,\MR{hadrons}$, is based on the measured distribution 
of the visible energy shown in  Figure~\ref{fig:hadron_evis}.
The Monte Carlo program PHOJET is used to simulate two-photon
collision processes.
The absolute cross section is derived by scaling the Monte Carlo to obtain the
best agreement with our data in the low end of the \Evis\ spectrum:
$0.32 \leq \Evis/\RS \leq 0.44$.
Consistently for all data sets, scale factors of $1.1$ are necessary.
In the signal region contaminations from $\ee\rightarrow \ee\,\MR{hadrons}$
between $11.6\ \pb$ and $13.0\ \pb$ 
are obtained for the different data sets.
No dependence on \RS\ is observed.
This is in agreement with results of a similar calculation performed
with the DIAG36 program.

Beam related background (beam-gas and beam-wall interactions) is small.
To the extent that the \Evis\ spectrum is similar to that of 
$\ee\rightarrow \ee\,\MR{hadrons}$, it is accounted for by determining the 
absolute normalisation from the data.

As a check, the non-resonant background is estimated by extrapolating an 
exponential dependence of the \Evis\ spectrum from the low energy part into
the signal region. 
This method yields consistent results.
Based on these studies we assign an error on the measured hadron cross 
section of $3\ \pb$ due to the understanding of the non-resonant background.
This error assignment is supported by our measurements of the hadronic
cross section at high energies ($130\ \GeV \leq \RS \leq 172\ \GeV$)
where the relative contribution of two-photon processes is much
larger~\cite{l3-90,l3-117}.
The extrapolation of these studies back to the \Zo\ peak yields a similar
result for the uncertainty.

\begin{figure}[htbp]
 \begin{center}
 \includegraphics[width=\figwidth]{\pltdir 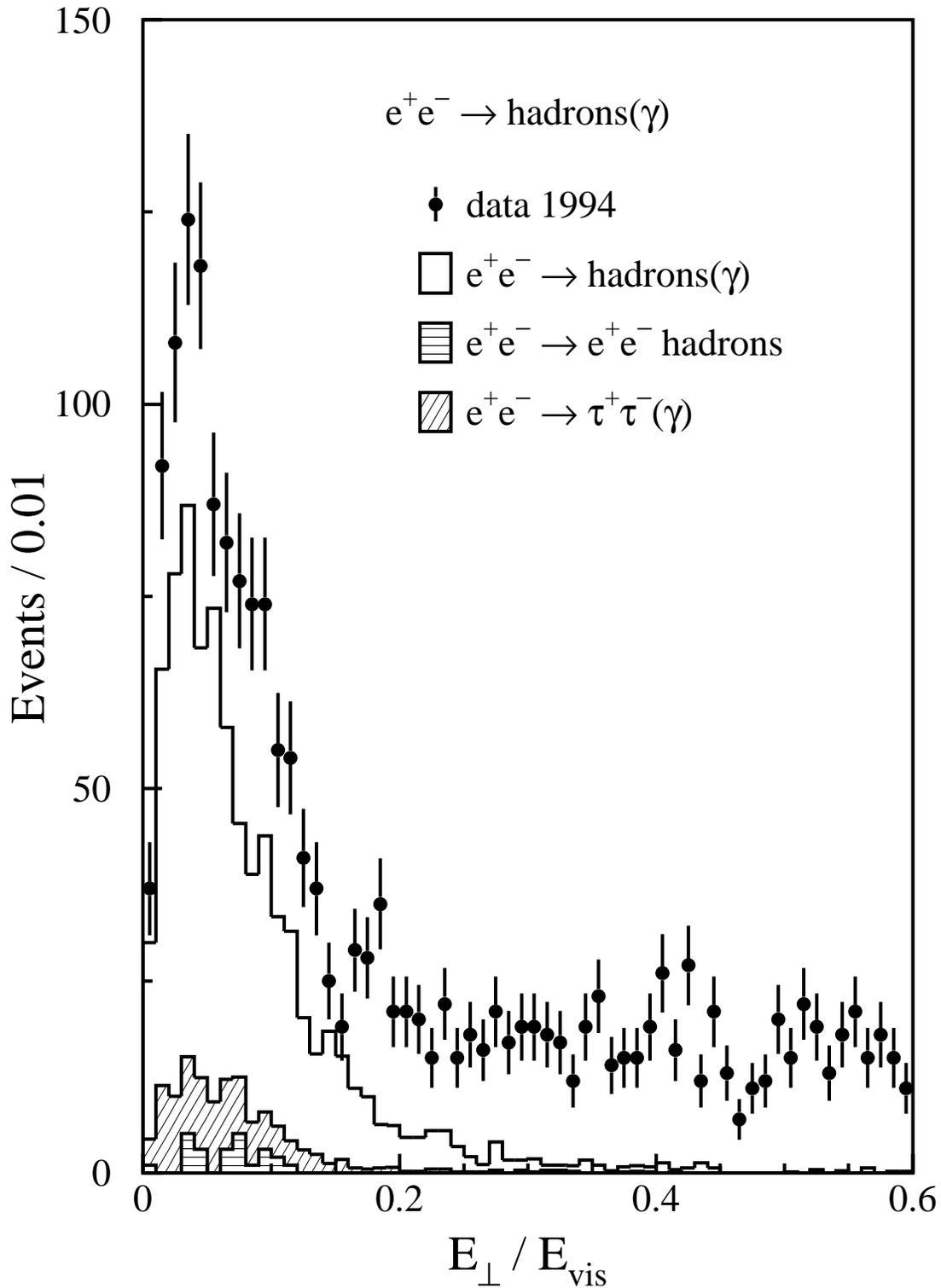}\vspace*{-5mm}
 \end{center}
 \caption{Measurement of the contamination from uranium noise and electronic
          noise in the \protect\EEHADG\ sample of 1994:
          The figure shows the transverse energy imbalance for event candidates
          with most of the energy observed either only in the electromagnetic
          or hadron calorimeter and with little matching between tracks in
          the \protect\TEC\ and the energy deposits.
         }
 \label{fig:hadron_noise}
\end{figure}

The contribution of random uranium noise and electronic noise in the detector
faking a signal event is determined from a subsample of the event 
candidates.
This subsample is obtained requiring that most of the observed energy stems
either from the electromagnetic or the hadron calorimeter and that 
there be little matching between individual energy deposits and
tracks.
The $\Etrans/\Evis$ distribution of this subsample shows an \EEHADG\ signal 
over a flat background (see Figure~\ref{fig:hadron_noise} for the 1994 data).
This background is consistent with a constant noise rate, from
which a background correction of 
$7.4\ \pb$ is derived.
An uncertainty of $1\ \pb$ on the hadron cross section is assigned to all
data sets from this correction. 
The absolute normalisation of the \EEHADG\ signal in
Figure~\ref{fig:hadron_noise} is not expected to be perfectly reproduced
by the Monte Carlo simulation.
However, this does not pose a serious problem as the noise rate
is determined from the tail of the spectrum.

The systematic error from event selection on the measured cross sections
is estimated by varying the selection cuts.
All cross section results are stable within $\pm 0.3$\permille.
The systematic errors to the cross section measurements \EEHADG\ are 
summarised in Table~\ref{tab:had_sys}.
Uncertainties which scale with the cross section and absolute uncertainties
are separated because they translate in a different way into errors on 
\Zo\ parameters, in particular on the total width.
The scale error is further split into a part uncorrelated among the data
samples, in this case consisting of the contribution of Monte Carlo statistics,
and the rest which is taken to be fully correlated and amounts to 
$0.39$\permille.

The results of the \EEHADG\ cross section measurements are discussed in
Section~\ref{sec:xsafbresults}.

%
%
%
\section{\protect\boldmath\EEMMG}\label{sec:muons}

\subsubsection*{Event Selection}

The selection of \EEMMG\ in the 1993 and 1994 data is similar to the selection 
applied in previous years described in Reference~\cite{l3-69}.
Two muons in the polar angular region $|\cos\theta| < 0.8$ are
required.
Most of the muons, 88\%, are identified by a reconstructed track in the muon 
spectrometer.
Muons are also identified by their minimum ionising particle (MIP)
signature in the inner sub-detectors, if less than two muon chamber layers
are hit.
A muon candidate is denoted as a MIP, if at least one of the following 
conditions is fulfilled:
\begin{enumerate}
  \item A track in the central tracking chamber must point within 
        $5^\circ$ in azimuth to a cluster in the electromagnetic calorimeter 
        with an energy less than 2~\GeV.
  \item On a road from the vertex through the barrel hadron calorimeter, 
        at least five out of a maximum of 32 cells must be hit, with an average 
        energy of less than 0.4~\GeV\ per cell.
  \item A track in the central chamber or a low energy electromagnetic 
        cluster must point within $10^\circ$ in azimuth to a muon chamber hit.
\end{enumerate}
In addition, both the electromagnetic and the hadronic energy in a 
cone of $12^\circ$ half-opening angle around the MIP candidate, corrected 
for the energy loss of the particle, must be less than 5~\GeV.

Events of the reaction \EEMMG\ are selected by the following criteria: 
\begin{enumerate}
 \item The event must have a low multiplicity in the calorimeters 
       $N_{\mathrm{cl}} \leq 15$. 
 \item If at least one muon is reconstructed in the muon chambers,
       the maximum muon momentum must satisfy
       $p_{\mathrm{max}} > 0.6\,\Ebeam$.
       If both muons are identified by their MIP signature there must be two 
       tracks in the \TEC\ with at least one with
       a transverse momentum larger than $3\ \GeV$.
 \item The acollinearity angle $\xi$ must be less than $90^\circ$, 
       $40^\circ$ or $5^\circ$ if two, one or no muons are reconstructed 
       in the muon chambers.
 \item The event must be consistent with an origin of an
       \ee-interaction requiring at least one time measurement of a 
       scintillation counter, associated to a muon candidate, to coincide
       within $\pm3\ \mathrm{ns}$ with the beam crossing.
       Also, there must be a track in the \TEC\  with a distance of closest 
       approach to the beam axis of less than $5\ \mathrm{mm}$.
\end{enumerate}

\begin{figure}[htbp]
 \begin{center}
  \includegraphics[width=\figwidth]{\pltdir 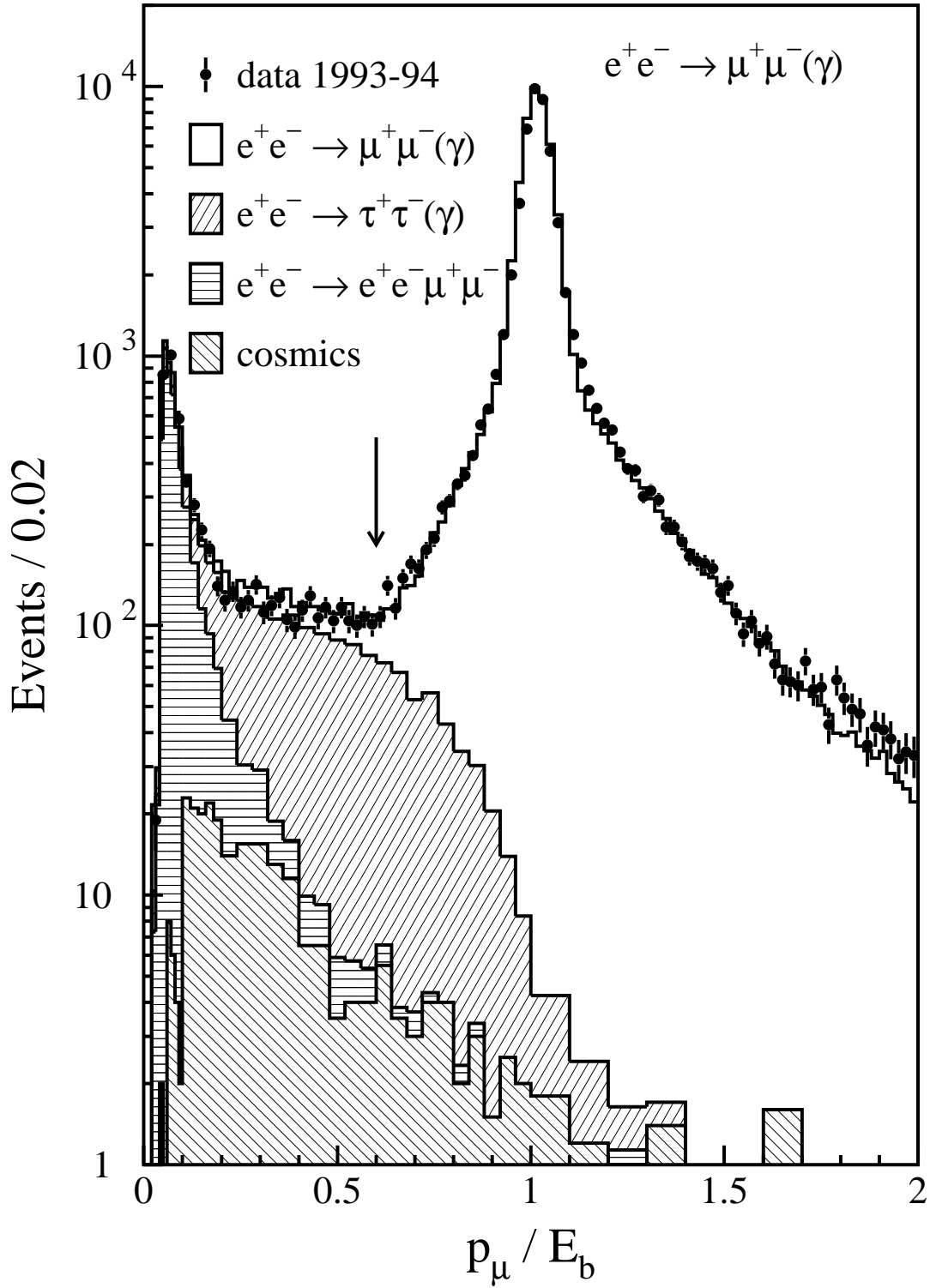}
 \end{center}
 \caption{The distribution of the maximum momentum measured in the
          muon chambers, $p_\mu$,  normalized to the beam energy
          for \protect\EEMMG\ candidates collected in $1993-94$.
          The contribution from cosmic ray muons is determined from our
          data as described in the text.
         }
 \label{fig:muon_pmax}
\end{figure}

\begin{figure}[htbp]
 \begin{center}
  \includegraphics[width=\figwidth]{\pltdir 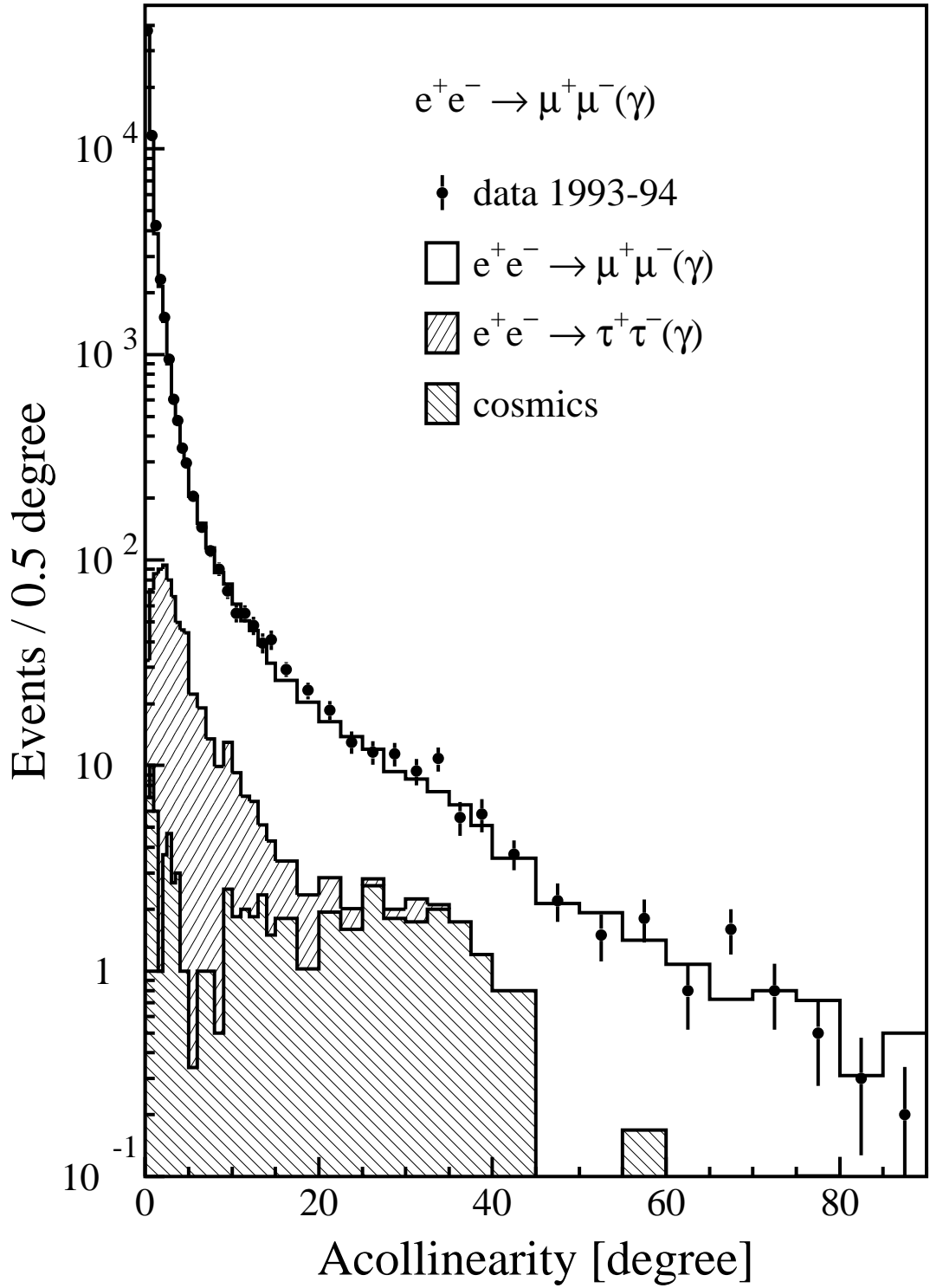}
 \end{center}
 \caption{The distribution of the acollinearity angle for 
          \protect\EEMMG\ candidates collected in $1993-94$.
         }
 \label{fig:muon_acol}
\end{figure}

As an example, Figure~\ref{fig:muon_pmax} shows the distribution of the
maximum measured muon momentum for candidates in the $1993-94$ data compared
to the expectation for signal and background processes.
The acollinearity angle distribution of the selected muon pairs is shown in
Figure~\ref{fig:muon_acol}.
The experimental angular resolution and radiation effects are well reproduced
by the Monte Carlo simulation.

\begin{figure}[htbp]
 \begin{center}
  \includegraphics[width=\figwidth]{\pltdir 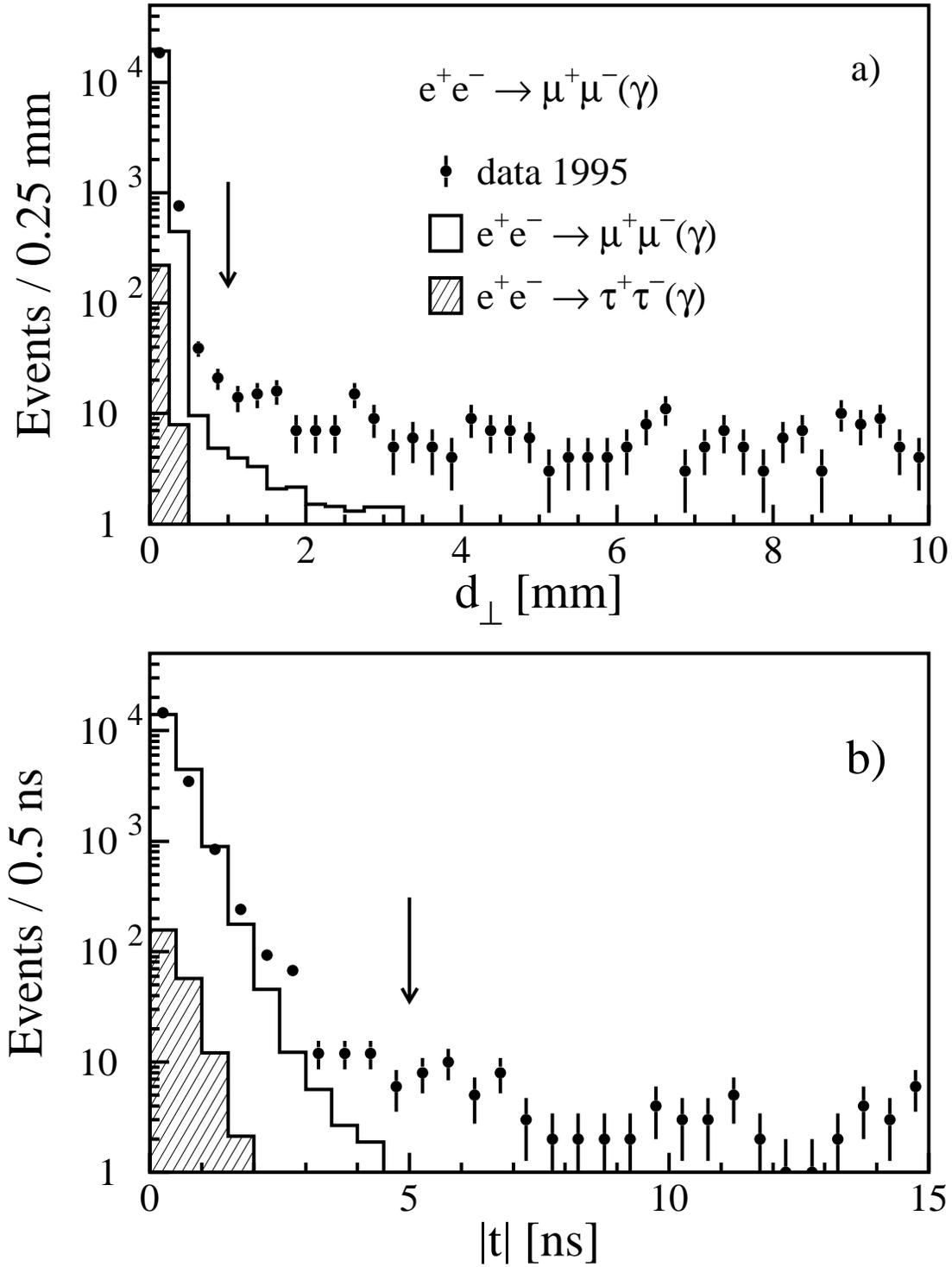}
 \end{center}
 \caption{Rejection of cosmic ray muons in the 1995 data: 
          Figure a) shows the distribution of the distance of closest 
          approach to the beam axis, $d_\perp$.
          Figure b) shows the absolute value of the time, $|t|$, closest
          to the beam crossing as measured by the scintillation counters 
          associated with the muon candidates.
         }
 \label{fig:muon_cosmic}
\end{figure}

The analysis of the 1995 data in addition uses the newly installed 
forward-backward muon chambers. 
The fiducial volume is extended to $|\cos\theta |<$ 0.9. 
Each event must have at least one track in the \TEC\ with a distance of 
closest approach in the transverse plane of less than $1\ \mathrm{mm}$ and 
a scintillation counter time coinciding within $\pm5\ \mathrm{ns}$ with the 
beam crossing.
The rejection of cosmic ray muons in the 1995 data is illustrated in 
Figure~\ref{fig:muon_cosmic}.

\begin{figure}[htbp]
 \begin{center}
  \includegraphics[width=\figwidth]{\pltdir 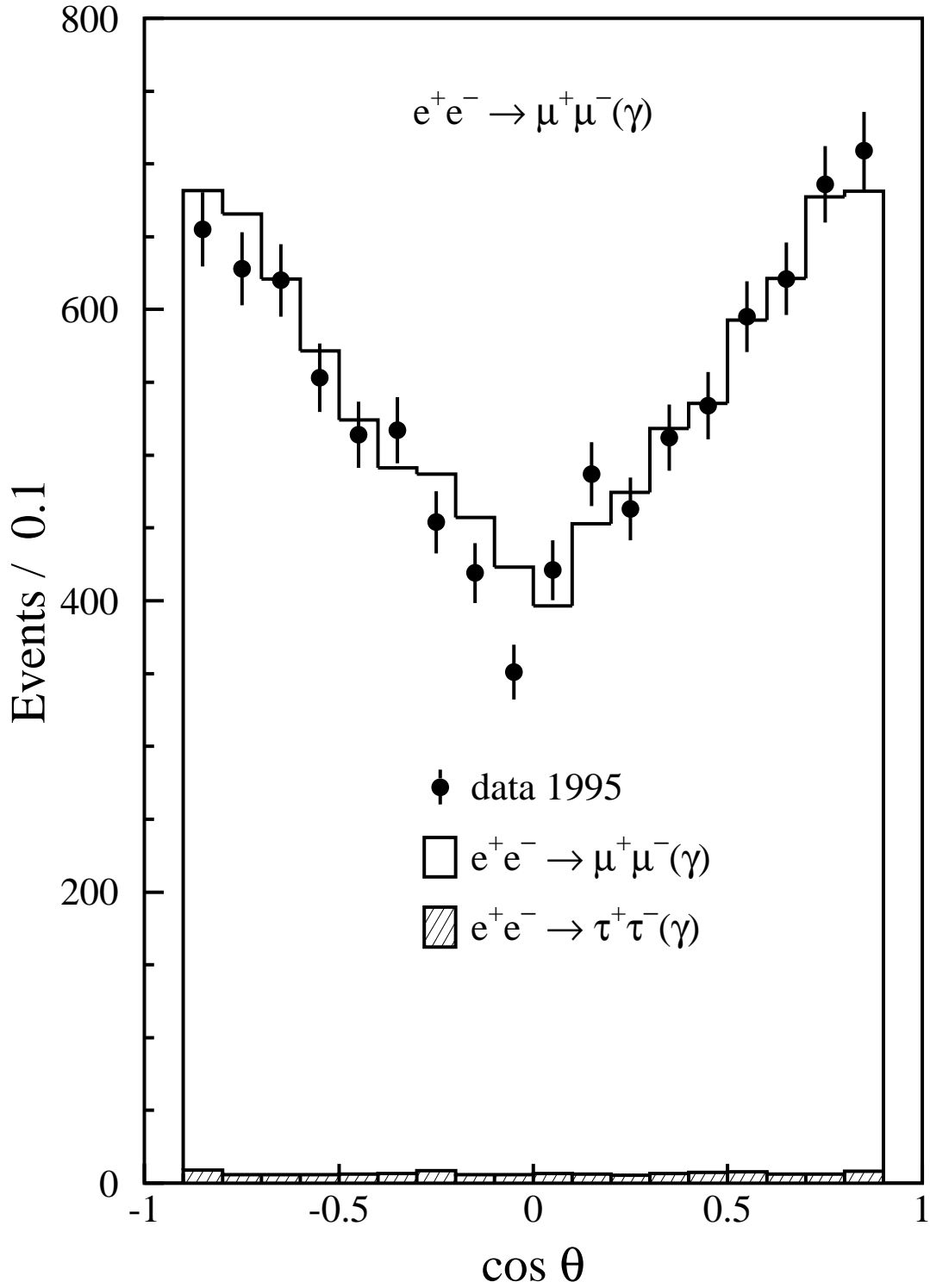}
 \end{center}
 \caption{Distribution of the polar angle, defined by the negative muon,
          of muon pairs collected in 1995.
          The pre-scan and peak data sets are combined.
         }
 \label{fig:muon_cost}
\end{figure}

For events with muons reconstructed in the muon chambers the maximum muon 
momentum must be larger than $\frac{2}{3}\,\Ebeam$.
Every muon without a reconstructed track in the muon chambers must have a 
transverse momentum larger than $3\ \GeV$ as measured in the \TEC.
The polar angle distribution of muon pairs collected in 1995 is shown in
Figure~\ref{fig:muon_cost}.

\subsubsection*{Total Cross Section}

The acceptance for the process \EEMMG\ in the fiducial volume
$|\cos\theta| < 0.8$ (0.9 for 1995 data) and for $\xi <90^\circ$ is determined
with events generated with the KORALZ program.
We obtain acceptances between $92.25\%$ and $93.04\%$, mainly depending on 
the \cms\ energy.
The systematic error on the cross section from imperfect description of
detector inefficiencies is estimated to be 
$2.7$\permille\ (3.2\permille\ for the 1995 data).
This number is calculated from a comparison with results obtained by removing 
events at the detector edges from the analysis and using different 
descriptions of time dependent detector inefficiencies.
Smaller contributions to the systematic error arise from the statistical 
precision of the Monte Carlo simulations performed for the different data 
samples. 

Muon pairs are mainly triggered by the muon and the central track
trigger.
The trigger efficiencies are studied as a function of the azimuthal angle
as inefficiencies are expected close to chamber boundaries.
For the 1995 data also the polar angular dependence of the trigger
efficiency is determined to account for effects in the forward region.
Events with both muons reconstructed in the muon chambers are triggered
with full efficiency.
The efficiency of the central track trigger is independently determined
using Bhabha events.
The overall trigger efficiency varies between 
$99.62\%$ and $99.90\%$ for the different years of data taking.
Systematic errors on the measured cross sections of less than 
$1$\permille\ are estimated from comparing a simulation of the 
central track trigger efficiency and its measurement with Bhabha events.

A background of $(1.35 \pm 0.03)\%$ remains in the sample arising from 
\EETTG\ events with both tau leptons decaying into muons.
The error reflects Monte Carlo statistics and the uncertainty of the branching
ratio $\mathrm{\tau^-\rightarrow\mu^-\bar{\nu}_\mu\nu_\tau}$~\cite{PDG98}.
Other backgrounds from \Zo\ decays are smaller than $0.1$\permille.
The contamination from the non-resonant two-photon process 
$\ee\rightarrow\ee\,\mm$ is $0.11\ \pb$, 
i.e. between $0.1$\permille\ and $0.3$\permille\ of the signal cross section,
as determined using the DIAG36 Monte Carlo program.

The residual contamination from cosmic ray muons in the event sample is 
determined from the sideband in the distribution of distance of closest
approach to the beam axis after all other selection cuts are 
applied (Figure~\ref{fig:muon_cosmic}).
Cosmic ray muons enter into the event sample at a rate of 
$(9.7 \pm 0.8)\cdot 10^{-4}$ per minute of data taking which translates 
to background contaminations between $1.9$\permille\ and $6.8$\permille\ for 
the different data sets depending on their average instantaneous luminosity and 
the signal cross section.
The statistical precision of the determination of the cosmic contamination
causes a systematic error of $0.3\ \pb$ on the total muon pair cross
section.

By varying the selection cuts we determine systematic errors on the
total cross section between $1.3$\permille\ and $2.2$\permille.
The systematic errors on the cross section measurements \EEMMG\ are 
summarised in Table~\ref{tab:muon_sys}.

Resonant four-fermion final states with a high-mass muon pair and a low-mass
fermion pair are accepted.
These events are considered as part of the signal if the invariant mass of the
muon pair exceeds $0.5\RS$.
This inclusive selection minimizes errors due to higher order radiative 
corrections.
Especially no cut is applied on additional tracks from low-mass fermion
pairs in the final state~\cite{Hoang}.

\subsubsection*{Forward-Backward Asymmetry}

The forward-backward asymmetry, \AFB, is defined as:
\begin{equation}
  \AFB = \frac{\sigma_{\mathrm{F}} - \sigma_{\mathrm{B}}} 
              {\sigma_{\mathrm{F}} + \sigma_{\mathrm{B}}} \, ,
  \label{eq:afb_def}
\end{equation}
where $\sigma_{\mathrm{F}}$ is the cross section for events with the fermion 
scattered into the hemisphere which is forward with respect to the 
$\mathrm{e^-}$ beam direction.
The cross section in the backward hemisphere is denoted by 
$\sigma_{\mathrm{B}}$.
Events with hard photon bremsstrahlung are removed from the sample by
requiring that the acollinearity angle of the event be less than 
$15^\circ$.
The differential cross section in the angular region 
$|\cos\theta| <0.9$ can then be approximated by the lowest order angular
dependence to sufficient precision:
\begin{equation}
  \frac{\mathrm{d}\sigma}{\mathrm{d}\!\cos\theta}
           \propto 
  \frac{3}{8} \left( 1 + \cos^2\theta \right)
         + \AFB \cos\theta   \, ,
 \label{eq:afb_born}
\end{equation}
with $\theta$ being the polar angle of the final state fermion
with respect to the $\mathrm{e^-}$ beam direction.

For each data set the forward-backward asymmetry is determined from 
a maximum likelihood fit to our data where the likelihood function is defined
as the product over the selected events labelled $i$ of the differential
cross section evaluated at their respective scattering angle
$\theta_i$:
\begin{equation}
  L = \prod_i \left(   \frac{3}{8} \left( 1 + \cos^2\theta_i \right)
                                 + \left( 1 - 2\kappa_i\right)
                                   \AFB \cos\theta_i
              \right) .
 \label{eq:loglike}
\end{equation}
The probability of charge confusion for a specific event, $\kappa_i$,
is included in the fit.
Only events with opposite charge assignment to the two muons are used
for this measurement.
The bias on the asymmetry measurement introduced by the use of the lowest
order angular dependence (Equation~\ref{eq:afb_born})
does not exceed 0.0003.

This method does not require an exact knowledge of the acceptance as a function
of the polar angle provided that the acceptance is independent of the 
muon charge. 
Events without a reconstructed muon in the muon chambers are included with the 
charge assignment obtained from the \TEC\ in a similar way as
for \ee\ final states~\cite{l3-69}.
This largely reduces effects of charge dependent acceptance in the
muon chambers.
The remaining asymmetry is estimated by artificially symmetrising the 
detector.
For each known, inefficient detector element, the element opposite
with respect to the centre of the detector is removed from the 
data reconstruction.
The event selection is applied again and, for the large 1994 data set,
the measured forward-backward asymmetry changes by $0.0011 \pm 0.0006$.
Half of this difference, $0.0006$, is assigned to all data sets as 
a systematic error on \AFB\ from a possible detector asymmetry.
In 1995 the forward-backward muon chambers did not contribute significantly
to the detector asymmetry.

The values of $\kappa_i$ are obtained from the fraction of events with 
identical charges assigned to both muons.
Besides its dependence on the transverse momentum, the charge measurement 
strongly depends on the number of muon chamber layers used in the 
reconstruction.
The charge confusion is determined for each event class individually.
The average charge confusion probability, almost entirely caused by muons 
only measured in the \TEC, is 
$(3.2 \pm 0.3)$\permille, $(0.8 \pm 0.1)$\permille\ and 
$(1.0 \pm 0.3)$\permille\ for the years 1993, 1994 and 1995, respectively, 
where the errors are statistical.
The improvement in the charge determination for 1994 and 1995 reflects
the use of the \SMD.

The correction for charge confusion is proportional to the forward-backward
asymmetry and it is less than $0.001$ for all data sets.
To estimate a possible bias from a preferred orientation of events with the
two muons measured to have the same charge we determine the forward-backward
asymmetry of these events using the track with a measured momentum closer 
to the beam energy.
The asymmetry of this subsample is statistically consistent with the
standard measurement.
Including these like-sign events in the 1994 sample would change the 
measured asymmetry by $0.0008$.
Half of this number is taken as an estimate of a possible bias of the asymmetry
measurement from charge confusion in the $1993-94$ data. 
The same procedure is applied to the 1995 data and the statistical precision
limits a possible bias to $0.0010$.

Differences of the momentum reconstruction in forward and backward events
would cause a bias of the asymmetry measurement because of the requirement
on the maximum measured muon momentum.
We determine the loss of efficiency due to this cut separately for forward
and backward events by selecting muon pairs without cuts on the reconstructed
momentum.
No significant difference is observed and the statistical error of this 
comparison limits the possible effect on the forward-backward asymmetry 
to be less than $0.0004$ and $0.0009$ for the $1993-94$ and 1995 data, 
respectively.

Other possible biases from the selection cuts on the measurement of the 
forward-backward asymmetry are negligible.
This is verified by a Monte Carlo study which shows that events not selected 
for the asymmetry measurement, but inside the fiducial 
volume and with $\xi <15^\circ$, do not have a different \AFB\ value.

The background from \EETTG\ events is found to have the same asymmetry as the 
signal and thus neither necessitates a correction nor causes a systematic 
uncertainty.
The effect of the contribution from the two-photon process 
$\mathrm{\ee\rightarrow\ee\,\mm}$, further reduced by the tighter
acollinearity cut on the measured muon pair asymmetry,
can be neglected.
The forward-backward asymmetry of the cosmic ray muon background is measured to
be $-0.02 \pm 0.13$ using the events in the sideband of the distribution
of closest approach to the interaction point.
Weighted by the relative contribution to the data set this leads to 
corrections of $-0.0007$ and $+0.0003$ to the peak$-2$ and peak$+2$
asymmetries, respectively.
On the peak this correction is negligible.
The statistical uncertainty of the measurement of the cosmic ray asymmetry
causes a systematic error of $0.0001$ on the peak and between $0.0003$ and 
$0.0005$ for the peak$-2$ and peak$+2$ data sets.

The systematic uncertainties on the measurement of the muon forward-backward
asymmetry are summarised in Table~\ref{tab:muon_sys_afb}.
In $1993-94$ the total systematic error amounts to $0.0008$ at the peak 
points and to $0.0009$ at the off-peak points due to the larger contamination
of cosmic ray muons.
For the 1995 data the determination of systematic errors is limited by the
number of events taken with the new detector configuration and the total
error is estimated to be  $0.0015$.

\begin{figure}[htbp]
 \begin{center}
  \includegraphics[width=\figwidth]{\pltdir 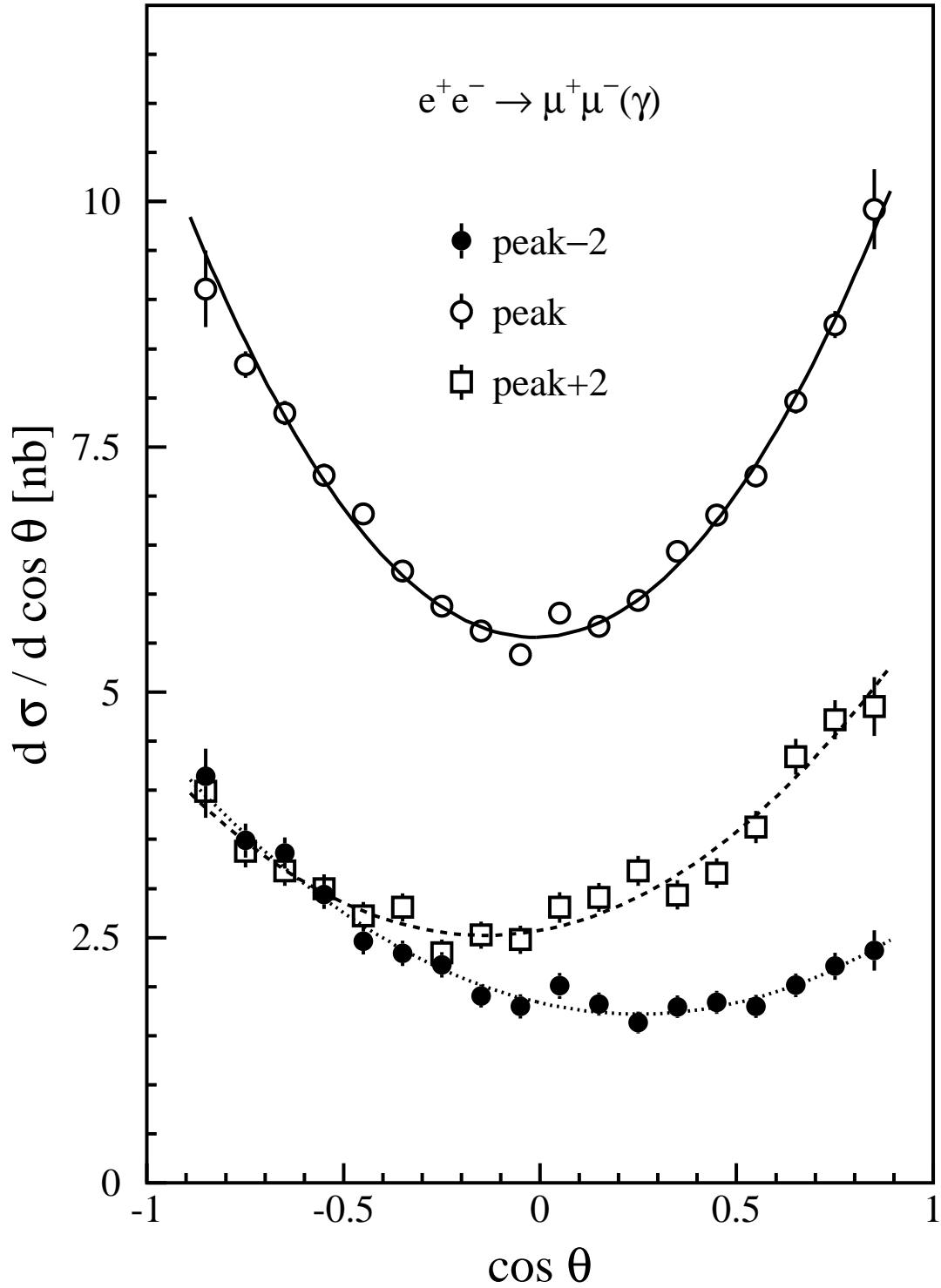}
 \end{center}
 \caption{The measured differential cross section \protect\EEMMG\ 
          combining the $1993-95$ data into three \protect\cms\ energy
          points.
          The lines show the result of a fit using the functional 
          form of Equation~\protect\ref{eq:afb_born}.
         }
 \label{fig:diffxs_muon}
\end{figure}

In Figure~\ref{fig:diffxs_muon} the differential cross sections 
$\mathrm{d}\sigma / \mathrm{d}\!\cos\theta$ measured from the $1993-95$ 
data sets are shown for three different \cms\ energies.
The data are corrected for detector acceptance and charge confusion.
Data sets with a \cms\ energy close to \MZ, as well as the 
data at peak$-2$ and the data at peak$+2$, are combined.
The data are compared to the differential cross section shape given in
Equation~\ref{eq:afb_born}.

The results of the total cross section and forward-backward asymmetry 
measurements in \EEMMG\ are presented in Section~\ref{sec:xsafbresults}.

%
%
\section{\protect\boldmath\EETTG}\label{sec:taus}

\subsubsection*{Event Selection}

The selection of \EETTG\ events aims to select all hadronic and leptonic
decay modes of the tau. 
\Zo\ decays into tau leptons are distinguished from other \Zo\ decays 
by the lower visible energy due to the presence of neutrinos and 
the lower particle multiplicity as compared to hadronic \Zo\ decays.
Compared to our previous analysis~\cite{l3-69} the selection of 
\EETTG\ events is extended to a larger polar angular range, 
$| \cos \theta_t | \leq 0.92$,
where $\theta_t$ is defined by the thrust axis of the event.

Event candidates are required to have a jet, 
constructed from calorimetric energy deposits~\cite{Cone_alg} and muon tracks, 
with an energy of at least $8\ \GeV$.
Energy deposits in the hemisphere opposite to the direction of this most
energetic jet are combined to form a second jet.
The two jets must have an acollinearity angle $\xi < 10^\circ$.
There is no energy requirement on the second jet.

High multiplicity hadronic \Zo\ decays are rejected by allowing at most 
three tracks matched to any of the two jets.
In each of the two event hemispheres there should be no track with an
angle larger than $18^\circ$ with respect to the jet axis.
Resonant four-fermion final states with a high mass tau pair and a low mass
fermion pair are mostly kept in the sample.
The multiplicity cut affects only tau decays into three charged particles with 
the soft fermion close in space leading to corrections of less than 
$1$\permille. 

If the energy in the electromagnetic calorimeter of the first jet exceeds 
$85\%$, or the energy of the second jet exceeds $80\%$, of the beam energy with
a shape compatible with an electromagnetic shower the event is classified as 
\EEEEG\ background and hence rejected.

Background from \EEMMG\ is removed by requiring that there be no isolated 
muon with a momentum larger than 80\% of the beam energy and that the sum
of all muon momenta does not exceed $1.5\Ebeam$.
Events are rejected if they are consistent with the signature of two MIPs.

To suppress background from cosmic ray events the time of scintillation counter
hits associated to muon candidates must be within $\pm 5\ \MR{ns}$
of the beam crossing.
In addition, the track in the muon chambers must be consistent with 
originating from the interaction point.

\begin{figure}[htbp]
 \begin{center}
  \includegraphics[width=\figwidth]{\pltdir 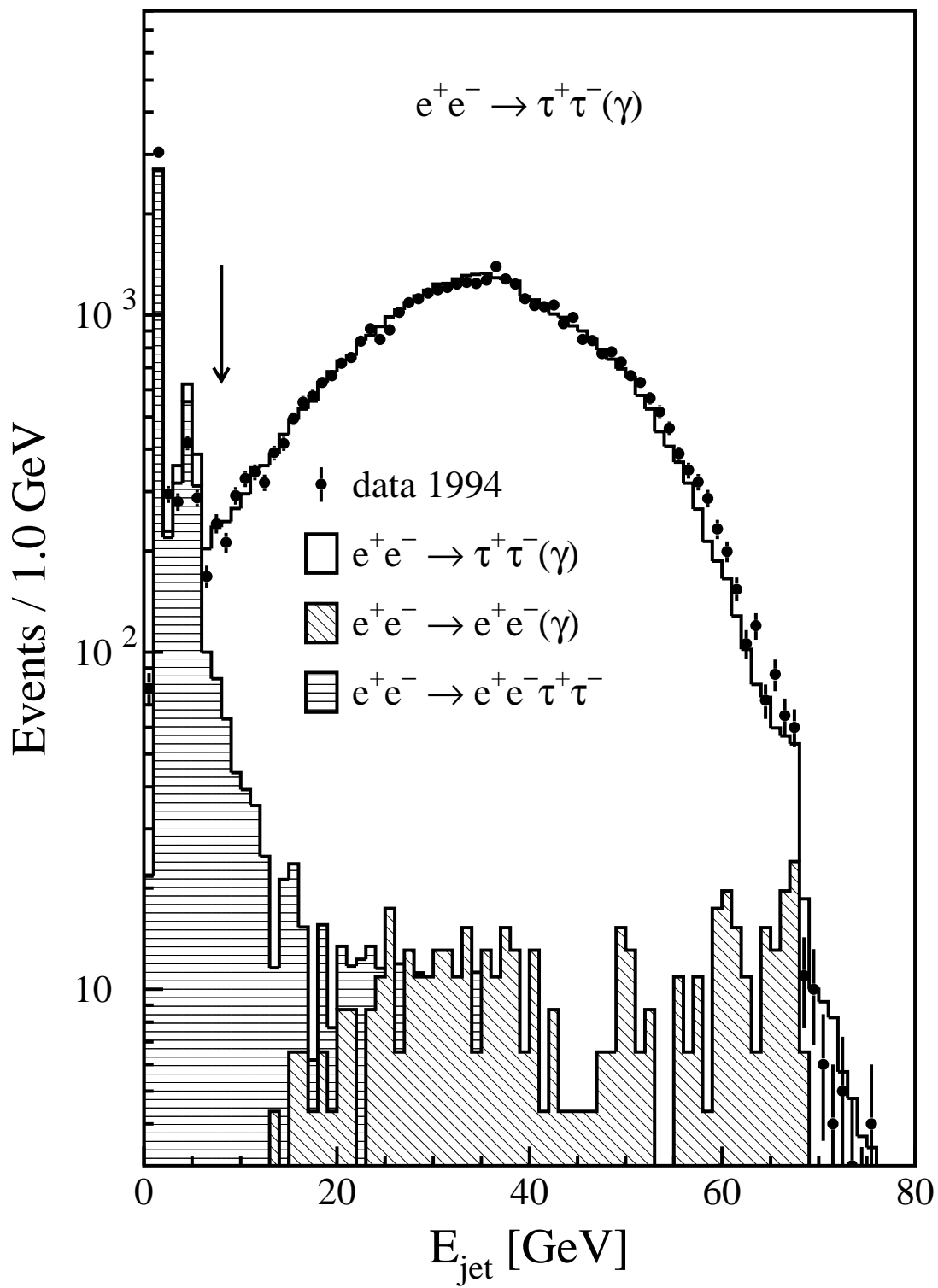}
 \end{center}
 \caption{Energy of the most energetic jet in \protect\EETTG\ event 
          candidates for 1994 data.
         }
 \label{fig:tau_ejet}
\end{figure}

\begin{figure}[htbp]
 \begin{center}
  \includegraphics[width=\figwidth]{\pltdir 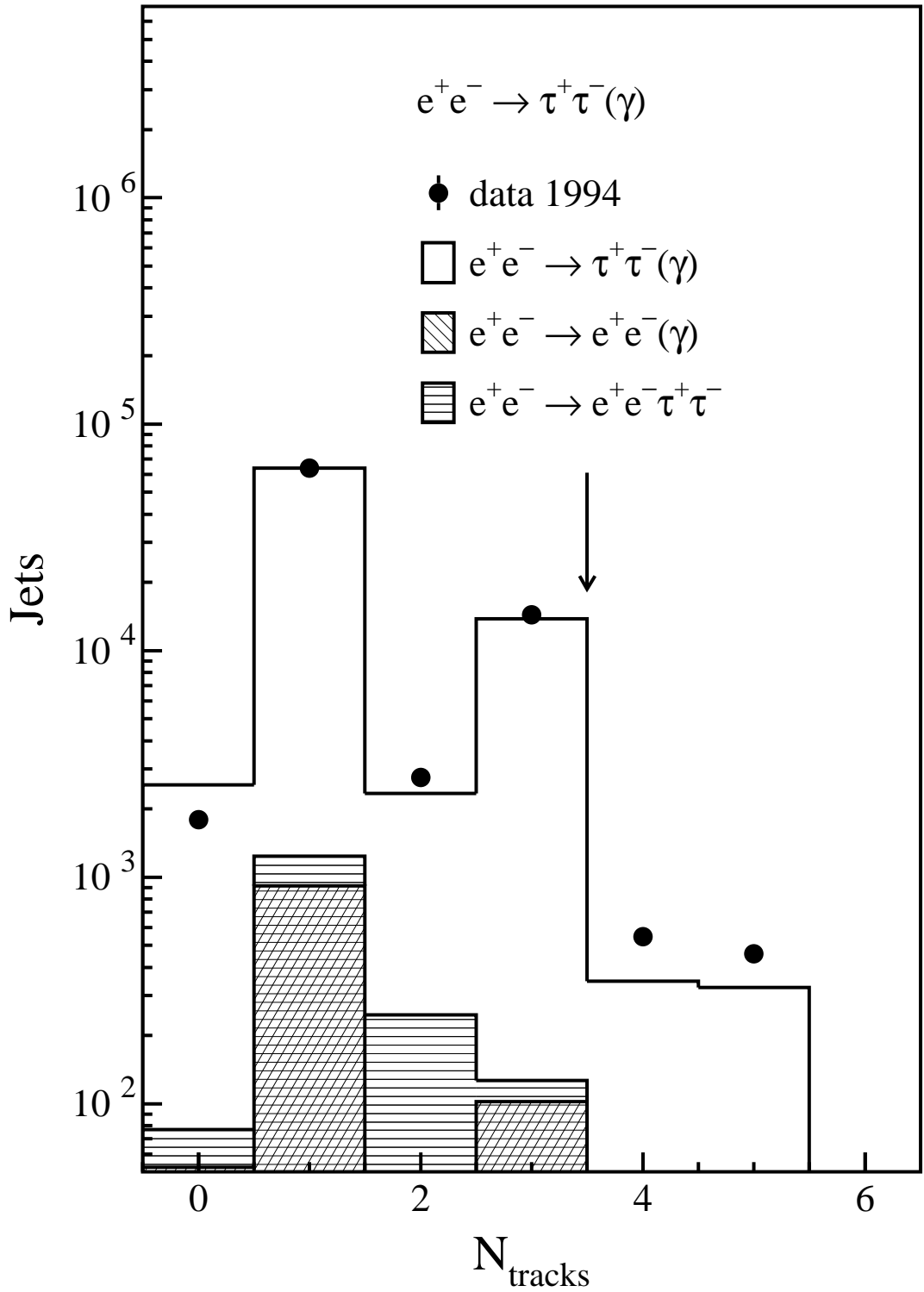}
 \end{center}
 \caption{Number of tracks associated to each of the two jets
         in \protect\EETTG\ candidate events.
         }
 \label{fig:tau_ntracks}
\end{figure}

\begin{figure}[htbp]
 \begin{center}
  \includegraphics[width=\figwidth]{\pltdir 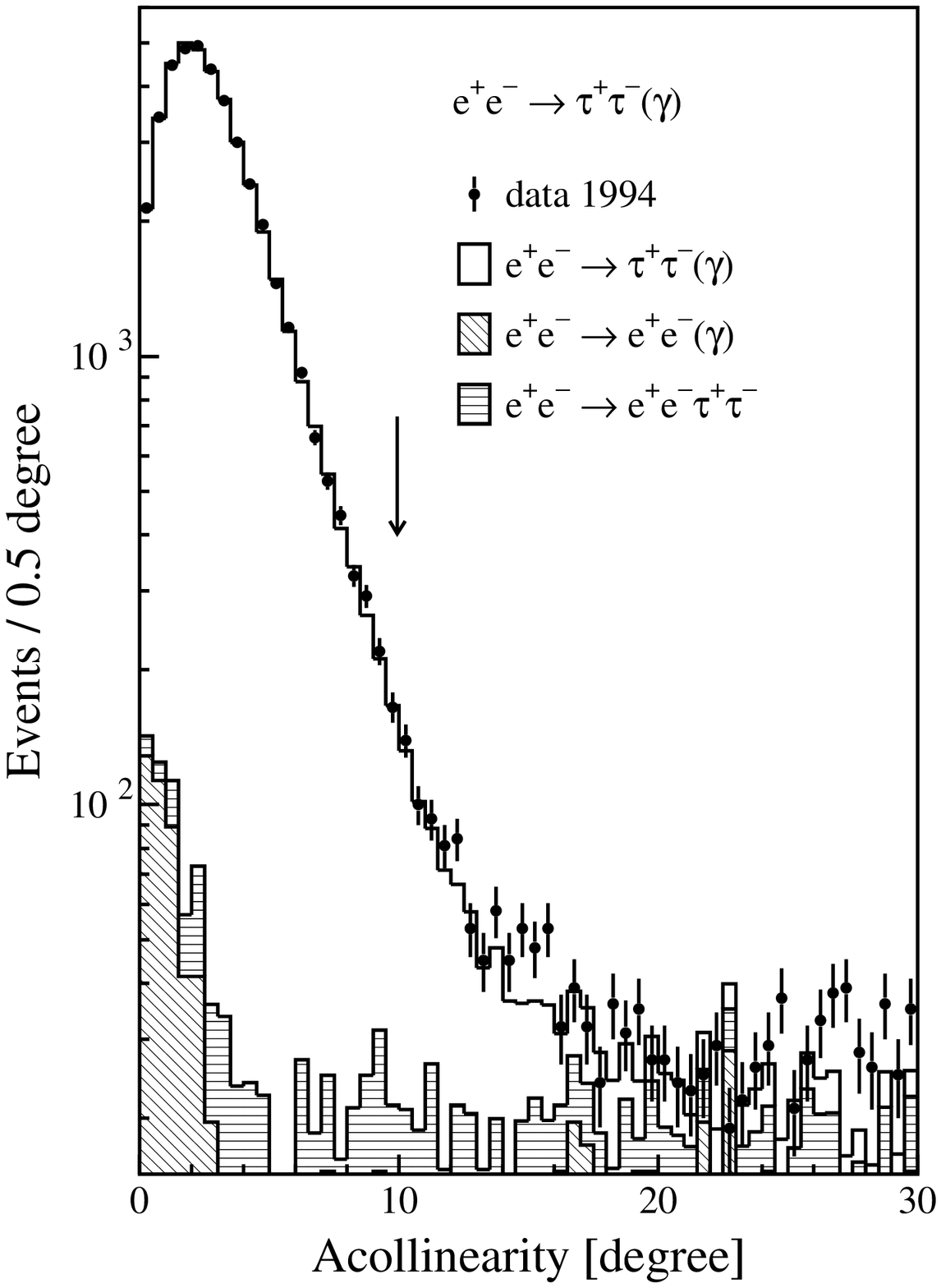}
 \end{center}
 \caption{The distribution of the acollinearity for 
          \protect\EETTG\ candidates.
         }
 \label{fig:tau_acol}
\end{figure}

\begin{figure}[htbp]
 \begin{center}
  \includegraphics[width=\figwidth]{\pltdir 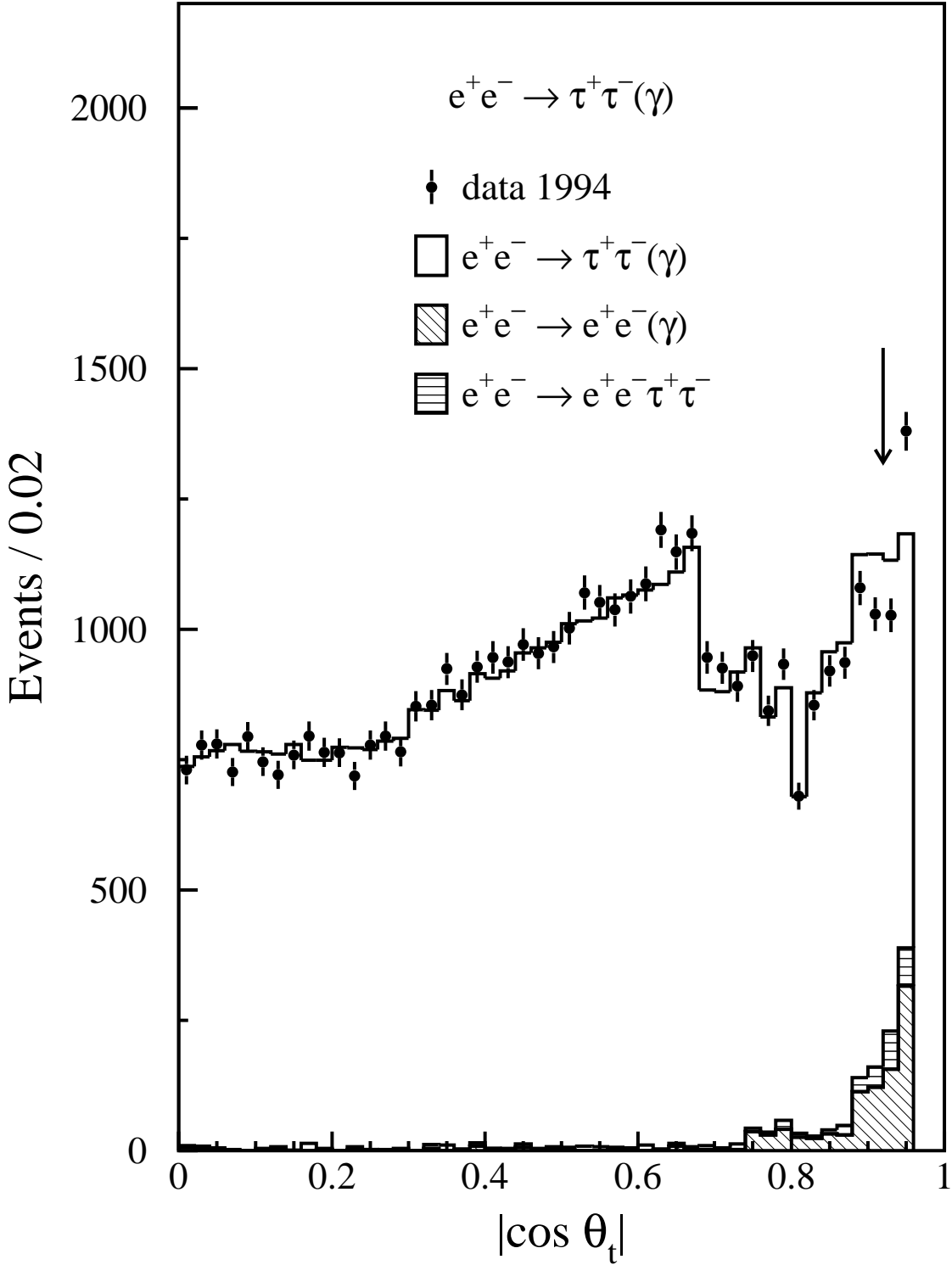}
 \end{center}
 \caption{Distribution of the polar angle of the event thrust axis
          for \protect\EETTG\ candidates collected in 1994.
          The structure seen for $| \cos \theta_t | > 0.65$
          reflects the modifications of the event selection 
          in the end-cap and the transition region. 
         }
 \label{fig:tau_cost}
\end{figure}

In Figures~\ref{fig:tau_ejet} to~\ref{fig:tau_cost}
the energy in the most energetic jet, 
the number of tracks associated to both jets,
the acollinearity between the two jets and
the distribution of $|\cos \theta_t |$ are shown for the 1994 data.
Data and Monte Carlo expectations are compared after all cuts are applied, 
except the one under study.
Good agreement between data and Monte Carlo is observed.
Small discrepancies seen in Figure~\ref{fig:tau_ntracks} are due to the imperfect
description of the track reconstruction efficiency in the central chamber.
Their impact on the total cross section measurement is small and is
included in the systematic error given below.

Tighter selection cuts must be applied in the region between barrel and
end-cap part of the BGO calorimeter and in the end-cap itself,
reducing the selection efficiency (see Figure~\ref{fig:tau_cost}).
This is due to the increasing background from Bhabha scattering.
Most importantly the shower shape in the hadron calorimeter is also used 
to identify candidate electrons and the cuts on the energy of the first and 
second jet in the electromagnetic end-cap calorimeter are tightened to 75\% 
of the beam energy.

\subsubsection*{Total Cross Section}

Between $70.21\%$ and $70.91\%$ of the signal events are accepted
inside the fiducial volume defined by $| \cos \theta_t | \leq 0.92$.
The acceptance for \EETTG\ events depends on the tau decay products. 
The experimental knowledge of tau branching fractions~\cite{PDG98} translates
to an uncertainty on the average acceptance of \EETTG\ events which 
contributes with $2$\permille\ to the systematic error on the cross section
measurement.
From the data the efficiency of the trigger system for selected 
\EETTG\ events is determined to be $(99.71 \pm 0.02)\%$.

The largest remaining background consists of Bhabha events,
$1.3\%$ to $3.7\%$, depending on the \cms\ energy, entering into the sample 
predominantly at low polar angles.
Background from \Zo\ decays into hadrons is determined to be between
$1.3$\permille\ and $2.7$\permille, depending on the data taking period,
and $7.5$\permille\ from \Zo\ decays into muons.
The statistical precision of the background determination by Monte Carlo 
simulations causes systematic errors between $1.0$\permille\ and $3.3$\permille.
Contaminations from non-resonant background are small:
$1$\permille\ to $2$\permille\ from two-photon collisions and 
$2$\permille\ to $3$\permille\ from cosmic ray muons,
depending on the \cms\ energy.
The systematic error from the subtraction of non-resonant background is 
estimated to be $1.2\ \pb$.

From variations of the above selection cuts contributions to the
systematic error on the total cross section between 
$5.3$\permille\ and $8.0$\permille\ are estimated for different years,
largely independent of the \cms\ energy.
The main contribution arises from the definition of the fiducial volume
by $| \cos \theta_t | \leq 0.92$, see Figure~\ref{fig:tau_cost}.
The systematic errors on the \EETTG\ cross section measurements are 
summarised in Table~\ref{tab:tau_sys}.

\subsubsection*{Forward-Backward Asymmetry}

The forward-backward asymmetry of \EETTG\ events is determined in the same
way as described for muon pairs (Equation~\ref{eq:loglike}).
The charge of a tau is derived from the sum of the charges of its decay 
products as measured in the central tracking and the muon chambers.
The event sample selected for the cross section measurement is used requiring
opposite and unit charge for the two tau jets.

The average probability for a mis-assignment of both charges as determined
from the ratio of like and unlike sign events is $(7.4 \pm 0.4)$\permille\ in
1993.
The use of the silicon microvertex detector reduced this mis-assignment to 
$(2.5 \pm 0.1)$\permille\ and $(1.3\pm 0.1)$\permille\ in 1994 and 1995.
Because the charge confusion probability is approximately independent of the
polar angle this average value is used in the fit for \AFB.
The systematic error on the forward-backward asymmetry from the uncertainty 
in the determination and the treatment of the charge confusion probability
is estimated to be less than $0.0001$ for all data sets.

The effect of a possible detector asymmetry, in particular at the edges
of the fiducial volume, is estimated from variation of the 
$\cos\theta_t$ cut.
The statistical accuracy of this test limits this uncertainty to $0.003$
which is taken as a systematic error.
The measured asymmetries are corrected for background contributions.
The uncertainty on the background contamination, in particular from \EEEEG, 
translates into an error of $0.001$ on the tau pair asymmetry.

Large Monte Carlo samples are used to study a possible bias on the measured 
asymmetry from the fit method and from the selection cuts.
In particular, energy and momentum requirements might preferentially select
certain helicity configurations leading to a bias in the determination of
\AFB.
The Monte Carlo simulation does not show evidence for such a bias 
and its statistical precision, $0.0004$, is taken as the systematic error.

During the 1995 data taking, large shifts of the longitudinal position of
the \ee-interaction point were observed caused by the reconfiguration
of the LEP radio frequency system~\cite{lep-09}.
However, they are found to have no sizeable effect on the
measurement of the forward-backward asymmetry.
The total systematic error assigned to the forward-backward asymmetry 
measurement of tau pairs is $0.0032$ (Table~\ref{tab:tau_sys_afb}).
It is fully correlated between the data sets.

\begin{figure}[htbp]
 \begin{center}
  \includegraphics[width=\figwidth]{\pltdir 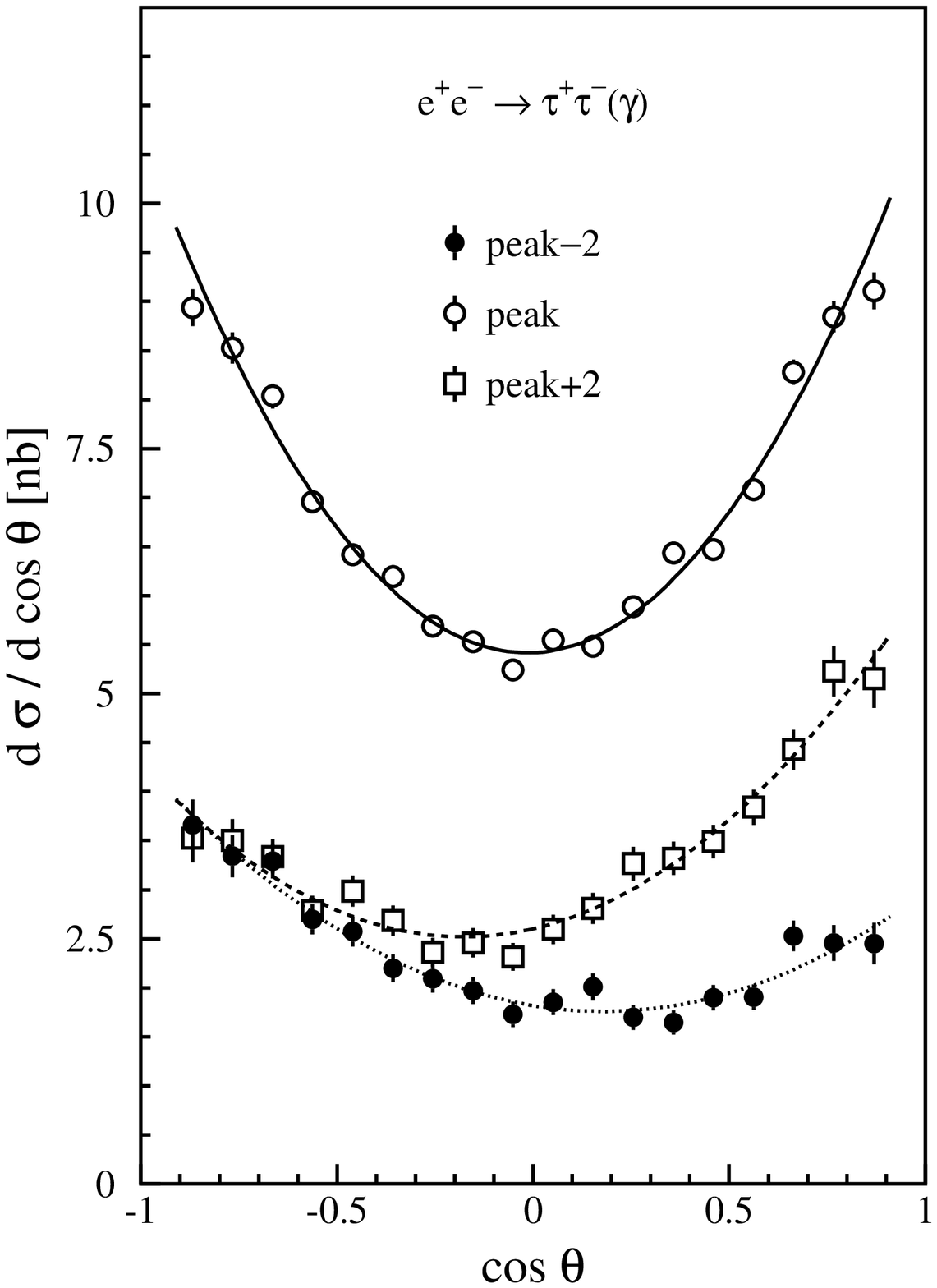}
 \end{center}
 \caption{Same as Figure~\protect\ref{fig:diffxs_muon}
          for \protect\EETTG.
         }
 \label{fig:diffxs_tau}
\end{figure}

The measured differential cross sections, combining the data into three
\cms\ energy points, are shown in Figure~\ref{fig:diffxs_tau}.
The lines show the results of fits to the data using the functional form of
Equation~\protect\ref{eq:afb_born}.

Section~\ref{sec:xsafbresults} presents the measurements of the total cross 
section and the forward-backward asymmetry in \EETTG.

%
%
\section{{\protect\boldmath\EEEEG}}
          \label{sec:electrons}

\subsubsection*{Event Selection}

The analysis of the reaction \EEEEG\ is restricted to the polar angular range 
$44^\circ < \theta < 136^\circ$ to increase the relative contribution of 
\Zo\ exchange to the measured cross section.
The signature of \ee\ final states is the low multiplicity high energy
deposition in the electromagnetic calorimeter with associated tracks in the
\TEC.

\begin{figure}[htbp]
 \begin{center}
   \includegraphics[width=\figwidth]{\pltdir 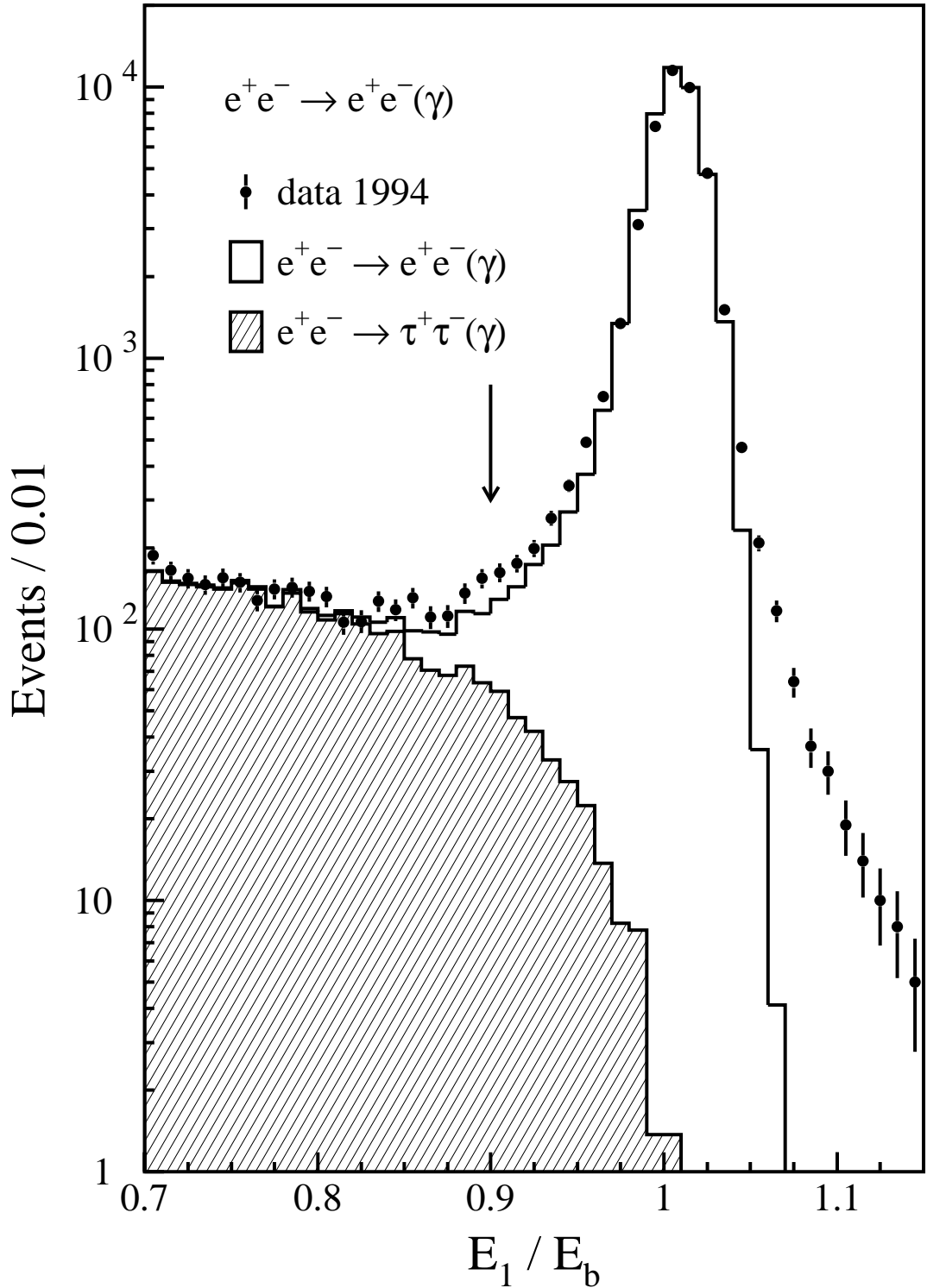}
 \end{center}
 \caption{Distribution of the energy of the highest energy cluster measured 
          in the electromagnetic calorimeter for \protect\EEEEG\ candidate 
          events (1994 data).
          Events below the cut value, indicated by the vertical arrow,
          can be selected by other criteria as described in the text.
         }
 \label{fig:electron_Emax}
\end{figure}

Most of the events are selected by requiring at least two clusters in the 
fiducial volume of the electromagnetic calorimeter, one with an energy greater
than $0.9\,\Ebeam$ and the other with more than $2\ \GeV$.
The polar angles are determined form the centre-of-gravity of the clusters in
the calorimeter and the interaction point.
Figure~\ref{fig:electron_Emax} shows the distribution of the highest energy
cluster, $E_1$, normalised to the beam energy for events which pass all cuts 
except the requirement on the most energetic cluster.

\begin{figure}[htbp]
 \begin{center}
   \includegraphics[width=\figwidth]{\pltdir 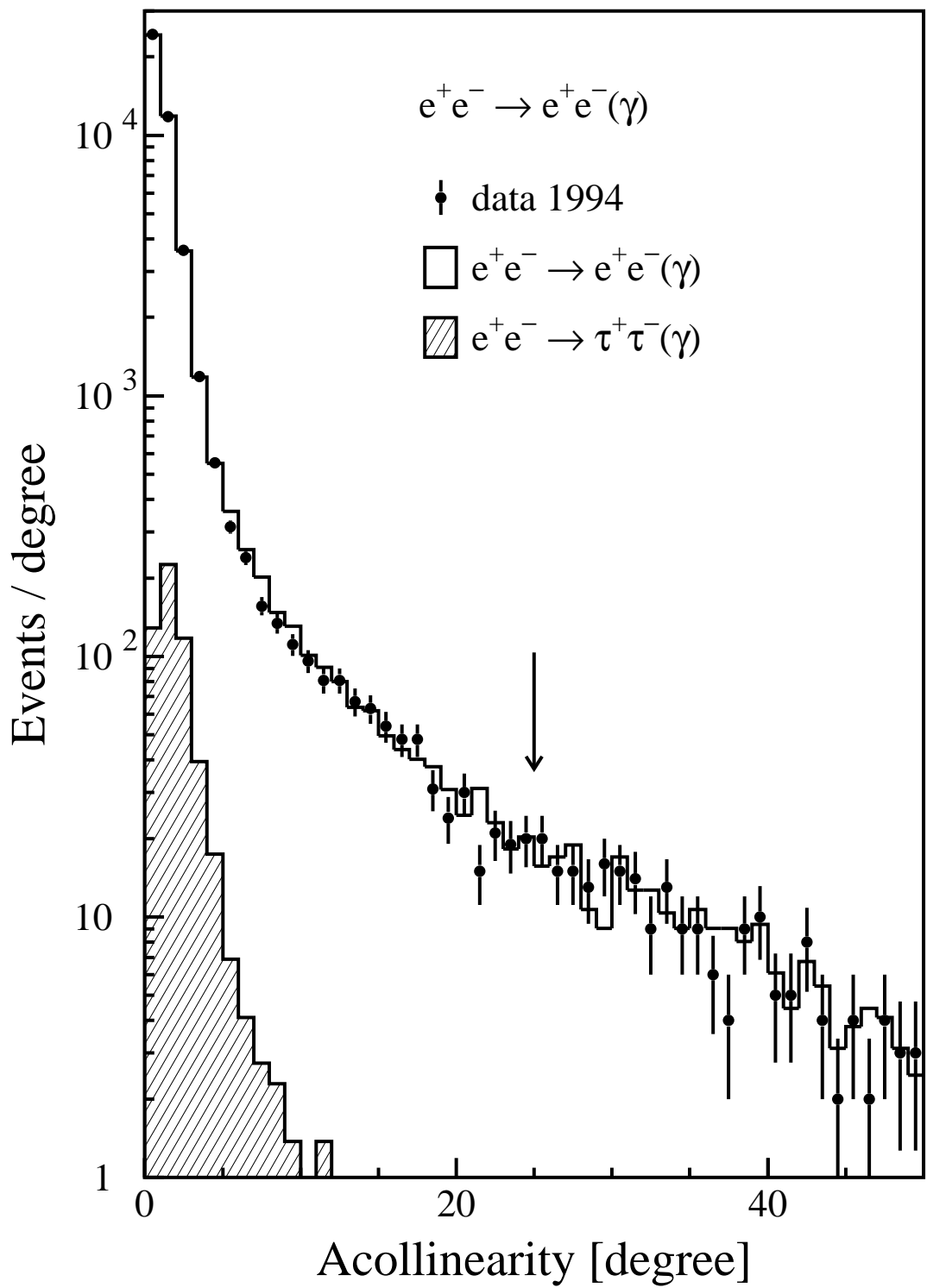}
 \end{center}
 \caption{Acollinearity angle between the two electron candidates in
          \protect\EEEEG\ events.
         }
 \label{fig:electron_acol}
\end{figure}

Electrons are discriminated from photons by requiring five out of 62 anodes of
the \TEC\ with a hit matching in azimuthal angle within 
$\pm 3^\circ$ with the cluster in the calorimeter.
Two electron candidates are required inside the fiducial volume and with an
acollinearity angle $\xi < 25^\circ$.
Figure~\ref{fig:electron_acol} shows the distribution of the acollinearity 
angle.
All other cuts except the one under study are applied.

The event selection depends on the exact knowledge of imperfections
of the electromagnetic calorimeter.
The impact of the discrepancies seen in Figure~\ref{fig:electron_Emax} around 
the cut value is significantly reduced by accepting also events without a second
cluster in the fiducial volume of the electromagnetic calorimeter. 
In this case a cluster in the hadron calorimeter is required consistent with 
an electromagnetic shower shape and at least $7.5\ \GeV$ opposite to the 
leading BGO cluster.
This recovers events, up to $4$\permille\ of the total sample, 
with electrons leaking through the BGO support structure.
Events failing the requirement on the most energetic cluster in the
electromagnetic calorimeter are accepted if the sum of the energies of the 
four highest energy clusters anywhere in the electromagnetic calorimeter 
is larger than $70\%$ of the \cms\ energy.
In addition this partially recovers radiative events.

For all event candidates the total number of energy deposits, 
$N_{\MR{cl}}$, must be less than $15$ ($12$ for 1995 data).

\subsubsection*{Total Cross Section}

The selection efficiency is determined using Monte Carlo events
generated with the program BHAGENE, which generates
up to three photons.
Efficiencies between 97.37\% and 98.53\% 
are obtained for the different samples,
where the differences originate from time dependent detector inefficiencies.
The use of high multiplicity hadron events allows to monitor the
status of each individual BGO crystal in short time intervals.
Inefficient crystals, typically 100 out of 8000 in the barrel part,
are identified and taken into account in the Monte Carlo simulation. 
This method, together with the redundancy of the selection
cuts, reduces the systematic error on the selection efficiency.
Limited Monte Carlo statistics causes systematic errors between 
$0.4$\permille\ and $1.0$\permille.

The calculation of the selection efficiency is checked using events generated 
with the programs BABAMC and BHWIDE.
The efficiencies calculated with the different event generators agree within 
$\pm 1$\permille\ which is taken as an estimate of the systematic error.

The efficiency of the electron and photon discrimination in the \TEC\ is 
determined using a subsample of data events selected by a tight acollinearity
cut ($\xi < 1^\circ$) and requiring two high energy clusters in the 
electromagnetic calorimeter ($E > 30\ \GeV$).
Here the contamination of $\ee\rightarrow\ee\Photon$ events with 
one electron and the photon inside, and the other electron outside the 
fiducial volume is expected to be very small.
In this sample, events with only one identified electron originate from 
mis-identified Bhabha events or from photon conversion of 
$\ee\rightarrow\Photon\Photon$ events.
The contamination of the latter in this sample is $0.4$\permille\ to 
$0.9$\permille\ as calculated from Monte Carlo.
After correction for this contamination, the probability that one of the 
electrons in \ee\ final states fails the electron-photon discrimination
is measured to be $(0.7 \pm 0.3)$\permille\ and $(1.0 \pm 0.1)$\permille\ for 
the 1993 and 1994 data, respectively.
We correct for this effect.

The method to determine this probability from the data is checked on fully 
simulated \EEEEG\ Monte Carlo events.
Firstly by not applying the electron-photon discrimination, the contamination of
events in the data used for the cross section measurement with one photon and 
only one electron in the fiducial volume is determined to be 
$(2.6 \pm 0.5)$\permille.
This is in reasonable agreement with the Monte Carlo prediction of 
$1.4$\permille.
Then we apply the above method to determine the probability that an electron
fails the electron-photon discrimination on the fully simulated events
and compare it to the value obtained using the generator information.
The result is consistent within $0.6$\permille\ which is assigned as a 
systematic error to the total cross section due to the simulation and 
determination of the electron-photon discrimination.

In 1995 the quality criteria on the status of the \TEC\ are relaxed to 
increase the data sample at the expense of a smaller efficiency on the electron
identification and a larger systematic error.
Between 1.9\permille\ and 2.8\permille\ of the electrons fail the 
electron-photon discrimination cuts as determined from Monte Carlo simulation.
We correct for this effect and a systematic error of 1.5\permille\ is assigned 
to the total cross section measurement.

Large angle Bhabha scattering events are triggered by the energy and the 
central track triggers. 
The overall trigger inefficiency is found to be $\leq 0.1$\permille\ and has a
negligible effect on the cross section measurement.

In the 1993 and 1994 data the longitudinal position of the \ee\ interaction 
point is stable within $\pm2\ \MR{mm}$.
The corresponding uncertainty on the definition of the fiducial volume 
translates to a systematic error of $0.5$\permille\ on the cross section
measurement.
Imperfections of the description of the BGO geometry and the shower shape
of electrons lead to a possible difference of the definition of the polar
angle between data and Monte Carlo simulation.
This difference is found to be less than $0.1^\circ$, translating
to a systematic error of $0.5$\permille\ on the cross section measurement.

The large movements of the interaction point in 1995 are determined from our 
\EEHADG\ data and the positions are used to calculate the scattering angle in 
\EEEEG\ events.
The remaining systematic uncertainty on the definition of the fiducial volume,
including the description of the BGO geometry, is estimated from a variation
of the cut on the polar angle to be $1.5$\permille.

The selected sample contains about $1\%$ background from the 
process \EETTG, only slightly depending on the \cms\ energy.
Contaminations from hadronic Z decays and the process 
$\ee \rightarrow \ee\,\ee$ are below $1$\permille\ and the remaining
background from $\ee\rightarrow\Photon\Photon$ is negligible.
The error on the total cross section from background subtraction is 
$0.4$\permille\ to $1.0$\permille\ originating from limited Monte Carlo
statistics.

The systematic uncertainty of the event selection, estimated from variations 
of the selection cuts around their nominal values, 
varies between $0.8$\permille\ and $2.7$\permille\ for the various data 
sets.
The systematic uncertainties contributing to the measurement of the cross
section \EEEEG\ are summarised in Table~\ref{tab:electron_sys}.

\subsubsection*{Forward-Backward Asymmetry}

The data sample for the forward-backward asymmetry measurement is obtained 
from the sample used for the measurement of the total cross section 
requiring in addition that each of the two electron candidates match
with a track within 25 mrad in azimuth.

The charge determination of the electrons is described in detail in 
reference~\cite{l3-69}.
The charge confusion is measured with the \EEMMG\ data sample which has an 
independent charge measurement from the muon spectrometer.
We obtain for the probability of a wrong event orientation 
values between 0.5\% and 4.6\%. 
Lower values are due to the exploitation of the \SMD\ in 1994 and 1995.
We determine the asymmetry of a subsample with much lower charge confusion by 
excluding events with tracks close to the cathode and anode planes of the 
\TEC.
Comparing these results to those obtained from the full sample we derive
a systematic error on \AFB\ of 0.002 from the uncertainty of the charge 
determination.

In the event sample used for the asymmetry measurement the main background
from \EETTG\ is reduced to about $4$\permille\ because the tight requirement
on the matching between tracks and clusters in the electromagnetic calorimeter
removes $\tau^- \rightarrow \rho^-\nu_\tau$ decays present in the 
cross section sample.
It induces a correction of $0.002$ on the asymmetry for the peak$-2$ and of 
less than $0.0005$ for the other data sets.
The effect is largest at peak$-2$ because of the difference of the 
\EEEEG\ and \EETTG\ asymmetries.
The uncertainty on the asymmetry measurement from background subtraction is 
estimated to be $0.0005$.

The asymmetry is determined from the number of events observed in the
forward and backward hemispheres, correcting for polar
angle dependent efficiencies and background.
The scattering angle is defined by the polar angle  of the electron, 
$\theta_{\mathrm{e^-}}$.
The determination of the asymmetry is repeated defining the angle by the
positron, $\theta_{\mathrm{e^+}}$, and taking the average of the two
\AFB\ values.
This reduces the sensitivity of the result to the size of the interaction 
region and its longitudinal offset.

Alternatively, we determine the forward-backward asymmetry using the
scattering angle in the rest system of the final state electron and positron:
\begin{equation}
\cos\theta^\star = 
   \frac{\sin(\theta_{\mathrm{e^+}} - \theta_{\mathrm{e^-}})}
        {\sin\theta_{\mathrm{e^-}} +\sin\theta_{\mathrm{e^+}}} \, .
 \label{eq:theta_star}
\end{equation}
This definition minimises the sensitivity to photon emission.
A Monte Carlo study shows that it differs by less than $0.0005$ from the 
above definition of \AFB\ due to different radiative corrections.
After correcting for this difference in the data the two approaches yield 
forward-backward asymmetries consistent within $0.0015$ which is taken as 
an estimate of the remaining uncertainty of the scattering angle from the
knowledge of the interaction point.
The contributions of the systematic error on the asymmetry measurement are
summarised in Table~\ref{tab:electron_sys_afb}. 

\begin{figure}[htbp]
 \begin{center}
   \includegraphics[width=\figwidth]{\pltdir 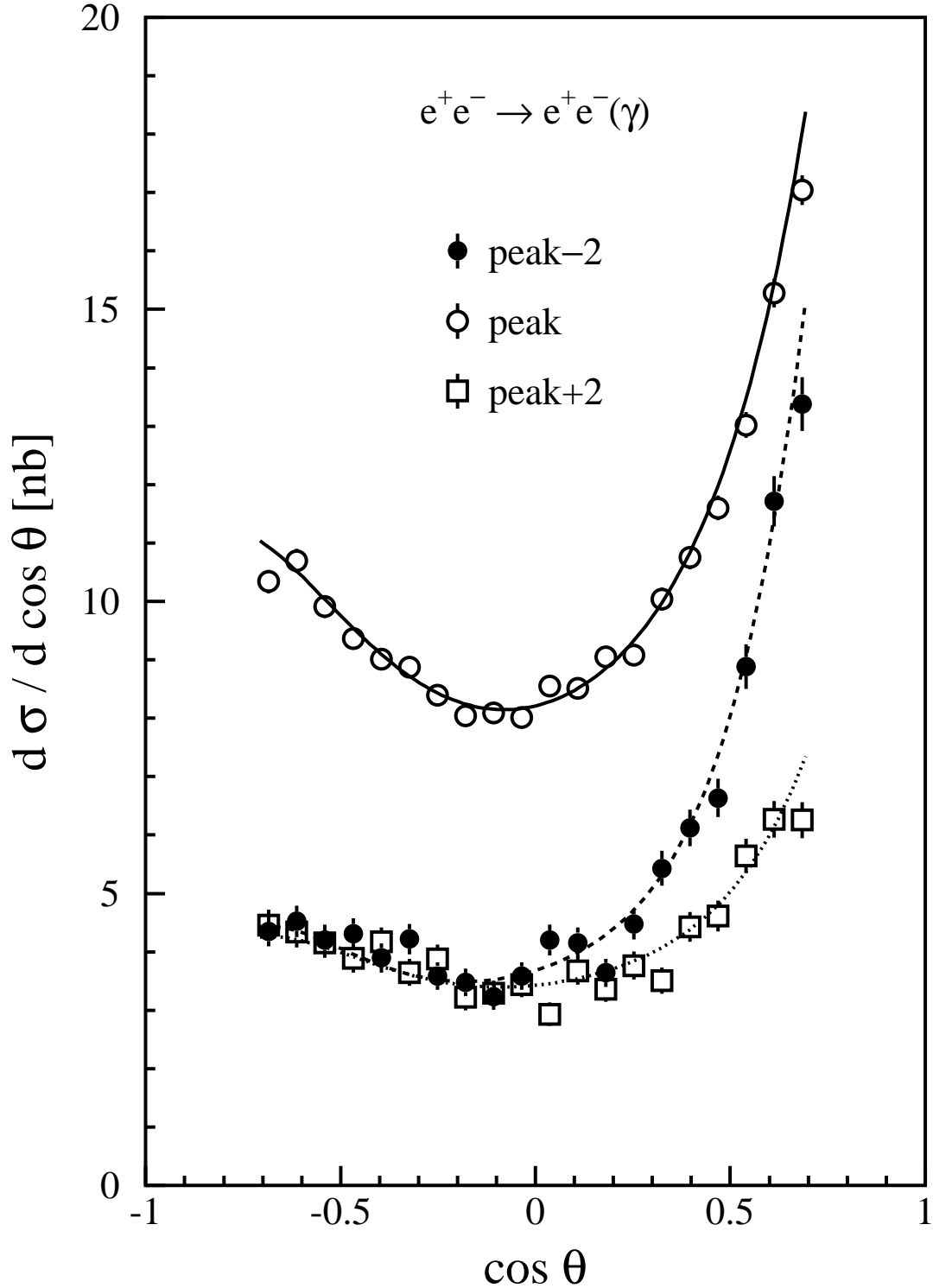}\vspace*{-5mm}
 \end{center}
 \caption{The measured differential cross section \protect\EEEEG\ for
          data collected between 1993 and 1995.
          The cross sections are calculated for an acollinearity angle
          $\xi <25^\circ$ and a mimimum energy of $1\ \GeV$ of each
          final state fermion.
          The data are compared to the SM predictions which are 
          shown as lines for three different \protect\cms\ energies.
         }
 \label{fig:diffxs_bhabha}
\end{figure}

The differential cross sections of the process 
\EEEEG\ at three different \cms\ energy points are shown in 
Figure~\ref{fig:diffxs_bhabha} together with the prediction 
of the ALIBABA program.

%
%
\section{Results on Total Cross Sections and Forward-Backward Asymmetries}
         \label{sec:xsafbresults}

The results of the measurements of the total cross section performed
between 1993 and 1995 in the four reactions
\EEHADG, \EEMMG, \EETTG\ and \EEEEG\ are listed in 
Tables~\ref{tab:had_xs} to~\ref{tab:electron_xs}.
The measured cross sections for \EEHADG\ are corrected to the full solid angle
for acceptance and efficiencies, keeping a lower cut on the effective 
\cms\ energy of $\RSprime > 0.1\RS$.
The measured cross sections for muon and tau pairs are extrapolated
to the full solid angle and the full phase space using ZFITTER.
The quoted Bhabha cross sections are for both final state leptons
inside the polar angular range $44^\circ < \theta < 136^\circ$, with
an acollinearity angle $\xi < 25^\circ$ and for a minimum energy of $1\ \GeV$
of the final state fermions.
In Table~\ref{tab:electron_xs} the $s$-channel contributions to the cross 
section extrapolated to the full phase space are also given.
Their calculation is described in Appendix~\ref{app:tchannel} and
they can be compared to the measurements of the other leptonic 
final states (Tables~\ref{tab:muon_xs} and~\ref{tab:tau_xs}).
Results of the measurements performed between 1990 and 1992 are presented
in Reference~\cite{l3-69}.

\begin{figure}
 \begin{center}
  \vspace*{-20mm}
  \includegraphics[width=\figwidth]{\pltdir 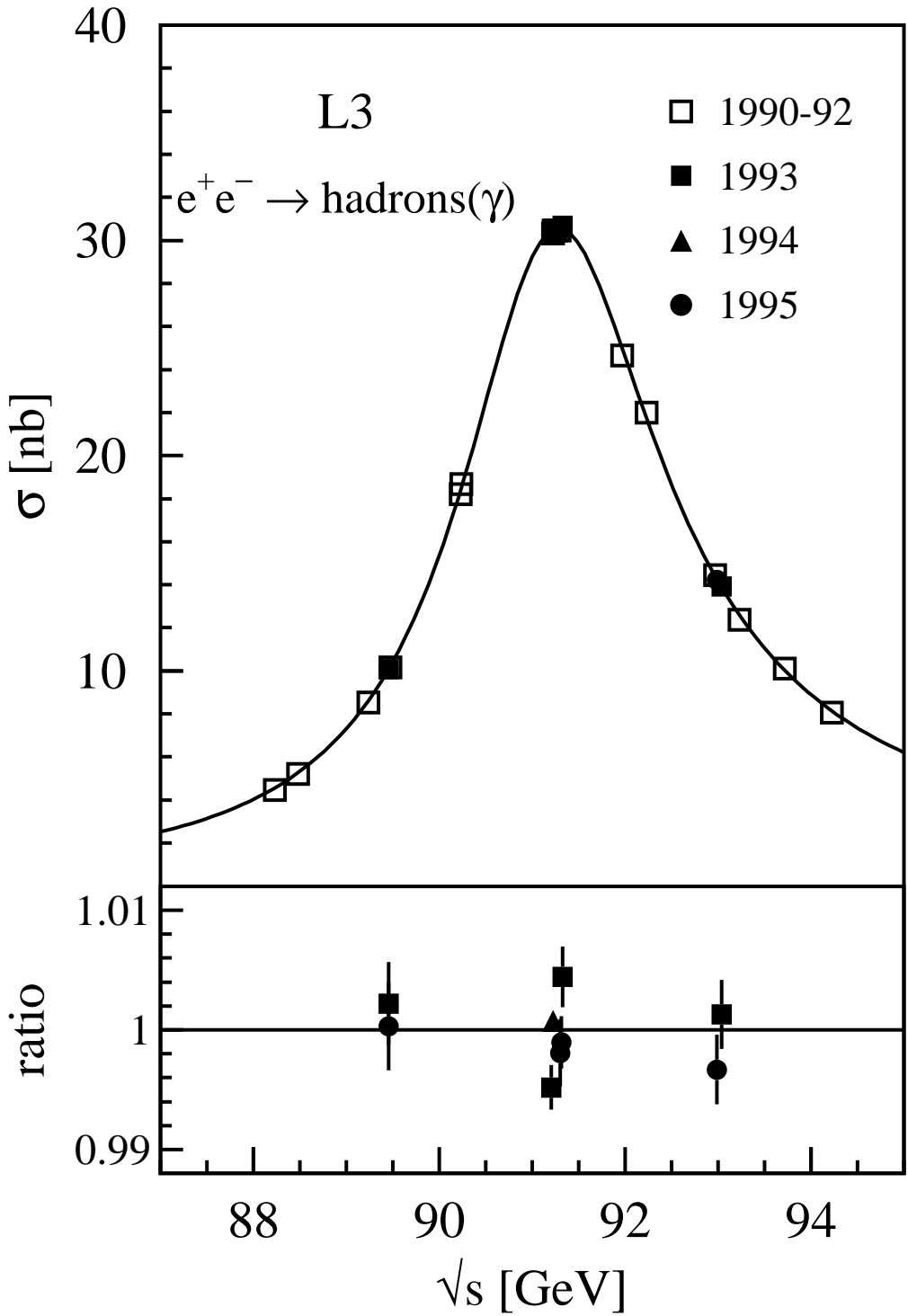}
 \end{center}
 \caption{The measured cross sections \protect\EEHADG\ as function of
          the \protect\cms\ energy.
          The solid line shows the result of the fit.
          At the bottom the ratio of the measured cross sections and the
          fit result for the data collected in $1993-95$ is shown.
          The errors are statistical only.}
\label{fig:hadron_xs}
\end{figure}

\begin{figure}
 \begin{center}
  \vspace*{-20mm}
   \includegraphics[width=\figwidth]{\pltdir 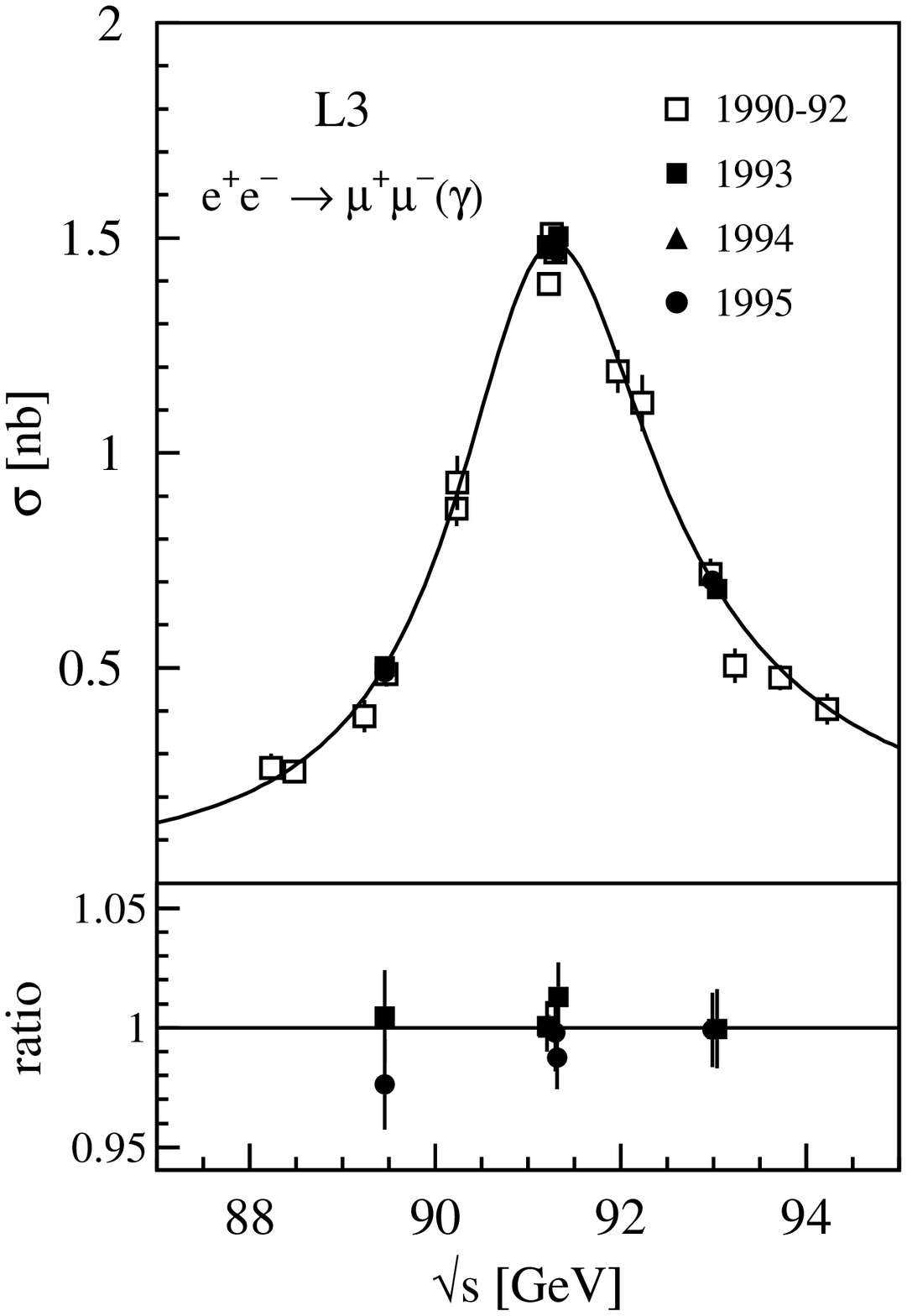}
 \end{center}
 \caption{Same as Figure~\protect\ref{fig:hadron_xs} for \protect\EEMMG.}
\label{fig:muon_xs}
\end{figure}

\begin{figure}
 \begin{center}
  \vspace*{-20mm}
  \includegraphics[width=\figwidth]{\pltdir 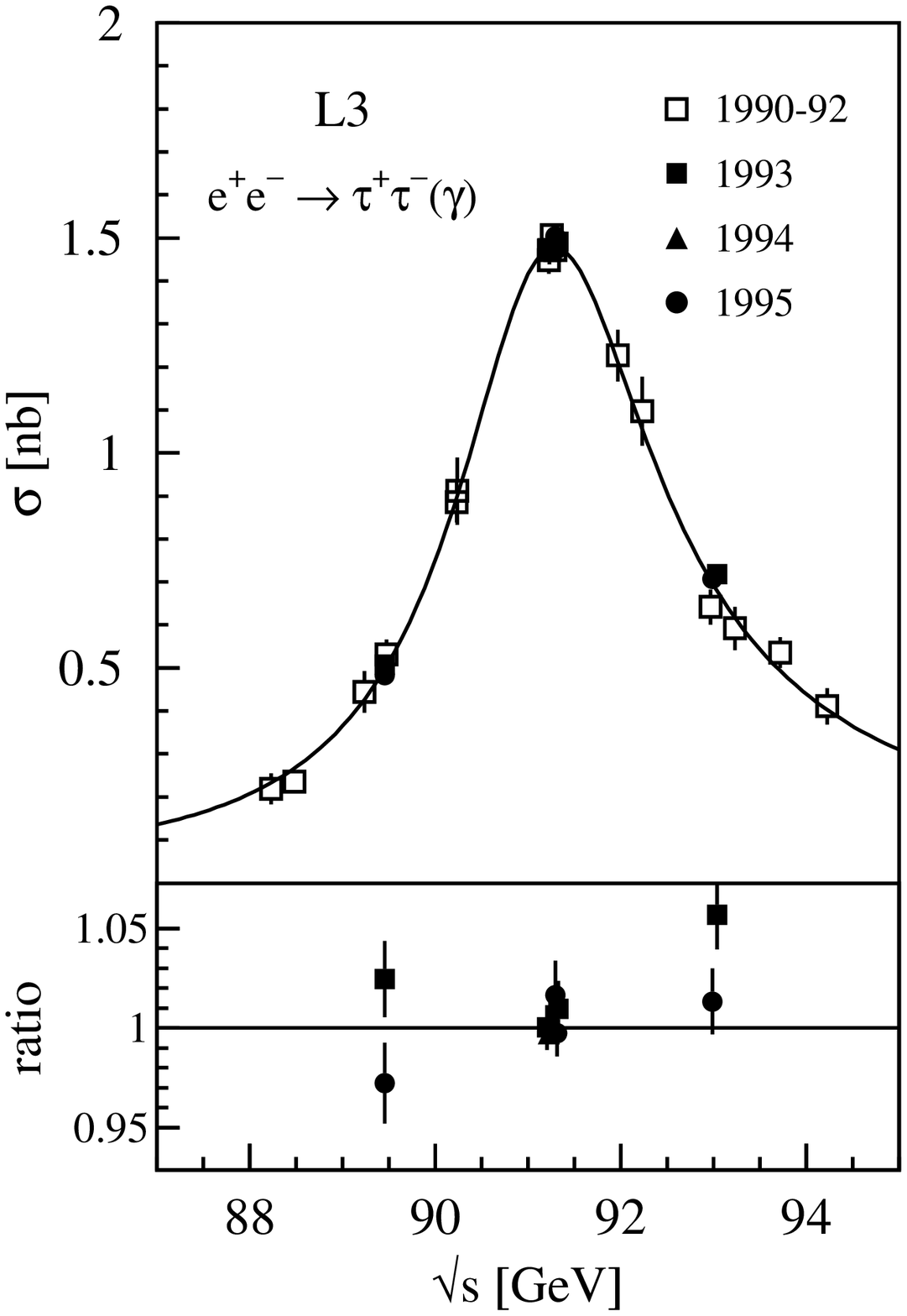}
 \end{center}
 \caption{Same as Figure~\protect\ref{fig:hadron_xs} for \protect\EETTG.}
 \label{fig:tau_xs}
\end{figure}

\begin{figure}
 \begin{center}
  \vspace*{-20mm}
  \includegraphics[width=\figwidth]{\pltdir 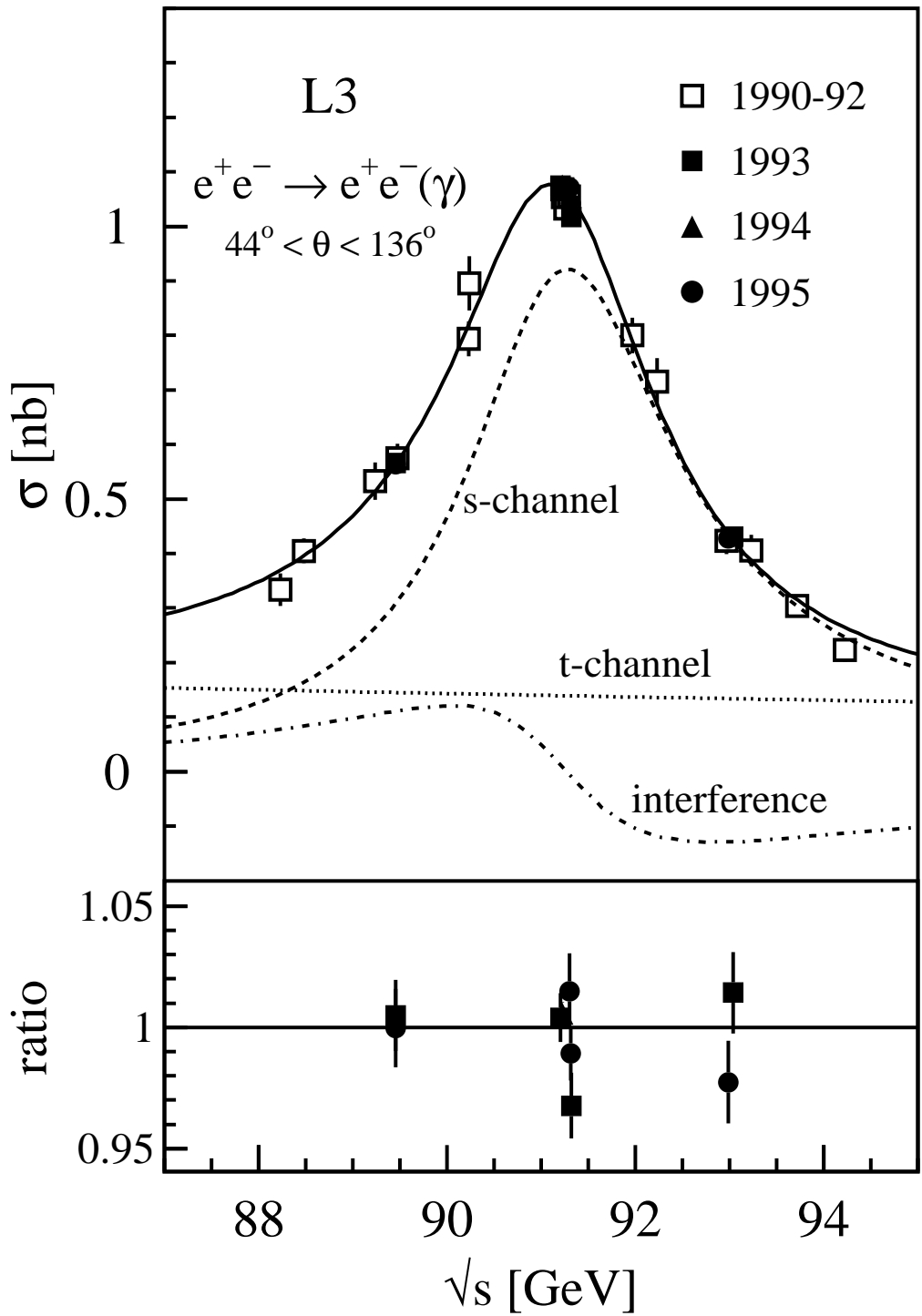}
 \end{center}
 \caption{Same as Figure~\protect\ref{fig:hadron_xs} for 
          \protect\EEEEG\ inside the fiducial volume
          $44^\circ < \theta < 136^\circ$
          for acollinearity angles $\xi<25^\circ$
          and a minimum energy of $1\ \GeV$ of the final state fermions.}
 \label{fig:bhabha_xs}
\end{figure}

Figures~\ref{fig:hadron_xs} to~\ref{fig:bhabha_xs}
compare the measurements of the total cross sections 
performed in $1990-95$ at the \Zo\ pole to the result of the fit to all cross 
section measurements imposing lepton universality 
described in section~\ref{sec:massfit}.
For Bhabha scattering the contributions from the $s$- and $t$-channels
and their interference are displayed separately.
Good agreement between measurements in different years is observed.

The measurements of the forward-backward asymmetry performed
between 1993 and 1995 in the leptonic reactions
\EEMMG, \EETTG\ and \EEEEG\ are listed in 
Tables~\ref{tab:muon_afb} to~\ref{tab:electron_afb}.
For muon and tau pairs the results are extrapolated to the full solid
angle keeping a cut on the acollinearity of $\xi < 15^\circ$ and
$\xi < 10^\circ$, respectively.
The measurements for the process \EEEEG\ apply to the same polar angular range
and cuts as the total cross section.
Table~\ref{tab:electron_afb} contains also the $s$-channel
contributions to the asymmetry (see Appendix~\ref{app:tchannel}) to be
compared to the measurements for muon and tau pairs.

\begin{figure}
 \begin{center}
  \vspace*{-20mm}
  \includegraphics[width=\figwidth]{\pltdir 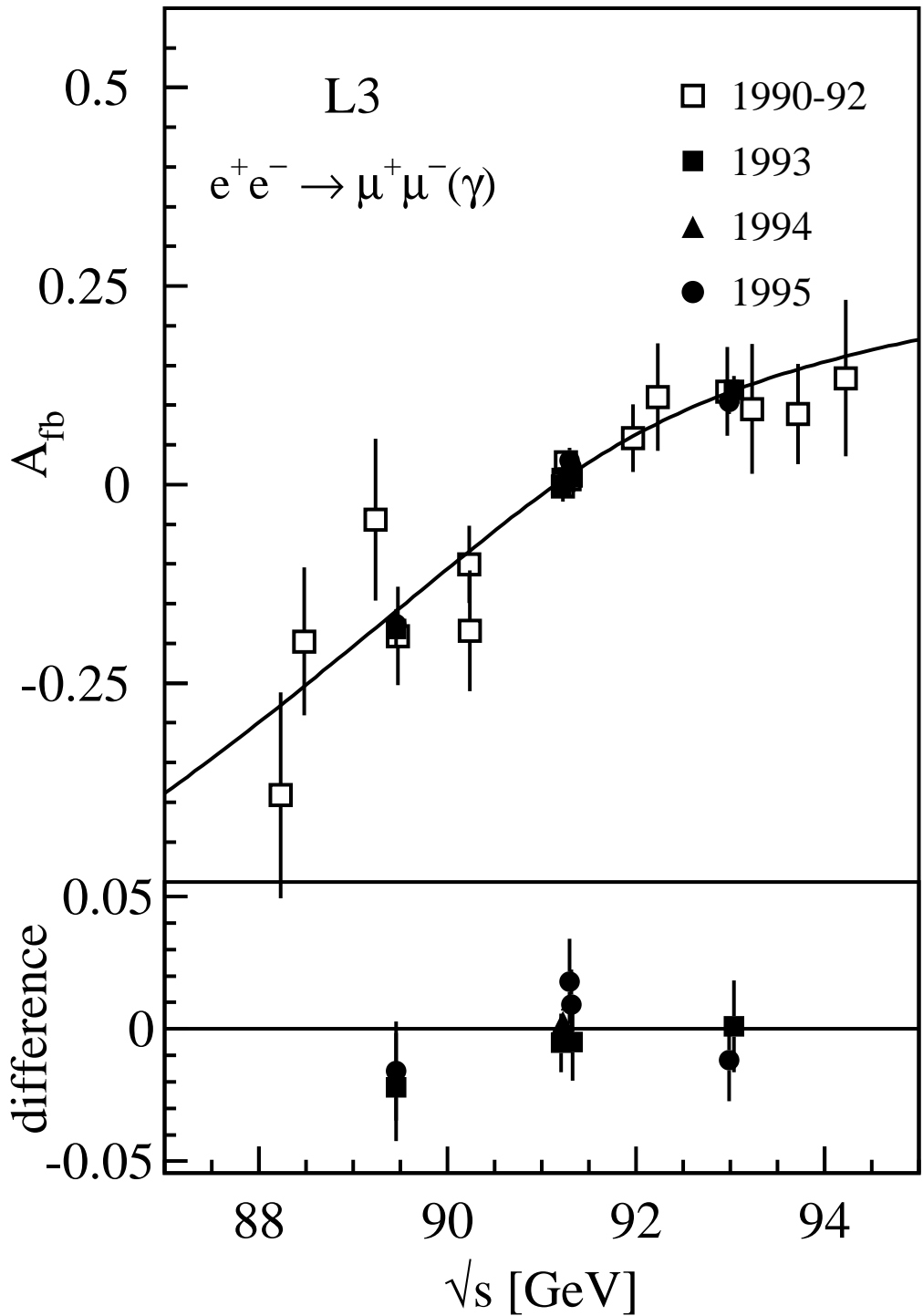}
 \end{center}
 \caption{The measured forward-backward asymmetry in \protect\EEMMG\ as 
          function of the \protect\cms\ energy.
          The solid line shows the result of the fit.
          At the bottom the difference of the measured asymmetry and the
          fit result for the data collected in $1993-95$ is shown.
          The errors are statistical only.}
 \label{fig:muon_afb}
\end{figure}

\begin{figure}
 \begin{center}
  \vspace*{-20mm}
  \includegraphics[width=\figwidth]{\pltdir 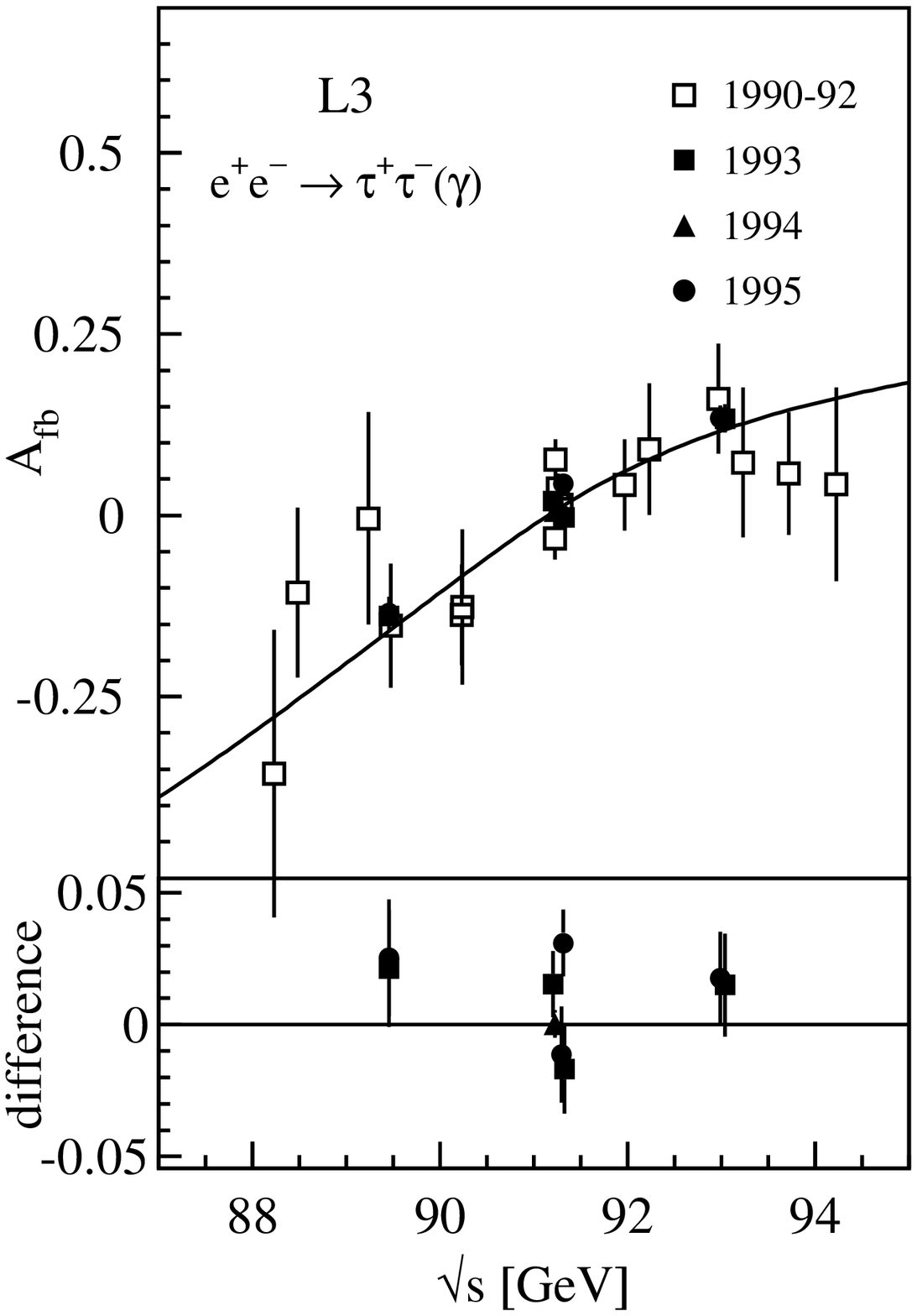}
 \end{center}
 \caption{Same as Figure~\protect\ref{fig:muon_afb} for \protect\EETTG.}
 \label{fig:tau_afb}
\end{figure}

\begin{figure}
 \begin{center}
  \vspace*{-20mm}
  \includegraphics[width=\figwidth]{\pltdir 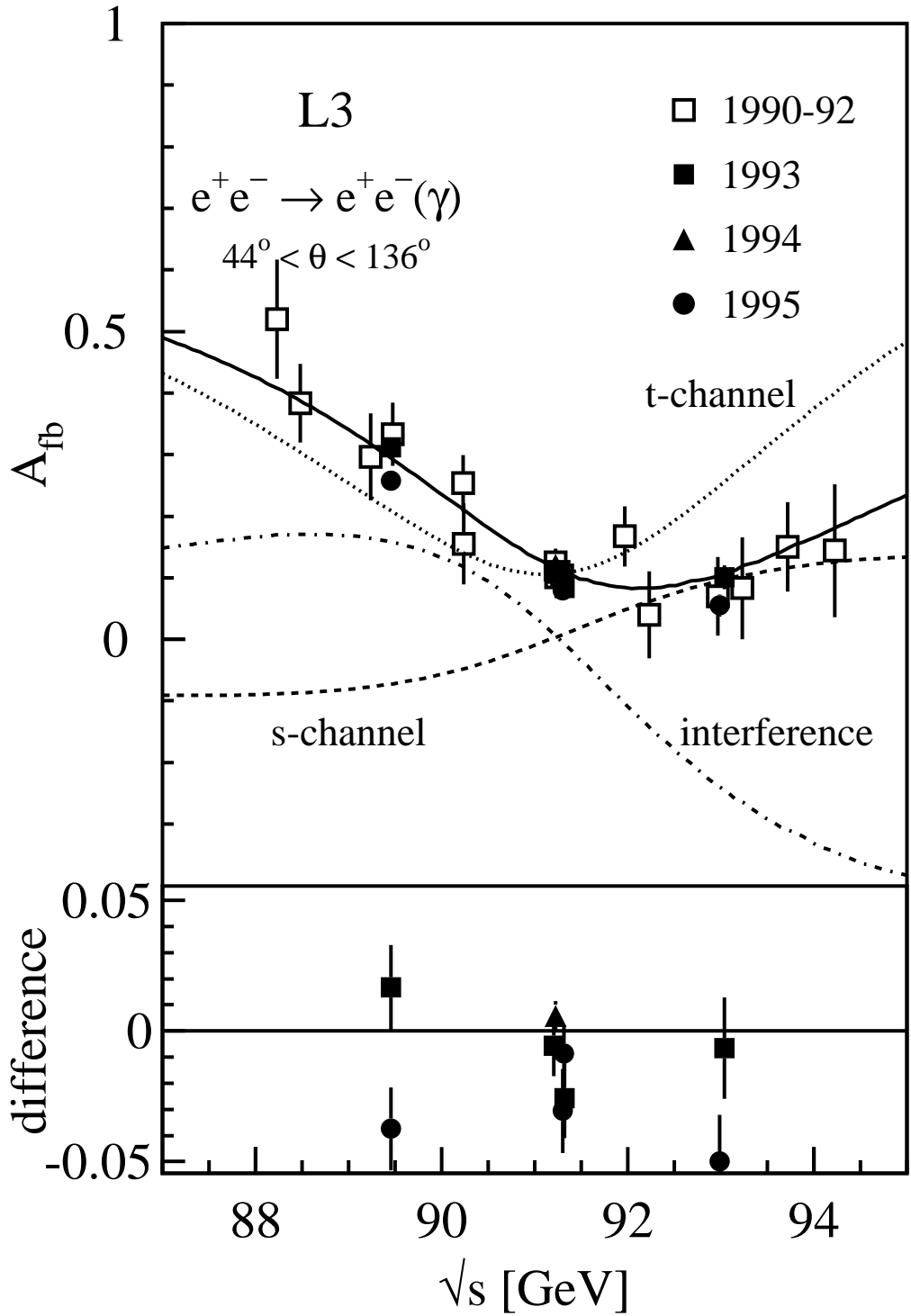}
 \end{center}
 \caption{Same as Figure~\protect\ref{fig:muon_afb} for 
          \protect\EEEEG\ in the fiducial volume
          $44^\circ < \theta < 136^\circ$.
          The same cuts as for the total cross section are applied.}
 \label{fig:bhabha_afb}
\end{figure}

Figures~\ref{fig:muon_afb} to~\ref{fig:bhabha_afb} compare these measurements 
to the results of the
fit to all hadronic and leptonic cross section and forward-backward
asymmetry measurements imposing lepton universality.
For the Bhabha scattering the difference of the forward and backward
cross sections in the $s$- and $t$-channels and in the interference,
all normalised to the total cross section,
are displayed separately.
Good agreement between measurements in different years is observed.

\begin{figure}
 \begin{center}
  \vspace*{-20mm}
  \includegraphics[width=\figwidth]{\pltdir 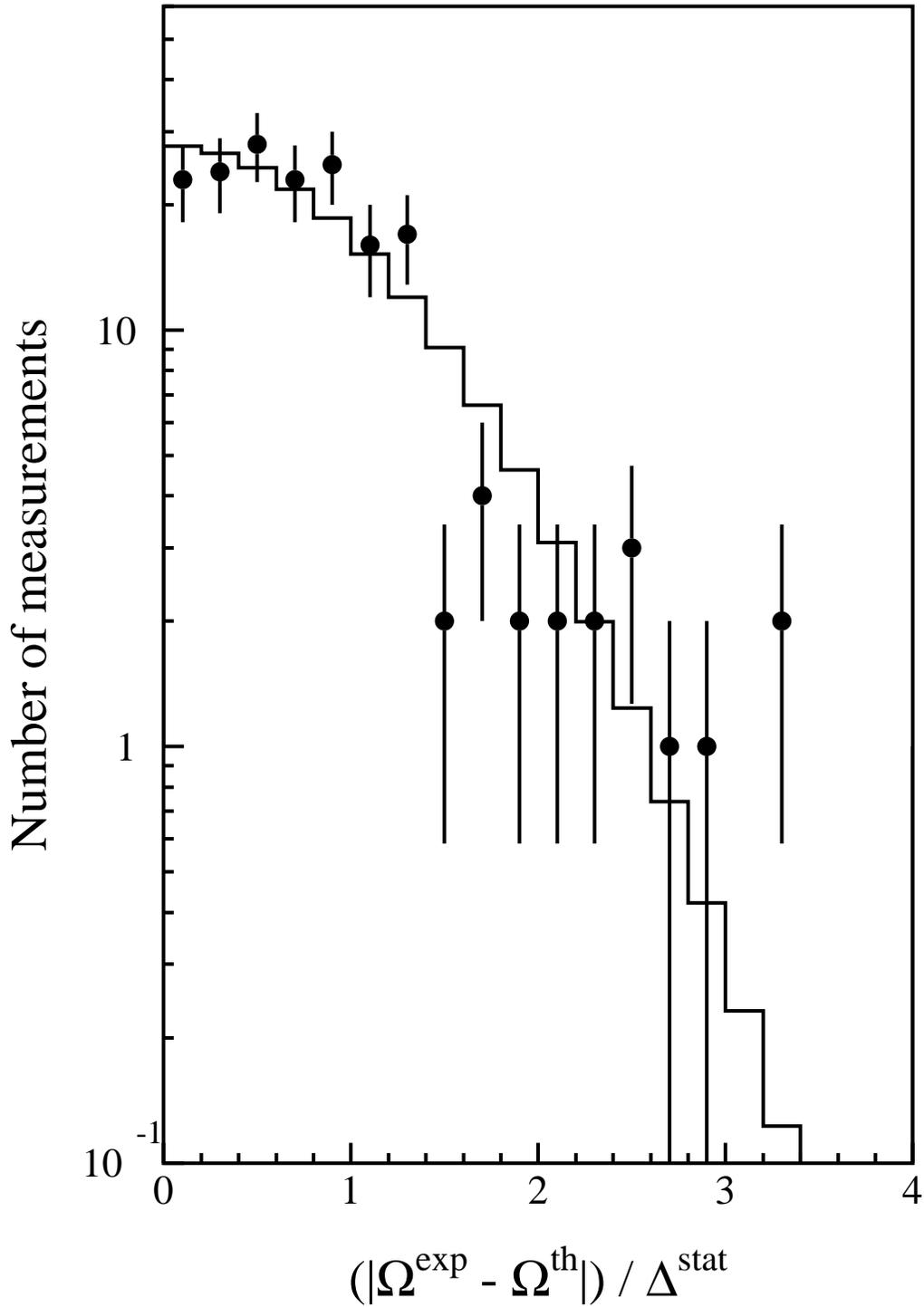}
 \end{center}
 \caption{Distribution of the absolute difference of the measured 
          cross section and forward-backward asymmetries ($1990-95$ data) 
          and the five parameter fit result (Table~\protect\ref{tab:fitpar59})
          divided by the statistical errors of the measurements.
          The histogram shows the expectation for a Gaussian
          distribution of the measurement in the absence of systematic
          errors.}
 \label{fig:prob}
\end{figure}

For the fits presented in the following sections we include the cross section 
and forward-backward asymmetry measurements from $1990-92$~\cite{l3-69}.
All our measurements at the Z resonance performed in the period $1990-95$ 
are self-consistent.
Qualitatively this can be seen from Figure~\ref{fig:prob} where for all
175 measurements the absolute difference between the measurements
and the expectations, divided by the statistical error of the measurements, 
is shown.
The expected cross sections and forward-backward asymmetries are calculated
from the result of the five parameter fit presented in 
Section~\ref{sec:fit_par59}.
The scattering of our measurements is compared with the one expected from 
a perfect Gaussian distribution.
The agreement is satisfactory considering that due to their complicated
correlations, systematic errors cannot be taken into account in 
this comparison.

%
%
\section{Fits for Electroweak Parameters}
         \label{sec:fit_intro}

Different analyses are used to extract electroweak parameters from the 
measured total cross sections and forward-backward asymmetries.

Firstly, we determine the electroweak parameters making a minimum of 
assumptions about any underlying theory, for example the \SM. 
The first analysis uses only the total cross section data to determine 
the parameters of the \Zo\ boson, its mass, the total and partial 
decay widths to fermion pairs.
The second analysis also includes the asymmetry data, which allows the 
determination of the coupling constants of the neutral weak current.
In a third analysis we fit the cross section and forward-backward 
asymmetry measurements in the S-Matrix ansatz~\cite{SMatrix} where all 
contributions from  \GZint\ are determined from the data.
Finally, all our measurements on electroweak observables are interpreted
in the framework of the \SM\ in order to determine its free parameters.

\subsubsection*{Lowest Order Formulae}

In all analyses, a Breit-Wigner ansatz is used to describe the 
\Zo\ boson. 
The mass, \MZ, and the total width, \GZ, of the Z boson are 
defined by the functional form of the Breit-Wigner denominator, which 
explicitly takes into account the energy dependence of the total 
width~\cite{swidth}. 
The total $s$-channel cross section to lowest order, $\sigma^\circ$, 
for the process  $\ee \rightarrow \ffbar$, is given by the sum of three terms, 
the \Zo\ exchange, 
$\sigma^\circ_{\mathrm{Z}}$, the photon exchange, $\sigma^\circ_\gamma$, 
and the \GZint, $\sigma^\circ_{\mathrm{int}}$:
\begin{eqnarray}
\sigma^\circ                & = & \sigma^\circ_{\mathrm{Z}} 
                                  + \sigma^\circ_{\gamma} 
                                  + \sigma^\circ_{\mathrm{int}}
                                   \nonumber \\
\sigma^\circ_{\mathrm{Z}}   & = & 
                        \frac{12\pi}{\MZ^2}\frac{\Gamma_{\MR{e}}\Gf}{\GZ^2}
                        \frac{s\GZ^2}{(s-\MZ^2)^2 + s^2\GZ^2/\MZ^2}    
                                   \nonumber \\ 
\sigma^\circ_{\gamma}       & = & \frac{4\pi\alpha^2}{3s}
                                q_{\MR{e}}^2q_{\mathrm{f}}^2 N_{\MR{C}}^{\MR{f}}
                                   \nonumber \\
\sigma^\circ_{\mathrm{int}} & = & \frac{4\pi\alpha^2}{3} J_{\MR{f}}
                             \frac{s-\MZ^2}{(s-\MZ^2)^2 + s^2\GZ^2/\MZ^2}, 
                                   \label{eq:xsloworder}
            \qquad \mathrm{f} = \mathrm{e},\,\mu,\,\tau,\,\mathrm{q}
\end{eqnarray}           
where $q_{\mathrm{f}}$ is the electric charge of the final-state fermion, 
$N_{\MR{C}}^{\MR{f}}$ its colour factor, and $\alpha$ the electromagnetic 
coupling constant. The pure photon exchange is determined by QED.
                   
The first analysis treats the mass and the total and partial widths 
of the \Zo\ boson as free and independent parameters. 
The interference of the Z exchange with the photon exchange adds another
parameter, the \GZint\ term, $J_{\MR{f}}$, besides those 
corresponding to mass and widths of the \Zo. 
Since in the \SM\ $|\sigma^\circ_{\mathrm{int}}(s)|\ll \sigma^\circ(s)$ 
for \cms\ energies close to \MZ, it is difficult to 
measure $J_{\MR{f}}$ accurately using data at the \Zo\ only. 
The \GZint\ term is usually taken from the 
\SM~\cite{l3-28,l3-69,l3-58}, 
thus making assumptions about the form of the electroweak unification. 
                         
The second analysis determines the vector and axial-vector coupling
constants of the neutral weak current to charged leptons, 
$g_{\mathrm{V}}^\ell$ and $g_{\mathrm{A}}^\ell$, 
by using the forward-backward asymmetries in addition to the 
total cross sections.
In lowest order, for $\RS=\MZ$ and neglecting the photon exchange,
the $s$-channel forward-backward asymmetry for the process 
\EELLG\ is given by:
\begin{eqnarray}
   \AFBZl & = &  \frac {3}{4} A_{\MR{e}} A_{\ell} \quad \MR{with\ }  
                                                                 \nonumber \\
   A_\ell & = &  \frac{2\, g_{\MR{V}}^\ell g_{\MR{A}}^\ell}
                            {(g_{\MR{V}}^\ell)^2 + (g_{\MR{A}}^\ell)^2} \, .
\label{eq:AFBloworder}
\end{eqnarray}
The energy dependence of the asymmetry distinguishes $g_{\MR{V}}^\ell$ and 
$g_{\MR{A}}^\ell$~\cite{YB_Boehm}.
The experimental precision on the coupling constants is improved by also
including information from tau-polarisation measurements which determine
$A_{\MR{e}}$ and $A_{\MR{\tau}}$ independently.

In Equation~\ref{eq:xsloworder}, the leptonic partial width, $\GL$, 
and the leptonic \GZint\ term, $J_\ell$,  
are now expressed in terms of $g_{\MR{V}}^\ell$ and 
$g_{\MR{A}}^\ell$:
\begin{eqnarray}
 \GL    & = & \frac{\GF\MZ^3}{6\sqrt{2}\pi} 
              \left[ (g_{\MR{V}}^\ell)^2 + (g_{\MR{A}}^\ell)^2 \right] 
                                                                \nonumber\\
 J_\ell & = & \frac{\GF\MZ^2}{\sqrt{2}\pi\alpha} 
              \, q_{\MR{e}} q_\ell 
              \, g_{\MR{V}}^{\MR{e}} g_{\MR{V}}^\ell,
\label{eq:Gammalepton}
\end{eqnarray}                                                          
where \GF\ is the Fermi coupling constant.
The hadronic cross section is given by the sum over the five kinematically
allowed flavours and their colour states.
Because no separation of quark flavours is attempted, this approach cannot be 
applied to the hadronic final state.
Therefore, the parameterisation of the first analysis is used to express the 
hadronic cross section in terms of \Gh\ and $J_{\mathrm{had}}$.

Our data are also interpreted in the framework of the S-Matrix 
ansatz~\cite{SMatrix}, which makes a minimum of theoretical assumptions.
This ansatz describes the hard scattering process of fermion-pair
production in \ee-annihilations by the $s$-channel exchange of two
spin-1 bosons, a massless photon and a massive \Zo\ boson.
The lowest-order total cross section, $\sigma_{\MR{tot}}^0$, and 
forward-backward asymmetry, \AFBZ, for $\ee\rightarrow\ffbar$ are given 
as:
\begin{eqnarray}
 \sigma_{\MR{a}}^0(s) &=& 
    \frac{4}{3}\pi\aqed^2 \left[    
             \frac{ g_{\MR{f}}^{\MR{a}} }{ s }
           + \frac{ s\,r_{\MR{f}}^{\MR{a}} 
                    + \left( s - \MZbar^2 \right) j_{\MR{f}}^{\MR{a}} }
                   { \left( s - \MZbar^2 \right)^2
                    + \MZbar^2 \GZbar^2  }       
      \right]
         \qquad\qquad \MR{for\ a = tot,FB}      \nonumber \\
 \AFBZ(s)             &=&
   \frac{3}{4} \frac{ \sigma_{\MR{FB}}^0(s) }{ \sigma_{\MR{tot}}^0(s) } \, .
 \label{eq:SMatrix}
\end{eqnarray}

The S-Matrix parameters $r_{\MR{f}}^{\MR{a}}$, $j_{\MR{f}}^{\MR{a}}$
and $g_{\MR{f}}^{\MR{a}}$ are real numbers which express
the size of the \Zo\ exchange, \GZint\  and photon exchange
contributions. 
Here, $r_{\MR{f}}^{\MR{a}}$ and  $j_{\MR{f}}^{\MR{a}}$ are treated as
free parameters while the photon exchange contribution, 
$g_{\MR{f}}^{\MR{a}}$, is fixed to its QED prediction. 
Each final state is thus described by four free parameters: two
for cross sections, 
$r_{\MR{f}}^{\MR{tot}}$ and  $j_{\MR{f}}^{\MR{tot}}$, 
and two for forward-backward asymmetries, 
$r_{\MR{f}}^{\MR{FB}}$ and  $j_{\MR{f}}^{\MR{FB}}$.
In models with only vector and axial-vector couplings of the \Zo\ boson, 
these four S-Matrix parameters are not independent of each other:
\begin{eqnarray}
  r_{\MR{f}}^{\MR{tot}} & \propto & 
         \left[ (g_{\MR{V}}^{\MR{e}})^2 +(g_{\MR{A}}^{\MR{e}})^2 \right] \,
         \left[ (g_{\MR{V}}^{\MR{f}})^2 +(g_{\MR{A}}^{\MR{f}})^2 \right]
                                                              \, , \nonumber \\
 j_{\MR{f}}^{\MR{tot}}  & \propto & g_{\MR{V}}^{\MR{e}} g_{\MR{V}}^{\MR{f}}
                                                              \, , \nonumber \\
 r_{\MR{f}}^{\MR{FB}}   & \propto & g_{\MR{A}}^{\MR{e}} g_{\MR{V}}^{\MR{e}} \, 
                                    g_{\MR{A}}^{\MR{f}} g_{\MR{V}}^{\MR{f}}
                                                              \, , \nonumber \\
 j_{\MR{f}}^{\MR{FB}}   & \propto & g_{\MR{A}}^{\MR{e}} g_{\MR{A}}^{\MR{f}}
                                                              \, .
\label{eq:SMatrixparam}
\end{eqnarray}
Under the assumption that only vector- and axial-vector couplings exist,
the S-Matrix ansatz corresponds to the second analysis discussed above 
without fixing $J_{\mathrm{had}}$ to the \SM.

The S-Matrix ansatz is defined using a Breit-Wigner denominator with 
$s$-independent width for the \Zo\ resonance.
To derive the mass and width of the \Zo\ boson for a Breit-Wigner with
$s$-dependent width, the following transformations are 
applied~\cite{SMatrix}:
$\MZ = \MZbar + 34.1\ \MeV$ and $\GZ = \GZbar + 0.9\ \MeV$.

\subsubsection*{Radiative Corrections}
                     
The QED radiative corrections to the total cross sections and 
forward-backward asymmetries are included by convolution and by the 
replacement $\alpha\rightarrow\alpha(s)=\alpha/(1-\Delta\alpha)$ to 
account for the running of the electromagnetic coupling 
constant~\cite{YB_berends,YB_Boehm}.

Weak radiative corrections are calculated assuming the validity of 
the \SM\ and as a function of the unknown mass of the Higgs boson.  
The coupling constants which are real to lowest order are modified by 
absorbing weak corrections and become complex 
quantities~\cite{YB_Consoli}.
Effective couplings, \gVbl\ and \gAbl, are defined which correspond to the 
real parts.  
When extracting \gVbl\ and \gAbl\ from the measurements, the small imaginary 
parts are taken from the SM.
Observables such as the leptonic partial widths (Equation~\ref{eq:Gammalepton})
and the leptonic pole asymmetry (Equation~\ref{eq:AFBloworder})
are redefined by replacing the vector and axial-vector coupling constants
by these effective couplings.

The effective couplings of fermions are expressed in terms of the effective 
electroweak mixing angle, \swsqb, and the effective ratio of the neutral 
to charged weak current couplings, 
$\bar{\rho} = 1/(1-\Delta\bar{\rho})$~\cite{YB_Consoli}:
\begin{eqnarray}
 \gVb^{\MR{f}} &=& \sqrt{\bar{\rho}}\, (I^{\MR{f}}_3 - 2\,q_{\MR{f}}\swsqb) 
                                                     \nonumber   \\
 \gAb^{\MR{f}} &=& \sqrt{\bar{\rho}}\, I^{\MR{f}}_3  \, ,
\label{eq:def_rho}
\end{eqnarray}
where $I^{\MR{f}}_3$ is the third component of the weak isospin of the 
fermion~f.
Due to weak vertex corrections, the definitions of $\bar{\rho}$ and 
\swsqb\ depend on the fermion.
However, except for the b-quark, these differences are small compared to the
experimental precision.
Therefore, we define \swsqb\ as the effective weak mixing angle for a
massless charged lepton.
It is related to the  on-shell definition of the weak mixing angle,
\swsq\ by the factor $\kappa$:
\begin{equation}
    \kappa = \frac{\swsqb}{\swsq} \qquad \mathrm{with\ }
    \swsq =  1 - \frac{\MW^2}{\MZ^2}\, .
\label{eq:def_kappa}
\end{equation}

\subsubsection*{Fits in the \SM}

The fourth analysis to determine electroweak parameters uses the 
framework of the \SM.
By comparing its predictions with the set of experimental measurements, 
it is possible to test the consistency of the \SM\ and to constrain the
mass of the Higgs boson.

The input parameters of the \SM\ are \aqed, the fermion masses, \MH, 
\MZ\ and the mass of the W boson, \MW. 
QCD adds one more parameter, the strong coupling constant, \as, 
which is relevant for hadronic final states.
The Cabbibo-Kobayashi-Maskawa matrix, relating electroweak and mass 
eigenstates of quarks, is not important for total hadronic cross sections
in neutral current interactions considered here.
Concerning the fermion masses, only the mass of the top quark is important
for \SM\ calculations performed below. 
All other masses are too small to play a significant role or are known to 
sufficient precision.

Generally, in \SM\ calculations for observables at the \Zo\ resonance,
the mass of the W is replaced by the Fermi coupling constant, $\GF$,
which is measured precisely in muon decay~\cite{GF_reference}.
These two parameters are related by
\begin{equation}
 \frac{\GF}{\sqrt{2}}  = 
     \frac{\pi\alpha}{2} \, \frac{1}{\MZ^2\swsq\cwsq} \, \nonumber \\
     \frac{1}{1 - \Delta r} \, ,
\label{eq:GFMW}
\end{equation}
where $\Delta r$ takes into account the electroweak radiative corrections.
These corrections can be split into QED corrections due to the
running of the QED coupling constant, $\Delta \aqed$, and pure weak corrections,
$\Delta r_{\MR{w}} $~\cite{YB_Burgers,Hollik1}:
\begin{eqnarray}
    \Delta r          &=& \Delta \aqed + \Delta r_{\MR{w}} \nonumber \\
    \Delta r_{\MR{w}} &=& - \ctwsq\,\Delta\bar{\rho} \, 
                          + \Delta r_{\MR{rem}}             \, .
 \label{eq:rweak}
\end{eqnarray}
The corrections $\Delta r_{\MR{rem}}$, not absorbed in the 
$\rho$-parameter, are smaller than the  main contributions discussed below 
but are nevertheless numerically important~\cite{Jegerlehner1,Hollik1} and
included in the calculations.

Weak radiative corrections originate mainly from loop corrections to the 
W propagator due to the large mass splitting in the top-bottom iso-spin doublet 
and Higgs boson loop corrections to the propagators of the heavy gauge 
bosons~\cite{Veltman}.
To leading order they depend quadratically on \Mt\ and 
logarithmically on \MH.
The $\Zo\bbar$ vertex receives additional weak radiative corrections which 
depend on the top mass.
Through the measurements of weak radiative corrections our results 
at the \Zo\ are sensitive to the mass of the top quark and the Higgs boson.
This allows to test the \SM\ at the one-loop level by comparing the top mass 
derived from our data with the direct measurement and to estimate the mass 
of the yet undiscovered Higgs boson which is one of the fundamental parameters
of the \SM.
With this procedure the relevant parameters in \SM\ fits are 
\MZ, \Mt, \MH, \aqedMZ\ and \as.

\subsubsection*{Fitting Programs and Methods}

The programs ZFITTER~\cite{ZFITTER621} and TOPAZ0~\cite{TOPAZ044}
are used to calculate radiative corrections and \SM\ predictions.
For computational reasons the fits are performed using ZFITTER.

Both programs include complete ${\cal O}(\alpha^2)$
and leading ${\cal O}(\alpha^3 \ln^3(s/m^2_{\mathrm{e}}) )$
QED calculations of initial state radiation~\cite{QED_calc}.
Final state corrections are calculated in ${\cal O}(\alpha)$ for QED and
${\cal O}(\alpha_{\mathrm{s}}^3)$~\cite{Larin} for QCD including also mixed 
terms ${\cal O}(\alpha\as)$.   
Interference of initial and final state radiation is included up to
${\cal O}(\alpha)$ corrections.
Pair production by initial state radiation is implemented~\cite{Arbuzov_ispp}.

Electroweak radiative corrections are complete at the one-loop level and are 
supplemented by leading ${\cal O} (\GF^2\Mt^4)$ and sub-leading
${\cal O} (\GF^2\Mt^2\MZ^2)$~\cite{Degrassi} two-loop corrections.
Complete mixed QCD-electroweak corrections of ${\cal O}(\alpha\as)$ with
leading ${\cal O}(\GF\Mt^2\alpha_s^2)$ terms are included~\cite{Kniehl}
together with a non-factorizable part~\cite{Czarnecki}.

For the reaction \EEEEG\ the contributions from the $t$-channel photon
and Z boson exchange and the \stint\ are calculated with the 
programs ALIBABA~\cite{ALIBABA} and TOPAZ0 (see Appendix~\ref{app:tchannel}).

Electroweak parameters are determined in $\chi^2$ fits using the 
MINUIT~\cite{MINUIT} program.
The $\chi^2$ is constructed from the theoretical expectations,
our measurements and their errors, including the correlations.
Apart from experimental statistical and systematic errors, and theoretical
errors, we take into account uncertainties on the LEP \cms\ energy. 
Technical details of the fit procedure are described in
Appendix~\ref{app:covariance}.

Theoretical uncertainties on \SM\ predictions of cross sections, asymmetries, 
Z decay widths and effective coupling constants are studied in detail in
Reference~\cite{LEP_PRWG,PCRP}.
Errors on the theoretical calculations of cross section and forward-backward
asymmetries based on total and partial \Zo\ widths, effective couplings
or S-Matrix parameters,
as used in the fits of Section~\ref{sec:fit_Zpara}, 
arise mainly from the finite precision of the QED convolution. 
They are found to be small compared to the experimental precision and
do not introduce sizeable uncertainties in the fit for \Zo\ parameters.
Residual \SM\ uncertainties in the imaginary part of the effective
couplings are even smaller.

The only exceptions are the theoretical uncertainty on the luminosity
determination, discussed in Section~\ref{sec:luminosity},
and the treatment of $t$-channel and \stint\ 
contributions to the \ee\ final state due to missing higher 
order terms and the precision of the ALIBABA program~\cite{Beenakker}.
Uncertainties on the Bhabha cross section and forward-backward asymmetry 
from calculations of the $t$-channel and \stint\ contributions 
are discussed in Appendix~\ref{app:tchannel}.

Additional uncertainties arise in the calculation of \SM\ parameters from 
the application of different re-normalisation schemes, momentum transfer scales
for vertex corrections and factorisations schemes~\cite{LEP_PRWG}.
By comparing different calculations as implemented in ZFITTER and TOPAZ0,
we find that the impact of these theoretical uncertainties on the fit results
for \SM\ parameters presented in Section~\ref{sec:fit_SMpara} is negligible 
compared to the experimental errors.

ZFITTER and TOPAZ0 calculations in the \SM\ framework are performed
based on five input parameters: 
the masses of the Z and Higgs bosons, the top quark mass, 
the strong coupling constant \as\ at \MZ\ and
the contribution of the five light quark flavours, \Da5had, to the running of 
the QED coupling constant to \MZ.
For comparison to the \SM\ we use the following set of values and 
uncertainties~\cite{mt_direct_pdg98,l3-178,MH1000,Jegerlehner_aqed,PDG98}:
\begin{equation}
 \begin{array}{r@{$\,=\,$}lr@{$\,=\,$}ll}
           \MZ & \mZqval \pm \mZqerr\ \MeV \, , &       
           \Mt & 173.8 \pm 5.2\ \GeV       \, , &       
           95.3\ \GeV \leq \MH \leq 1\ \TeV      \, , \\      
      \Da5had  & 0.02804 \pm0.00065        \, , &       
           \as & 0.119 \pm 0.002           \, .         
 \end{array}
\label{eq:smpara}
\end{equation}
The central values are calculated for $\MH = 300\ \GeV$.
This arbitrary choice is motivated by the logarithmic dependence of
electroweak observables on \MH\ and it leads to approximately symmetric
theoretical errors.

We use the default settings of ZFITTER which provide the most accurate 
calculations.
Exceptions are that in all calculations we allow for the variation of the 
contribution of the five light quarks to the running of the QED coupling 
constant, \Da5had.
Secondly, as recommended by the authors of ZFITTER, the corrections of
Reference~\cite{Czarnecki} 
are explicitly calculated for \SM\ expectations and in fits in the 
\SM\ framework (Section~\ref{sec:fit_SMpara}).
In all other cases they are absorbed in the definitions of the parameters.

%
%

\section{Determination of \protect\boldmath\Zo\  Parameters}
         \label{sec:fit_Zpara}

\subsection{Mass, Total and Partial Widths of the \protect\boldmath\Zo}
\label{sec:massfit}

We determine the mass, the total width and the partial decay widths of the
\Zo\ into hadrons, electrons, muons and taus in a fit to the measured total 
cross sections.
These parameters describe the contribution of the \Zo\ exchange to the total
cross section.
The photon exchange and \GZint\ contributions are fixed to their 
\SM\ expectations.
Two fits are performed: one assuming and one not assuming lepton universality,
where in the first one a common leptonic width is defined as the decay width of
the \Zo\ into a pair of massless charged leptons.
The results of both fits are summarised in Table~\ref{tab:fitwidth} 
and the correlation coefficients for the parameters determined in the two 
fits are given in Tables~\ref{tab:fitcor6} and~\ref{tab:fitcor4},
respectively.
The partial decay widths into the three charged lepton species are found to be 
consistent within errors.
It should be noted that due to the mass of the tau lepton, \GT\ is expected
to be $0.19\ \MeV$ smaller than \GL.

Our new results with significantly reduced errors are in agreement with the
\SM\ expectations and our previous measurements~\cite{l3-69}.
For the mass \MZ\ and the total width \GZ\ we obtain:
\begin{eqnarray}
   \MZ &=& \mZqval \pm \mZqerr\ \MeV \, ,\nonumber \\
   \GZ &=& \GZqval \pm \GZqerr\ \MeV \, .
\label{eq:mzgz_result}
\end{eqnarray}
These are measurements of the mass of the \Zo\ boson
with an accuracy of $3.4 \cdot 10^{-5}$ and of its total decay width
of $1.7 \cdot 10^{-3}$.
The contribution to the total errors on \MZ\ and \GZ\ from the LEP energy
is estimated by  performing fits to the $1993-95$ data with and without 
taking into account LEP energy errors.
From a quadratic subtraction of the errors of the fitted parameters we find
$\Delta\MZ(\mathrm{LEP}) = 1.8\ \MeV$ and
$\Delta\GZ(\mathrm{LEP}) = 1.3\ \MeV$,
in agreement with the estimates given in Reference~\cite{lep-09}.

The impact of the uncertainties on \SM\ parameters on the fit results is 
negligible.
The largest effect is an uncertainty on the \Zo\ mass of $\pm 0.2\ \MeV$ caused
by the calculation of the \GZint\ contribution when varying the Higgs and top
masses and $\Da5had$ in the ranges given in Equation~\ref{eq:smpara}.

Motivated by the different methods used to obtain the absolute scale of the
LEP energy in the years $1990-92$, $1993-94$ and 1995, resulting in different 
uncertainties, we determine the mass of the \Zo\ for these three periods
independently.
The mass values obtained are consistent within their statistical errors.

To check our results on \MZ\ and \GZ\ the fit assuming lepton universality
is repeated twice: i) using only the leptonic cross sections and 
ii) using only the \EEEEG\ data.
The results for the mass and total width obtained this way are
$\MZ = 91\,198.7 \pm 8.2\ \MeV$, $\GZ = 2\,508.1 \pm 13.4\ \MeV$ 
using all three lepton species and
$\MZ = 91\,177 \pm 16\ \MeV$,    $\GZ = 2\,497 \pm 26\ \MeV$ 
when using only Bhabha scattering data.
Within the errors, dominated by the statistical errors of the measurements,
these values are in agreement 
with those given in Table~\ref{tab:fitwidth} where the \EEHADG\ cross section
measurements contribute most.
Also, we conclude that there is no significant bias introduced in the 
determination of the mass and the total width of the Z boson
by the treatment of the $t$-channel in Bhabha scattering.

From the difference of the total width and the partial widths into hadrons
and charged leptons, including their correlations, the decay width 
of the \Zo\ into invisible particles is derived to be
\begin{equation}
 \Ginv = \Ginvqval \pm \Ginvqerr\ \MeV\, . 
\label{eq:ginv}
\end{equation}
This number is determined in the fit assuming lepton universality
and it is in agreement with our direct determination of \Ginv\ from 
cross section measurements of the reaction \EENNG~\cite{l3-141} which yields
$ \Ginv = 498 \pm 12\MR{(stat)} \pm 12\MR{(sys)}\ \MeV$.

In the \SM, the invisible width is exclusively given by the \Zo\ decays
into neutrinos and the result can be interpreted as the number of
neutrino generations $N_{\MR{\nu}}$.
Using the \SM\ prediction $\GL/\GN = 0.5021 \pm 0.0002$ 
for the ratio of the \Zo\ decay width into charged leptons and neutrinos
we obtain:
\begin{equation}
 N_{\MR{\nu}} = \frac{\Ginv}{\GL}\, 
                   \left( \frac{\GL}{\GN} \right)^{\MR{SM}}
              = \nnqval \pm \nnqerr \, .
\label{eq:nnuresult}
\end{equation}
This formula is used because the experimental precision on the ratio 
$\Ginv/\GL$ is better than that on \Ginv.

\subsection{Limits on Non-Standard Decays of the \protect\boldmath\Zo}
\label{sec:nonSM}

From the measurements of total and partial \Zo\ decay widths presented
in the previous section we derive experimental limits on additional
\Zo\ decay widths not accounted for in the \SM.
These limits take into account experimental and theoretical errors
added in quadrature.
The latter are derived from adding in quadrature the changes in the
theoretical predictions when varying the \SM\ input parameters by their 
errors as given in Equation~\ref{eq:smpara}.
This is motivated by the fact that these parameters are determined in
independent experiments with the exception of the mass of the Higgs 
boson.
A value of $\MH=1\ \TeV$ is used here to calculate the Z widths which results 
in the lowest \SM\ predictions and therefore in conservative limits.

The 95\% confidence level (C.L.) limits on non-standard decay widths, 
$\Gamma^{\MR{NP}}_{\MR{95}}$, are calculated using the formula~\cite{PDG96}:
\begin{eqnarray}
 1-0.95 &=& \frac{\int_{-\infty}^{\Gamma^{\MR{exp}}} \MR{d}\Gamma\,
               G (\Gamma;
                 \Gamma^{\MR{SM}} + \Gamma^{\MR{NP}}_{\MR{95}},
                  \Delta) }
               {\int_{-\infty}^{\Gamma^{\MR{exp}}} \MR{d}\Gamma\,
                G (\Gamma;
                  \Gamma^{\MR{SM}},
                  \Delta) }
    \nonumber \\
\MR{with\ } G(\Gamma;\mu,\Delta) &=& \frac{1}{\sqrt{2\pi}\Delta}
                       \exp{ \left[-\frac{(\Gamma-\mu)^2}{2\Delta^2} \right]}
            \, ,
\label{eq:gamma95}
\end{eqnarray}
where $\Gamma^{\MR{exp}}$ is our experimental result,
$\Gamma^{\MR{SM}}$ the \SM\ expectation for $\MH = 1\ \TeV$ and
$\Delta$ the combined experimental and theoretical error.

The limits obtained for the total, hadronic, leptonic and invisible
widths, as well as for the three lepton species, are summarised in
Table~\ref{tab:gamma_limits}.
Also listed are the differences of our measurements and the \SM\ expectations
together with their experimental and theoretical 68\% C.L. errors.
It should be noted that the results on the total and partial widths are
correlated; hence the limits derived in this section cannot be applied 
simultaneously.

\subsection{Fits to Total Cross Sections and Forward-Backward Asymmetries}
\label{sec:fit_par59}

The measured leptonic forward-backward asymmetries are included
in the fits.
Besides \MZ\ and \GZ\ the measurements are fitted to the 
hadronic pole cross section, \s0had, the ratios of hadronic to leptonic
widths, \RL, and the leptonic pole asymmetries, \AFBZl, which are defined 
as:
\begin{equation}
 \s0had = \frac{12\pi}{\MZ^2} \, \frac{\GE\Gh}{\GZ^2} \, ,
          \quad
 \RL    = \frac{\Gh}{\GL} \, ,
          \quad
 \AFBZl = \frac{3}{4} \, A_{\MR{e}} \, A_{\MR{\ell}} \\
          \qquad (\ell = \MR{e},\,\mu\,,\tau)\, .
\label{eq:s0had_def}
\end{equation}
The advantage of this parameter set is that the parameters are less
correlated than the partial widths.
Two fits are performed, one with and one without assuming lepton 
universality.
The results are listed in Table~\ref{tab:fitpar59} and the correlation
matrices are given in Tables~\ref{tab:fitcor9} and \ref{tab:fitcor5}.

\begin{figure}
 \begin{center}
  \includegraphics[width=\figwidth]{\pltdir 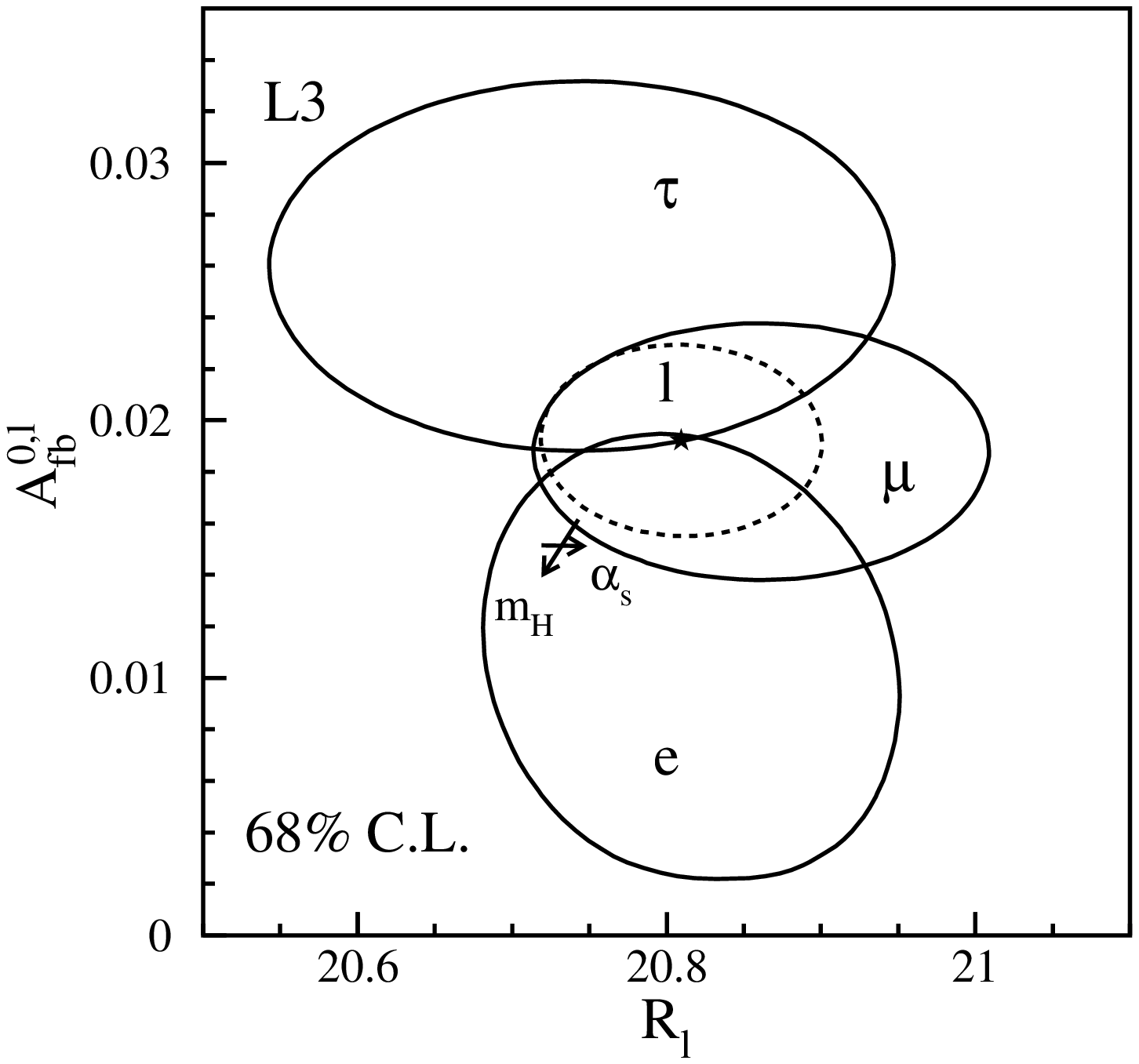}
 \end{center}
 \caption{Contours in the $\protect\AFBZl-\protect\RL$ plane for
          electrons, muons and taus obtained from a fit to total cross 
          sections and forward-backward asymmetries.
          The dashed line shows the contour assuming lepton universality
          and the star indicates the central value.
          The arrows show the change in the \protect\SM\ prediction
          when varying the input parameters \protect\as\ and \protect\MH\
          in the ranges defined in Equation~\protect\ref{eq:smpara}.
          The uncertainty on the \protect\SM\ prediction due to the other
          parameters is small.
         }
 \label{fig:AfbRl}
\end{figure}

The 68\% C.L. contours in the $\AFBZl-\RL$ plane are derived from these
fits for the three lepton species separately and for all leptons 
combined (Figure~\ref{fig:AfbRl}).
In this plot the contour of $\AFBZt-\RT$ is shifted by the difference in 
expectation for \RT\ due to the tau mass to facilitate the comparison with 
the other leptons.
Also for the forward-backward asymmetries good agreement among the
lepton species is observed.
Our results are in agreement with the \SM\ expectations.

From the measurements of the forward-backward pole asymmetries
the polarisation parameter, $A_\ell$, can be derived for the three 
individual lepton types as well as the average value.
The results are listed in Table~\ref{tab:Alepton}.
Because of their relation to the measured pole asymmetry (Equation~\ref{eq:AFBloworder})
the results for $A_{\MR{e}}$, $A_\mu$ and $A_\tau$ are highly correlated.
They are compared to $A_{\MR{e}}$ and $A_\tau$ derived from our measurements 
of the average and the forward-backward tau-polarisation~\cite{l3-140}.
All measurements are in good agreement and yield an average value of
\begin{equation}
 A_\ell = \Alepval \pm \Aleperr \, .
\label{eq:Allepton_result}
\end{equation}

\subsection{Vector- and Axial-Vector Coupling Constants of Charged Leptons}
           \label{sec:gagv}

The effective coupling constants, \gVbl\ and \gAbl, are obtained 
from a fit to cross section and forward-backward asymmetry measurements,
and including our results from tau-polarisation.
We use the results from tau-polarisation on $A_{\MR{e}}$ and $A_\tau$ 
as given in Table~\ref{tab:Alepton} together with a 8\% correlation 
of the errors.
The inclusion of tau-polarisation results significantly improves the 
determination of the effective coupling constants. 

Fits with and without assuming lepton universality are performed and
the vector and axial-vector coupling constants so obtained are listed
Table~\ref{tab:gagv}.
The axial-vector coupling constant of the electron is taken to be negative,
in agreement with the combination of results from neutrino-electron
scattering and low energy \AFB\ measurements~\cite{CHARM_II}.
All other signs are unambiguously determined by our measurements.

\begin{figure}
 \begin{center}
  \includegraphics[width=\figwidth]{\pltdir 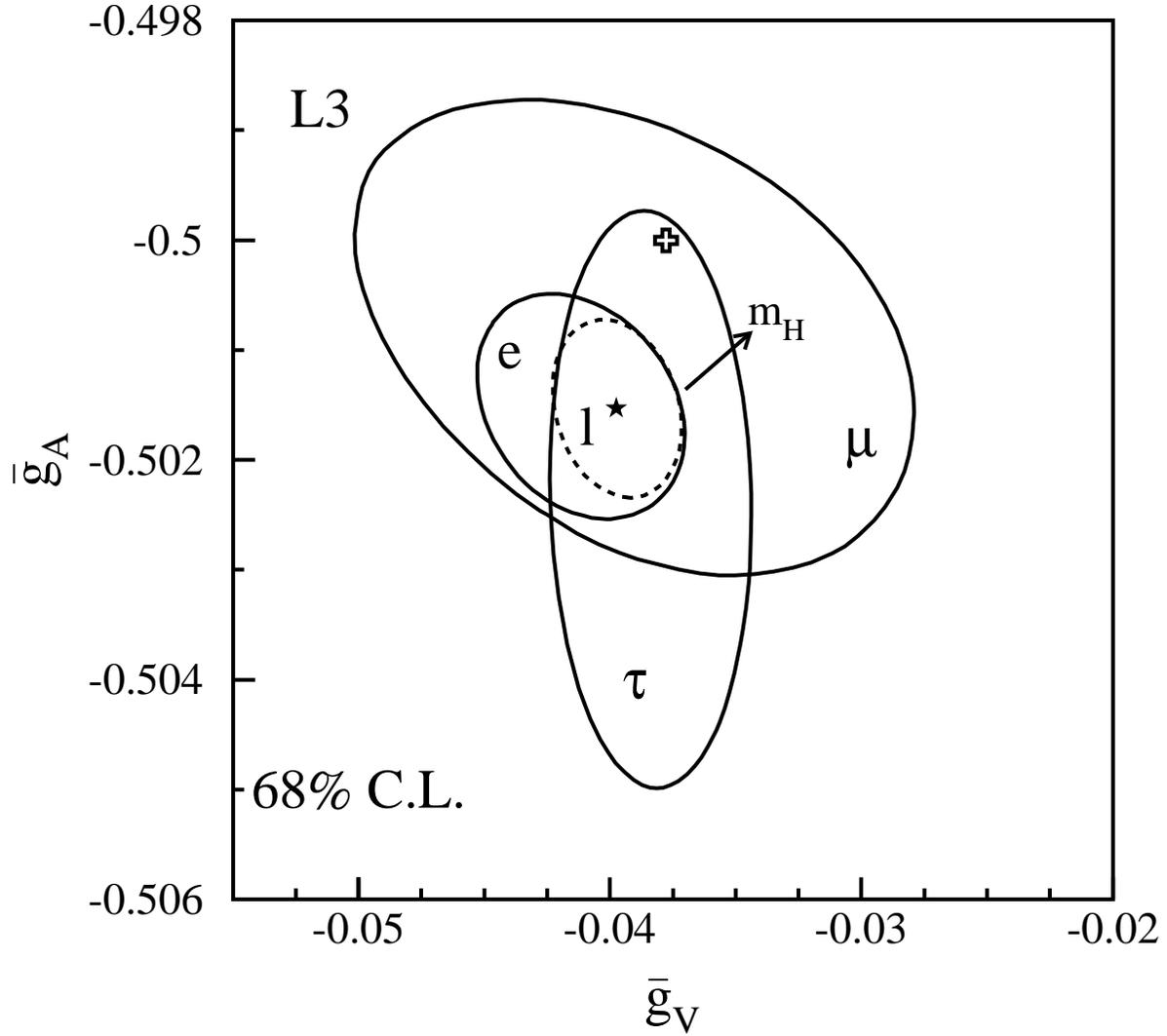}
 \end{center}
 \caption{Contours in the \protect\gAb-\protect\gVb\ plane for
          electrons, muons and taus obtained from a fit to total cross 
          sections, forward-backward and tau polarisation asymmetries.
          The dashed line shows the contour assuming lepton universality
          and the star indicates the central value.
          The arrow shows the change in the \protect\SM\ prediction
          when varying \MH\ in the range defined in 
          Equation~\protect\ref{eq:smpara}.
          The uncertainty on the \protect\SM\ prediction due to the other
          parameters is small.
          The hollow cross indicates the \protect\SM\ expectation 
          without weak radiative corrections.
          }
 \label{fig:gagv}
\end{figure}

The 68\% C.L. contours in the \gVb-\gAb\ plane are shown in 
Figure~\ref{fig:gagv}, revealing good agreement among the three lepton species
and thus supporting lepton universality in neutral currents.
This is quantified by calculating the ratio of muon and tau to 
electron coupling constants, taking into account their correlations 
(see Table~\ref{tab:gagvratio}).

The average vector and axial-vector coupling constants of charged leptons
are found to be
\begin{equation} 
 \gVbl = \gVqval \pm \gVqerr \, , \qquad
 \gAbl = \gAqval \pm \gAqerr \, .
\label{eq:gagv_result}
\end{equation}
The resulting axial-vector coupling constant \gAbl\ is significantly different 
from its lowest order \SM\ expectation $-1/2$.
This is interpreted as proof for the existence 
of weak radiative corrections from higher order processes and corresponds to a 
measurement of the $\rho$-parameter of
\begin{equation} 
 \bar{\rho} = 1.0061 \pm 0.0021 \, .
\label{eq:rho_result}
\end{equation}
The evidence for the existence of weak radiative corrections is illustrated
in Figure~\ref{fig:gagv}.

Measurements of forward-backward asymmetries and tau-polarisation can be
compared in terms of the effective weak mixing angle, \swsqb, defined by the 
ratio of the coupling constants (Equation~\ref{eq:def_rho}).
From the average leptonic pole asymmetry and the tau polarisation, 
the values listed in Table~\ref{tab:sweff} are obtained.
Our results obtained from the measurement of the forward-backward 
asymmetry of b-quarks~\cite{l3-163} and the measurement of the
quark charge asymmetry, \QFB,~\cite{l3-157} are also shown.
All four measurements of the weak mixing angle are in agreement
with each other and the average yields:
\begin{equation}
 \swsqb =  \swsqval \pm \swsqerr \, .
\label{eq:sweff_asym}
\end{equation}

\subsection{Fits in the S-Matrix Framework}
            \label{sec:fit_Smatrix}

The programs SMATASY~\cite{SMATASY621} together with ZFITTER, ALIBABA and TOPAZ0
are used for the calculation of the theoretical predictions, 
including QED radiative corrections, of total cross sections and 
forward-backward asymmetries.
Further details can be found in Reference~\cite{l3-62}.

The results of the fits in the S-Matrix framework with and without imposing
lepton universality to the cross sections and forward-backward asymmetries
measured at the Z resonance are shown in Table~\ref{tab:smatrix}.
The fitted parameters for electrons, muons, taus and hadrons are in agreement
with each other and with the expectations from the \SM.
The correlations of the parameters as obtained in the two fits are shown in
Tables~\ref{tab:cor16} and~\ref{tab:cor8}, respectively.

\begin{figure}
 \begin{center}
  \vspace*{-20mm}%
  \hspace*{-15mm}
  \includegraphics[width=\figwidth]{\pltdir 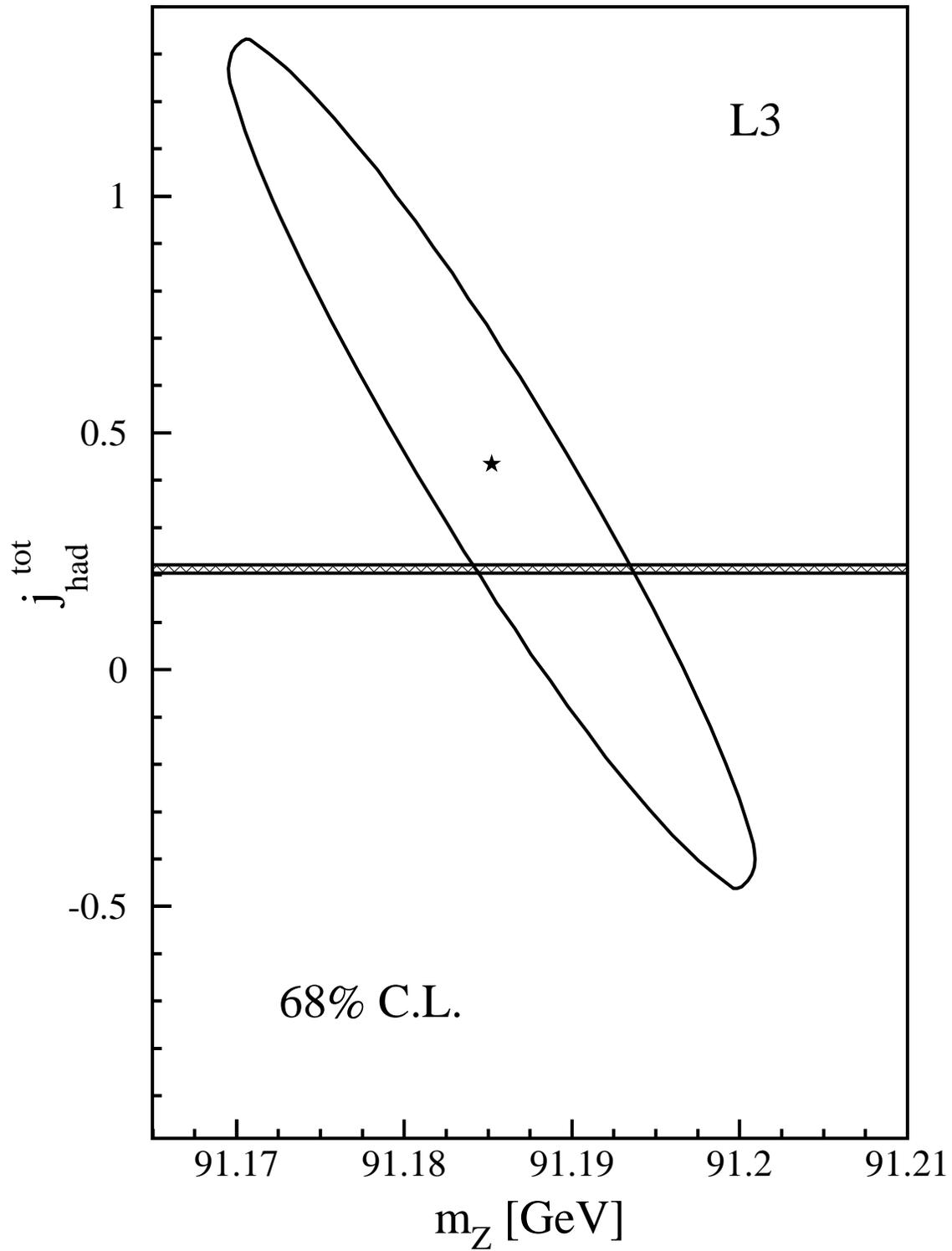}
  \end{center}
 \caption{Contour in the \protect\jtxsha-\protect\MZ\ plane 
          obtained from the S-Matrix fit assuming lepton universality.
          The horizontal band shows the \protect\SM\ predition for
          \protect\jtxsha\ with its uncertainty.
         }
 \label{fig:mzjhad}
\end{figure}

Large correlations among the parameters are observed.
Of particular importance is the correlation of \mZjhadcor\ between the mass of 
the \Zo\ boson and the hadronic interference term,  \jtxsha.
It causes an increase of the error on \MZ\ with respect to the 
fits performed in 
Sections~\ref{sec:massfit} and~\ref{sec:fit_par59}
where the \GZint\ terms are 
fixed to their \SM\ expectations.
The correlation between \MZ\ and \jtxsha\ is illustrated in
Figure~\ref{fig:mzjhad}.

Comparing the results on the \Zo\ boson mass obtained with the two analyses
(Table~\ref{tab:fitwidth} or~\ref{tab:fitpar59}
and Table~\ref{tab:smatrix}) good agreement is found.
From a quadratic subtraction we estimate the additional error on the \Zo\ mass 
arising from the experimental uncertainty on the hadronic interference term  
to be:
\begin{equation}
     \MZ =  \mZSmatrixqval \pm \mZqerr \pm  9.8 (\jtxsha)\ \MeV \, .
 \label{eq:MZ_SMatrix}
\end{equation}

The interference between the photon and the \Zo\ is measured to much better
precision at \cms\ energies below or above the \Zo\ resonance.
By adding our measurements of hadronic and leptonic cross sections and
forward-backward asymmetries above the \Zo\ resonance
this contribution to the error on \MZ\ is significantly reduced.
This will be reported in a forthcoming publication.

%
%

\section{Results on \protect\SM\ Parameters}
         \label{sec:fit_SMpara}

We interpret our measurements in the framework of the \SM\ to check its 
consistency by comparing our results to other measurements.
The strategy will be to test at first QCD radiative corrections in terms of
\as\ before we verify weak radiative corrections by comparing the top mass
derived from our data at the \Zo\ resonance to the direct measurement.
From our measurements at the \Zo, the W mass is determined and compared to
our result obtained above the W-pair threshold.
Finally we use all our measurements of electroweak parameters and include
the direct measurement of \Mt\ to estimate the mass of the \SM\ Higgs boson.

Fits are performed to our data to determine the set of \SM\ parameters given 
in Equation~\ref{eq:smpara}.
The program ZFITTER is used for \SM\ calculations.
In all fits the QED coupling constant at the mass of the \Zo\ is
calculated using the constraint on the contribution from the five light quark
flavours to the  running as obtained in Reference~\cite{Jegerlehner_aqed}.

The input data for the \SM\ fits are our measurements of total cross sections
and forward-backward asymmetries at the \Zo\ resonance performed between 
1990 and 1995.
In addition, our results from tau polarisation (Table~\ref{tab:Alepton}),
the effective weak mixing angles from b-quark forward-backward and
form quark charge asymmetry (Table~\ref{tab:sweff}),
as well as the partial decay width into b-quarks
$R_{\MR{b}} = \GB/\Gh = 0.2174 \pm 0.0032$~\cite{l3-185}, 
are included.

Firstly, the sensitivity of our data to QCD radiative corrections
is exploited to determine the strong coupling constant at the mass of the
Z boson:
\begin{eqnarray}
 \as &=& 0.1226\,^{+0.0066}_{-0.0060} \, .
\label{eq:mtas}
\end{eqnarray}
In this fit the result of Section~\ref{sec:massfit} is obtained again 
for \MZ\ and \Da5had\ remains within the imposed constraint.
The masses of the top quark and the Higgs boson are free parameters and
their uncertainties are included in the error on \as.

The value for \as\ is in good agreement with our determination of the 
strong coupling constant from hadronic event topologies
$\as = 0.1216 \pm 0.0017 \pm 0.0058$~\cite{l3-162}.
We use this measurement as an additional constraint and obtain for the 
top mass:
\begin{equation}
   \Mt = \mtlval\,^{+\mtlerrup}_{-\mtlerrlo}\ \GeV  \, .
\label{eq:fitmt}
\end{equation}

This result for the top quark mass is based on our measurements of weak 
radiative corrections and their interpretation in the \SM\ framework.
The agreement with the direct \Mt\ measurements by the CDF and D0
experiments, 
$\Mt = 173.8 \pm 5.2\ \GeV$~\cite{mt_direct_pdg98},
means that the bulk of weak radiative corrections indeed originates from
the large mass of the top quark. 
The result for the Higgs boson mass obtained in this fit,
$ \log_{10}\MH/\GeV  = 1.99\,^{+0.98}_{-0.66}$,  
is in agreement with the range allowed by the direct search and 
the \SM\ (Equation~\ref{eq:smpara}).

\begin{figure}
 \begin{center}
  \includegraphics[width=\figwidth]{\pltdir 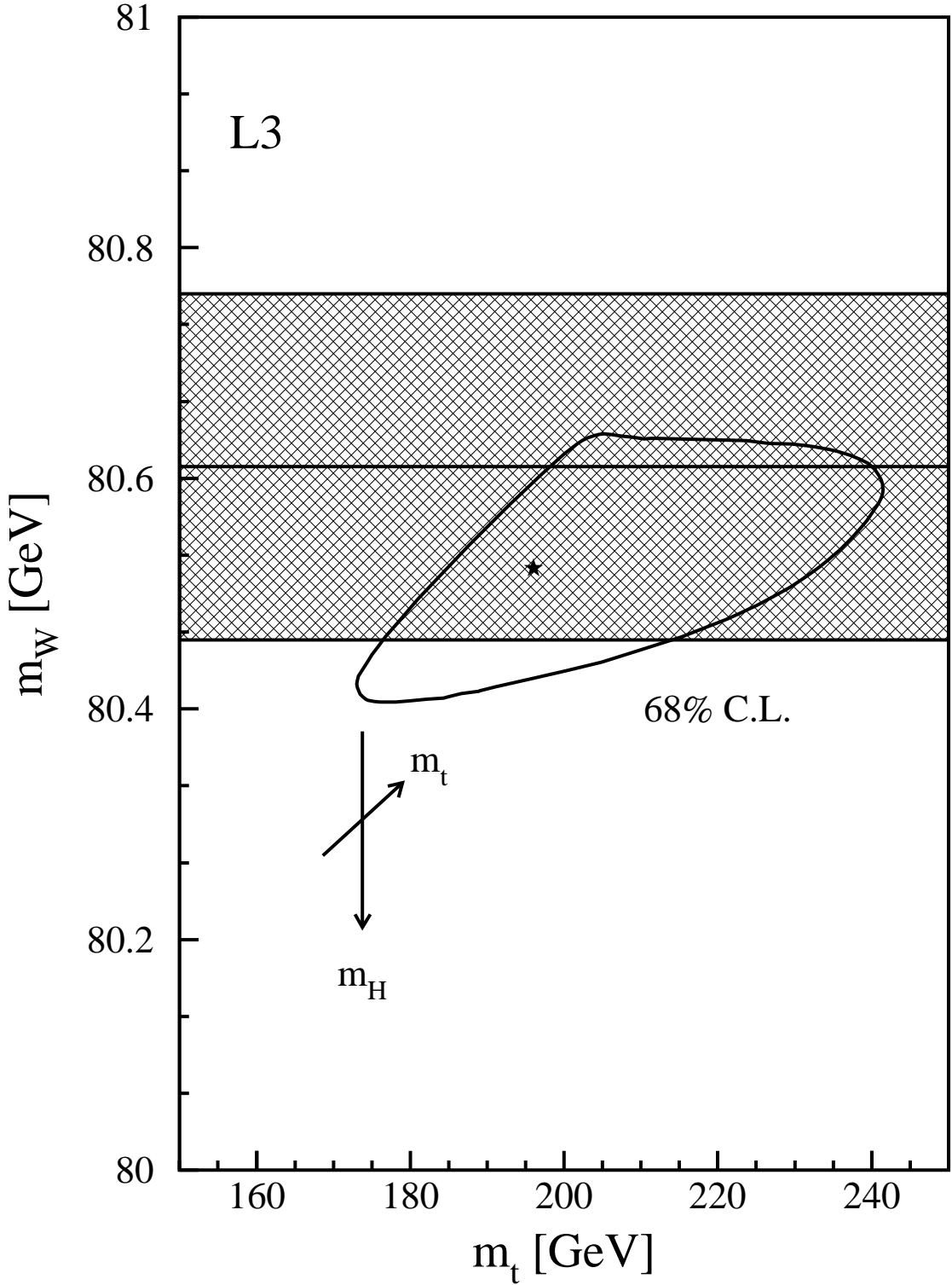}
 \end{center}
 \caption{Contour in the \protect\Mt-\protect\MW\ plane obtained from the 
          SM fit to our data.
          The fit result is compared to our direct measurement
          of the W mass indicated by the hatched band and 
          the \protect\SM\ expectation.
          }
 \label{fig:mWmt}
\end{figure}

From the result of this fit which is based on measurements at the Z resonance
a value for the mass of the W boson is derived:
\begin{equation}
 \MW = 80.523 \pm  0.079\ \GeV \, .
\label{eq:Wmass_result}
\end{equation}
The 68\% C.L. contour in the \Mt-\MW\ plane obtained in this fit is shown
in Figure~\ref{fig:mWmt}.
This result for the W mass agrees well with our direct measurements of 
\MW\ performed at \cms\ energies between $161\ \GeV$ and 
$183\ \GeV$~\cite{l3-111,l3-130,l3-171} which yield a combined value of 
$\MW = 80.61 \pm 0.15\ \GeV$.

We include the direct measurement of \MW\ to determine the electroweak 
radiative corrections $\Delta r$, the parameters $\rho$ and $\kappa$, 
the effective weak mixing angle \swsqb\ and the on-shell definition \swsq\ as 
well as a combined result for \MW:
\begin{equation}
 \begin{array}{r@{$\,=\,$}lr@{$\,=\,$}lr@{$\,=\,$}l}
     \Delta r  & 0.0257 \pm 0.0043       \, , &
     \rho      & 1.0078 \pm 0.0017       \, , &
     \kappa    & 1.0493 \pm 0.0053       \, ,    \\
     \swsqb    & 0.23075\pm 0.00054      \, , &
     \swsq     & 0.2199 \pm 0.0013       \, , &
     \MW       & 80.541 \pm 0.069\ \GeV  \, .
 \end{array}
\label{eq:dquant}
\end{equation}
This value for the effective weak mixing angle \swsqb\ derived in a \SM\ fit
is in good agreement with the result obtained, in a less model-dependent way,
from measurements of asymmetries (Table~\ref{tab:sweff}).

\begin{figure}
 \begin{center}
  \includegraphics[width=\figwidth]{\pltdir 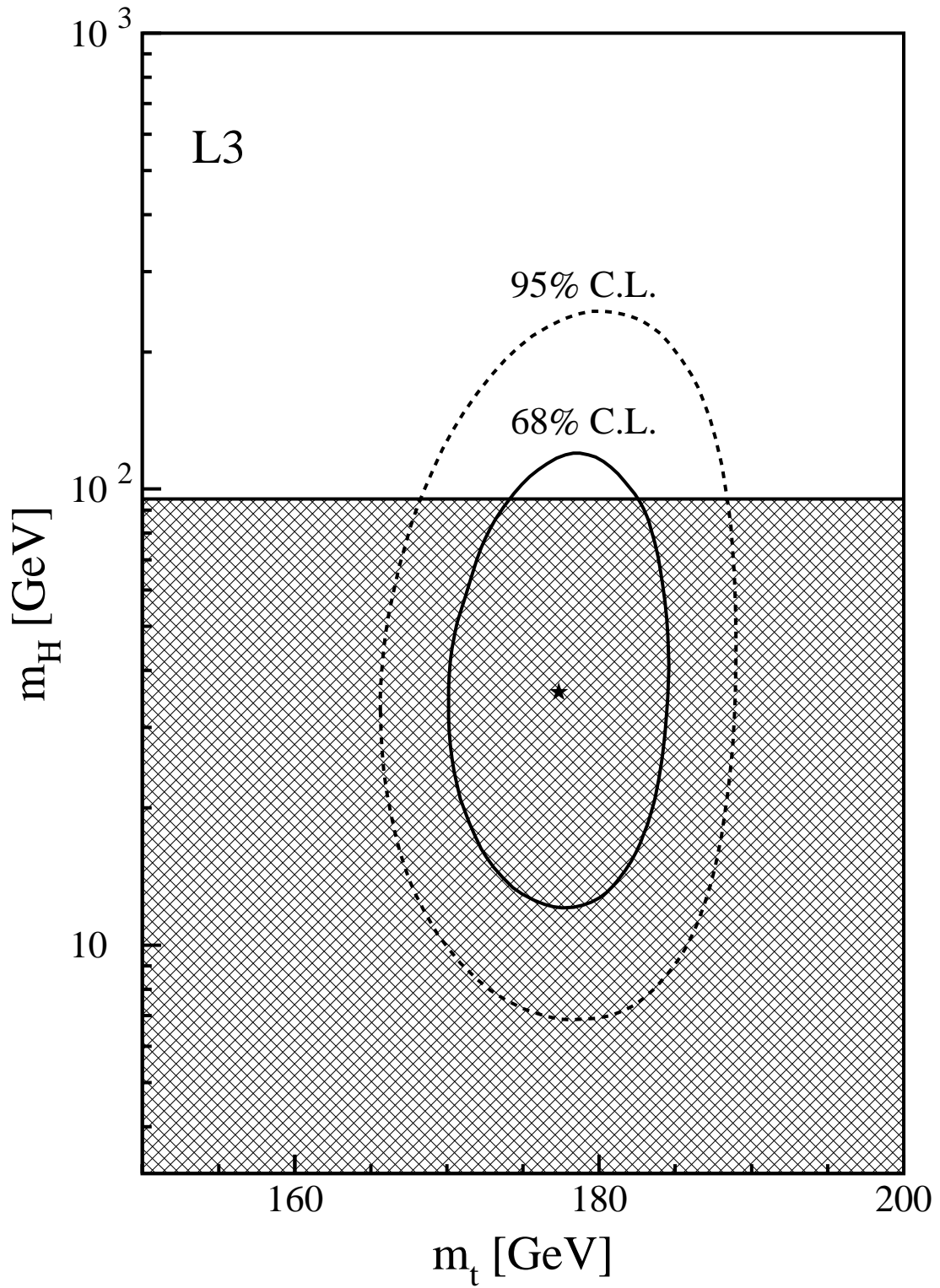}
 \end{center}
 \caption{Contours in the \protect\Mt-\protect\MH\ plane obtain from the 
          SM fit.
          The hatched area indicates values of the Higgs mass excluded by our
          direct search result~\protect\cite{l3-178}. 
          }
 \label{fig:mhmt}
\end{figure}

\begin{figure}
 \begin{center}
  \includegraphics[width=\figwidth]{\pltdir 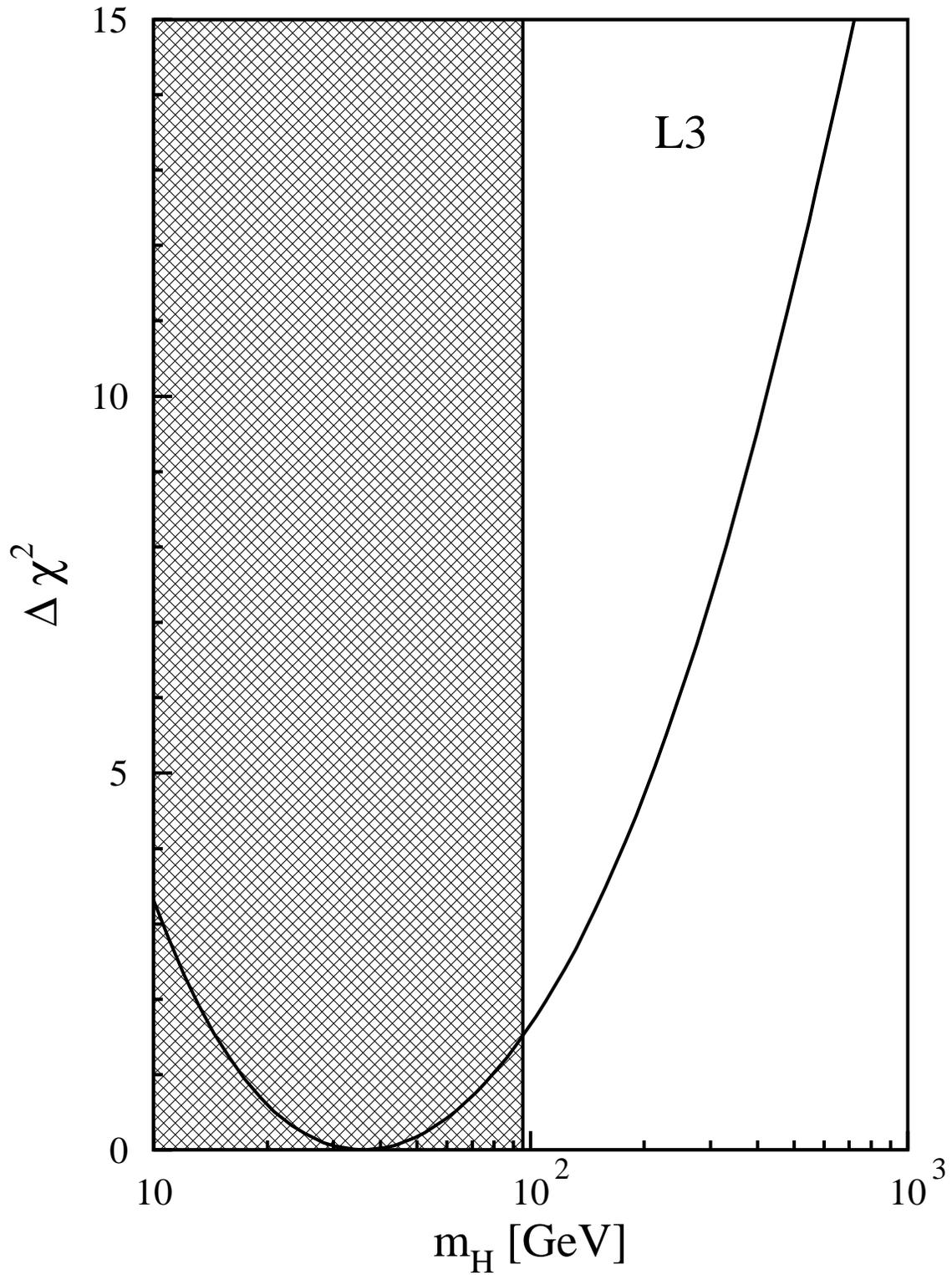}
 \end{center}
 \caption{The $\chi^2$ dependence of the \SM\ fit as function of the
          Higgs boson mass.
          The shaded area indicates the mass range excluded by the
          direct search.
          }
 \label{fig:mhchi2}
\end{figure}

Finally, we constrain the mass of the top quark to the combined value 
from the direct measurement of D0 and CDF.
The five \SM\ parameters and their correlations obtained in this fit
are summarised in Tables~\ref{tab:Hfit} and~\ref{tab:Hfitcor}, respectively.
In particular, for the mass of the yet undiscovered \SM\ Higgs boson, we obtain
a value and an upper limit:
\begin{eqnarray}
 \MH &=& 36^{+43}_{-19}\ \GeV                        \, ,    \nonumber \\
     &<& \mhlimit\ \GeV \qquad 95\%\ \mathrm{C.L.}
\label{eq:mhiggs}
\end{eqnarray}
Figure~\ref{fig:mhmt} shows the 68\% and 95\% C.L. contours in the 
\Mt-\MH\ plane
and Figure~\ref{fig:mhchi2} the dependence of the $\chi^2$ of the fit
on the Higgs mass from which the upper mass limit is derived.
The result is compatible with the result of our direct search for the 
\SM\ Higgs boson $\MH > 95.3\ \GeV$~\cite{l3-178}.

%
%
\section{Summary and Conclusion}
         \label{sec:conclusion}

We report on the precise measurements of total cross sections and
forward-backward asymmetries of the reactions 
\EEHADG, \EEMMG, \EETTG\ and \EEEEG\ 
at \cms\ energies at the peak and the wings of the \Zo\ resonance
performed in the years $1993-95$.
A total luminosity of $103\ \pbinv$ corresponding to 2.5 million hadronic 
and 250 thousand leptonic decays of the \Zo\ was collected which 
significantly improve our measurements of the resonance curve.
Including the data samples collected in previous years, the total number
of \Zo\ decays observed by the L3 detector during the first phase of
LEP amounts to 4 million which are used to determine the properties of
the \Zo\ and other \SM\ parameters.

All our measurements are consistent with the hypothesis of lepton universality.
From the measured total hadronic and leptonic cross sections we obtain:
\begin{eqnarray*}
   \MZ = \mZqval \pm  \mZqerr\ \MeV \, ,   & &
   \GZ = \GZqval \pm  \GZqerr\ \MeV \, ,   \\
   \Gh = \Ghqval \pm  \Ghqerr\ \MeV \, ,   & &
   \GL = \Glqval \pm  \Glqerr\ \MeV \,.
 \label{eq:Zpara_conclusion}
\end{eqnarray*}
%
From these results, the decay width of the \Zo\ into invisible particles
is derived to be $\Ginv = \Ginvqval \pm \Ginvqerr\ \MeV$, which in the 
\SM\ corresponds to a  number of light neutrino species of:
\begin{equation}
    N_{\MR{\nu}} = \nnqval \pm \nnqerr \, .
 \label{eq:nnu_conclusion}
\end{equation}

Including our measurements of leptonic forward-backward asymmetries and 
tau polarisation the effective vector and axial-vector coupling constants
of charged leptons to the \Zo\ are determined to be:
\begin{equation}
   \gVbl = \gVqval \pm \gVqerr \, , \qquad
   \gAbl = \gAqval \pm \gAqerr \, .
 \label{eq:gagv_conclusion}
\end{equation}
For the effective weak mixing angle we obtain:
\begin{equation}
    \swsqb = \swsqval \pm \swsqerr \, ,
 \label{eq:sintheta_conclusion}
\end{equation}
including our measurements of the b-quark forward-backward and quark charge
asymmetries.

Our measurements are sensitive to higher order weak radiative corrections which
depend on the masses of the top quark and the Higgs boson.
Using in addition our measurements of the partial width 
$\Zo\rightarrow\bb$ and \as, we derive in the \SM\ framework a top quark 
mass 
\begin{equation}
 \Mt = \mtlval\,^{+\mtlerrup}_{-\mtlerrlo}
              \ \GeV   \, ,
\label{eq:mt_conclusion}
\end{equation}
which is in agreement with the direct measurements of \Mt.
Using our direct measurement of \MW\ and the knowledge of \Mt\ 
our data constrain the mass of the
Higgs boson to 
\begin{equation}
  \MH < \mhlimit\ \GeV \qquad 95\%\ \MR{C.L.}
 \label{eq:mH_conclusion}
\end{equation}
%

%
\section*{Acknowledgments}

We congratulate the CERN accelerator divisions for their achievements
in the precise calibration of the beam energy and we express our gratitude 
for the excellent performance of the LEP machine.
We like to thank W.\ Beenakker, D.\ Bardin, G.\ Passarino and their collaborators
for many fruitful discussions and their help in theory questions.
We acknowledge the contributions of the engineers 
and technicians who have participated in the construction 
and maintenance of this experiment. 

%
%
\clearpage
\appendix
\addcontentsline{toc}{section}{Appendix}
\noindent {\bf \LARGE Appendix}  \\[1mm]

\section{\boldmath Treatment of Contributions related to the $t$-channel}
          \label{app:tchannel}

In the case of the process \EEEEG, the existence of the $t$-channel exchange 
of photons and \Zo\ bosons and its interference with the $s$-channel 
exchange lead to additional complications. 
Analytical programs to calculate this process, such as ALIBABA and TOPAZ0, 
are not directly suited for fitting purposes, as computationally they are very
time consuming. 
Thus, the following procedure is adopted. 
During the initialisation of a fit, ALIBABA is used once to calculate 
the predictions of the $t$-channel and \stint\ contributions to 
the measured \EEEEG\ cross sections and forward-backward asymmetries. 
Calculations are performed at several \cms\ energy values in the vicinity of 
the data points allowing for a reduction of the statistical error of the
Monte Carlo integration used by ALIBABA and the calculation of derivatives 
needed to construct the covariance matrix (see Appendix~\ref{app:covariance}).
ZFITTER is employed during the fits to calculate the corresponding 
s-channel contributions as a function of the varying electroweak 
parameters. 

The contribution from \stint\ depends on the fit parameter, most importantly
on the Z boson mass.
This dependence is taken into account by converting the difference between the
Z mass used in the initialisation and the current fit value into an equivalent 
shift in the  \cms\ energy at which the $t$-channel and \stint\ contributions 
are calculated.
The dependence of the \stint\ on \MZ\ is responsible for a correlation of the 
results for \MZ\ and the electron $s$-channel cross
section which amounts to $+11\%$ between \MZ\ and \RE\ in the nine parameter
fit (Table~\ref{tab:fitcor9}).

In the analytical program ZFITTER polar angular cuts can only be applied on
the positron while in the experimental cross section measurements both electron
and positron are required to lie within the fiducial volume.
Correction factors are calculated with TOPAZ0 which allows for both types of
cuts.
Finally, cross sections of \EEEEG\ to be compared to the experimental
measurement are calculated as:
\begin{eqnarray}
 \sigma^{\MR{th}} &=& \sigma^{t+s/t, \MR{AL}}
                      (44^\circ<\theta_{\MR{e^-}}, \theta_{\MR{e^+}}<136^\circ)
                     + \sigma^{s, \MR{ZF}}(44^\circ<\theta_{\MR{e^+}}<136^\circ)\, \, R
                  \nonumber \\
 \MR{with\ }     R & = & \frac{\sigma^{s, \MR{T0}}
                       (44^\circ<\theta_{\MR{e^-}}, \theta_{\MR{e^+}}<136^\circ)}
                              {\sigma^{s, \MR{T0}}
                              (44^\circ<\theta_{\MR{e^+}}<136^\circ)}
 \label{eq:Bhabha_calc}
\end{eqnarray}
All cross sections are defined for an acollinearity angle cut 
of $\xi < 25^\circ$.
The indices AL, T0 and ZF label cross sections calculated with the ALIBABA,
TOPAZ0 and ZFITTER programs, respectively.
Cuts on the polar angle of the electron and positron are given in parentheses.
This procedure is applied because it combines the most accurate treatment
of electroweak radiative corrections as available in ZFITTER and TOPAZ0
with the most complete calculations of $t$-channel contributions in ALIBABA.
In the case of the forward-backward asymmetry,
$\AFB = (\sigma_{\mathrm{F}}-\sigma_{\mathrm{B}})/
(\sigma_{\mathrm{F}}+\sigma_{\mathrm{B}})$, the analogous calculations for the
forward, $\sigma_{\mathrm{F}}$, and backward cross sections,
$\sigma_{\mathrm{B}}$, are performed.

The cross sections calculated with the ALIBABA and TOPAZ0 programs for all 
$1990-95$ data sets are listed in in Table~\ref{tab:tchann_numbers}.
The theoretical uncertainties on the calculation of the $t$-channel contributions
are listed in Table~\ref{tab:tchann_errors}.
These errors are applied to the measured cross sections and forward-backward
asymmetries in the fits (see Appendix~\ref{app:covariance}).

Due to the contribution of the $t$-channel the results from the 
Bhabha channel cannot be directly compared to the measurements of the
other leptonic final states. 
To permit such a comparison, Equation~\ref{eq:Bhabha_calc} is used to calculate
the $s$-channel contributions, replacing $\sigma^{\MR{th}}$ by the measurements
and using ZFITTER for the extrapolation to the full solid angle.
The $s$-channel cross sections obtained this way, 
$\sigma^{\mathrm{s}}_{\mathrm{e}}$, without
any cuts, and $s$-channel asymmetries, 
$\AFB^s$, with an acollinearity angle cut of 
$\xi<25^\circ$, are given in 
Tables~\ref{tab:electron_xs} and~\ref{tab:electron_afb}.

\section{Construction of the Covariance Matrix}
         \label{app:covariance}

All fits for electroweak parameters described in this article are performed 
in the error matrix analysis.
They consist of minimising a $\chi^2$ function defined as
\begin{equation}
 \chi^2 = D^T \, V^{-1} \, D  \, ,
\label{eq:def_chisquare}
\end{equation}
where $D$ is a column vector with elements defined by the difference between
measurements $\Omega_i^{\MR{exp}}$ and theoretical expectations
$\Omega_i^{\MR{th}}$ which are calculated during the fit as a function of 
the fit parameters:
\begin{equation}
  D_i = \Omega_i^{\MR{exp}} - \Omega_i^{\MR{th}} \, .
\label{eq:def_Dvector}
\end{equation}
The index $i$ runs over all cross section ($\Omega_i = \sigma_i$)
and forward-backward asymmetry ($\Omega_i = {\AFB}_{,i}$)
measurements considered in the fit.
There are 100 cross section and 75 forward-backward asymmetry measurements
at the \Zo\ resonance taken in $1990-95$.

The covariance matrix $V$ is constructed in the following way from all 
experimental and theoretical errors affecting the measurements.
The diagonal elements, $V_{ii}$, are obtained adding all individual errors
of measurement $i$ in quadrature
\begin{eqnarray}
 V_{ii} &=& 
               \left( f_i\, \Delta_i^{\MR{stat}} \right)^2 
            +  \left( f_i\, \Delta_i^{\MR{unc}}  \right)^2
            +  \left( \Delta_i^{\MR{cor}}  \right)^2
            +  \left( \Delta_i^{\MR{abs}}  \right)^2 
     \nonumber \\
        & & + V_{ii}^{\MR{lum}} 
            + V_{ii}^{\MR{LEP}}
            + V_{ii}^{\epsilon_{\MR{cms}}}
            + V_{ii}^t                              \, ,
\label{eq:Vdiagonal}
\end{eqnarray}
where $\Delta_i^{\MR{stat}}$ is the statistical error of the measurement,
$\Delta_i^{\MR{unc}}$ the uncorrelated part of the systematic error,
$\Delta_i^{\MR{cor}} = \delta_i^{\MR{cor}} \, \sigma_i^{\MR{th}}$ the 
correlated systematic error which scales with the expected value
and $\Delta_i^{\MR{abs}}$ the correlated part of the systematic error which
does not scale.
In case of the forward-backward asymmetry the systematic error can simply
be expressed in terms of $\Delta_i^{\MR{abs}}$ only.
The statistical and uncorrelated systematic errors which are derived from
the measurements (Tables~\ref{tab:had_xs} to~\ref{tab:electron_afb})
are scaled during the fit with factors $f_i$ to the expected errors using 
the theoretical expectations:
\begin{equation}
 f_i =  \sqrt{\frac{\sigma_i^{\MR{th}}}{\sigma_i^{\MR{exp}}} } 
        \quad (\MR{for}\ \sigma_i) \, ,
  \qquad
 f_i =  \sqrt{\frac{1 - \left( {\AFB}^{\MR{th}}_{,i}  \right)^2}
                   {1 - \left( {\AFB}^{\MR{exp}}_{,i} \right)^2} \,
          \frac{\sigma_i^{\MR{exp}}}{\sigma_i^{\MR{th}}} }
                \quad (\MR{for}\ {\AFB}_{,i}) \, .
\label{eq:scale_factor}
\end{equation}

For cross section measurements there is an additional contribution
of the luminosity measurement, $V_{ii}^{\MR{lum}}$.
Its calculation as well as the contribution from the 
uncertainty on the LEP \cms\ energy, $V_{ii}^{\MR{LEP}}$, 
and its spread, $V_{ii}^{\epsilon_{\MR{cms}}}$, both applied to cross 
sections and \AFB\ measurements, and the theoretical uncertainty
on the subtraction of the $t$-channel and \stint\ contribution to
Bhabha scattering, $V_{ii}^t$, are described below.

The off-diagonal elements are constructed from correlated error sources:
\begin{eqnarray}
 V_{ij} &=&    \Delta_i^{\MR{cor}} \, \Delta_j^{\MR{cor}} 
             + \Delta_i^{\MR{abs}} \, \Delta_j^{\MR{abs}}
     \nonumber \\
        & & + V_{ij}^{\MR{lum}} 
            + V_{ij}^{\MR{LEP}} 
            + V_{ij}^{\epsilon_{\MR{cms}}}
            + V_{ij}^t
     \qquad\qquad (i \neq j) \, .
\label{eq:Voffdiagonal}
\end{eqnarray}
Experimental systematic errors, $\Delta^{\MR{cor}}$ and $\Delta^{\MR{abs}}$,
are only applied to elements connecting the same observable, either cross 
section or asymmetry, of the same reaction.
As above, for forward-backward asymmetry measurements contributions from
$\Delta^{\MR{cor}}$ and $V^{\MR{lum}}$ are not applicable.
Contributions from $V_{ij}^{\MR{lum}}$, $V_{ij}^{\MR{LEP}}$,
$V_{ij}^{\epsilon_{\MR{cms}}}$ and $V_{ij}^t$ enter also into off-diagonal 
elements connecting measurements of different observables or reactions.

All statistical errors and the individual contributions to systematic 
errors for the measurements performed in $1993-95$ are listed in 
Tables~\ref{tab:had_xs} to~\ref{tab:electron_afb}.
For the measurements performed in $1990-92$ systematic errors are quoted
in Reference~\cite{l3-69} as relative errors, $\delta_i$, 
for the cross section and absolute errors, $\Delta_i$, for
the forward-backward asymmetries.
The correlation among these systematic errors is treated by using 
$V_{ij} = (\min(\delta_i, \delta_j))^2 \,
  \sigma_i^{\MR{th}}\sigma_j^{\MR{th}}$
for the cross sections and 
$V_{ij} = (\min(\Delta_i, \Delta_j))^2$ for the asymmetries.
Contributions from uncertainties on the luminosity and the LEP energy 
are added, where applicable.

Correlations between experimental systematic errors in the $1990-92$ and 
$1993-95$ data sets are estimated in the same way by using the smaller values of
$1993-95$.
Exceptions are the measurements of the \EETTG\ cross section where due to the
completely revised analysis in $1993-95$ an additional factor of $0.72$ is 
applied to $\delta_i^{\MR{cor}}$.
Other contributions to elements connecting measurements of $1990-92$ and 
$1993-95$ are from $V_{ij}^{\epsilon_{\MR{cms}}}$, $V_{ij}^t$ and the 
theoretical uncertainty in $V_{ij}^{\MR{lum}}$.

\subsubsection*{Uncertainty on the Luminosity}

For the $1993-95$ cross sections the contributions to the covariance matrix
from errors on the luminosity measurement are obtained from the sum of the 
total experimental errors, including their correlations, and the theoretical 
uncertainty:
\begin{equation}
 V_{ij}^{\MR{lum}} = 
  \left( \delta_k^{\MR{lum,exp}} \,\sigma_i^{\MR{th}} \right) \,
  \left( \delta_l^{\MR{lum,exp}} \,\sigma_j^{\MR{th}} \right) \,
         \rho_{kl}^{\MR{lum,exp}}
    + \left( \delta_m^{\MR{lum,stat}} \right)^2 \, 
        \sigma_i^{\MR{th}}\sigma_j^{\MR{th}}     
    + \left( \delta^{\MR{lum,th}} \right)^2     \, 
       \sigma_i^{\MR{th}}\sigma_j^{\MR{th}}     \, .
\label{eq:Vlum}
\end{equation}
The indices $k$ and $l$ label the years of the data sets $i$ and $j$.
Total experimental errors on the luminosity, $\delta^{\MR{lum,exp}}$,
are given in Table~\ref{tab:lumi_sys} and their correlations,
$\rho_{kl}^{\MR{lum,exp}}$, in Table~\ref{tab:lumi_correlations}.
The statistical error on the luminosity measurement, 
$\delta^{\MR{lum,stat}}$, is given in Table~\ref{tab:lumi_statistics}
for the nine data taking periods.
It applies only to cross section measurements $i$ and $j$  performed in the
same period $m$.

The combined experimental and theoretical error on the luminosity 
determination in $1990-92$ is $\delta^{\MR{lum}} = 6$\permille.
It is treated as fully correlated and the corresponding terms in the 
covariance matrix are calculated as 
$V_{ij}^{\MR{lum}} = (\delta^{\MR{lum}})^2 \, 
 \sigma_i^{\MR{th}}\sigma_j^{\MR{th}}$.
For the $1990-92$ data the statistical error of the luminosity measurement
is included in the quoted statistical errors of the hadron cross section
measurements and it is neglected for the leptonic  cross sections.
The theoretical uncertainty, $\delta^{\MR{lum,th}} = 0.61$\permille, is 
fully correlated among all cross section measurements in $1990-95$.

The measurement of the luminosity, and hence of $\sigma_i^{\MR{exp}}$,
depends on the \cms\ energy $E$ due to the approximate $1/E^2$ dependence
of the Bhabha cross section:
\begin{equation}
    \frac{\MR{d}\sigma_i^{\MR{exp}}}{\MR{d}E}
    = - \kappa_m \frac{\sigma_i^{\MR{exp}}}{E_i} \, .
\label{eq:dsigmadE}
\end{equation}
Because of the \GZint\ and higher order contributions the exponent 
$\kappa_m$ differs from 2 and it is calculated with BHLUMI to be
$\kappa_m = 1.95,\ 2.27$ and $1.97$ for the peak$-2$, peak and peak$+2$
data sets, respectively.
This dependence causes a small contribution to the uncertainty on the
cross section measurements and it is taken into account for the $1993-95$
data together with the error on the LEP energy.

\subsubsection*{Uncertainty on the Centre-of-Mass Energy}

The errors on the LEP \cms\ energy are transformed into equivalent errors
of cross section and asymmetry measurements using the partial derivatives of 
the theoretical cross sections and forward-backward asymmetries with respect 
to the \cms\ energy, $\partial\Omega_i^{\MR{th}}/ \partial E$.
The dependence of the measured cross section, via the luminosity, on the 
\cms\ energy is included where applicable.

For the $1993-95$ data sets the terms in the covariance matrix are
determined as
\begin{eqnarray}
  V_{ij}^{\MR{LEP}} &=& v_{kl}^{\MR{LEP}}\,
                        \frac{\MR{d}\Omega_i}{\MR{d} E} \,
                        \frac{\MR{d}\Omega_j}{\MR{d} E} \, , \nonumber \\
 \MR{with\ } \frac{\MR{d}\Omega_i}{\MR{d} E} 
                    &=& -\frac{\partial\sigma_i^{\MR{th}}}{\partial E} 
                        - \kappa_m \frac{\sigma_i^{\MR{exp}}}{E_i}
                        \qquad \MR{for\ 1993-95\ cross\ sections,}\nonumber \\
                    &=& -\frac{\partial\Omega_i^{\MR{th}}}{\partial E}
                         \hspace*{27mm} \MR{otherwise.}
\label{eq:VLEP}
\end{eqnarray}
where $k$ and $l$ are the indices of the LEP energy calibration
periods corresponding to the data sets $i$ and $j$, with $\kappa_m$ the
appropriate factor defined in Eq.~\ref{eq:dsigmadE}.
For a pair of data sets taken in any of the seven periods with precise LEP 
energy calibration, $v_{kl}^{\MR{LEP}}$ is the corresponding element of the 
LEP energy error matrix as given in Table~\ref{tab:leperror1}.
For data sets taken during the 1993 or 1995 pre-scans the LEP energy
error is treated as uncorrelated with any other period.
Hence, only elements connecting data sets taken during the same pre-scan 
receive contributions $v_{kk}^{\MR{LEP}} = (\delta_k^{\MR{LEP}})^2$, where 
$\delta_k^{\MR{LEP}}$ are the uncertainties on the \cms\ energy 
of $ 18\ \MeV$ and $ 10\ \MeV$ in 1993 and 1995, respectively.

The uncertainties on the LEP \cms\ energy in $1990-92$ are listed in
Table~\ref{tab:leperror9092}.
They are uncorrelated to the energy errors in $1993-95$.

The correction factors for cross sections, $f^i_{\epsilon_{\MR{cms}}}$,
and the absolute corrections for forward-backward asymmetries,
$\alpha^i_{\epsilon_{\MR{cms}}}$,
applied to the measurements to account for the spread of the \cms\ energy,
$\epsilon_{\MR{cms}}$,
can be calculated to sufficient precision from Taylor expansions:
\begin{equation}
 f^i_{\epsilon_{\MR{cms}}} = 1 - \frac{1}{2}\frac{1}{\sigma_i^{\MR{th}}}
                       \frac{\partial^2 \sigma_i^{\MR{th}}}{\partial E^2} \,
                       \epsilon_{\MR{cms}}^2
 \, , \qquad
 \alpha^i_{\epsilon_{\MR{cms}}} = - \frac{1}{2}
    \left[ 
              \frac{\partial^2 {\AFB}^{\MR{th}}_{,i} }{\partial E^2}
          + 2\,\frac{1}{\sigma_i^{\MR{th}}}
               \frac{\partial \sigma_i^{\MR{th}}}{\partial E}
               \frac{\partial {\AFB}^{\MR{th}}_{,i} }{\partial E}
    \right] \,
                       \epsilon_{\MR{cms}}^2 
     \, .
\label{eq:bspread}
\end{equation}
The contributions to the covariance matrix from the uncertainty on 
the \cms\ energy spread, $\Delta \epsilon_{\MR{cms}}$, are then
given by
\begin{eqnarray}
 V_{ij}^{\epsilon_{\MR{cms}}} &=& \Delta_i^{\epsilon_{\MR{cms}}}
                                  \Delta_j^{\epsilon_{\MR{cms}}}
                                                           \nonumber \\
 \MR{with\ } \Delta_i^{\epsilon_{\MR{cms}}} &=& 
                2\, \sigma_i^{\MR{th}}\,
                \left( f^i_{\epsilon_{\MR{cms}}} - 1 \right)\,
                \frac{\Delta \epsilon_{\MR{cms}}}{\epsilon_{\MR{cms}}}
                \qquad\,\ \MR{for\ cross\ sections}
                                                      \, ,  \nonumber \\
                                             &=& 
                2\, \alpha^i_{\epsilon_{\MR{cms}}} \,
                \frac{\Delta \epsilon_{\MR{cms}}}{\epsilon_{\MR{cms}}}
                \qquad\qquad\qquad\quad \MR{for\ \AFB} \, .
\label{eq:Vbspread}
\end{eqnarray}
The spread of the \cms\ energy and its error for the $1993-95$ data sets
are given in Table~\ref{tab:cms_spread}.
The \cms\ energy spread used in Reference~\cite{l3-69} for the $1990-92$ 
data has been revised and the values and errors given in 
Reference~\cite{lep-09} are used.
The error on the \cms\ energy spread is fully correlated between all data sets
of $1990-95$, hence all elements of the covariance matrix receive a 
contribution.

\subsubsection*{\boldmath Error on $t$-channel Contributions}

The last term in Equations~\ref{eq:Vdiagonal} and~\ref{eq:Voffdiagonal}
applies to \EEEEG\ cross section and \AFB\ measurements only.
It accounts for the uncertainty on the calculation of $t$-channel and
\stint\ contributions:
\begin{equation}
   V_{ij}^t = \Delta_k^t \, \Delta_{l}^t \, .
\label{eq:Vtchannel}
\end{equation}
The indices $k$ and $l$ refer to the \cms\ energies of the data set $i$ and $j$
and the corresponding errors on cross sections and asymmetries,
$\Delta^{\MR{t}}$, are listed in 
Table~\ref{tab:tchann_errors}~\cite{Beenakker}.
Off-diagonal elements receive a contribution only if the two data sets are both
below, on or above the peak.
Cross section and asymmetry measurements are also connected this way.

\subsubsection*{Constraints}

In fits using other results, $\Omega_m^{\MR{c}}$, than cross section and 
\AFB\ measurements these additional measurements are added with their
errors, $\Delta_m^{\MR{c}}$, to the $\chi^2$ function.
This applies to our measurements of tau-polarisation, b-quark
and quark charge asymmetries, $R_{\MR{b}}$ and \as, as well as to the value of 
\Da5had\ and the measurement of \Mt\ used in \SM\ fits.
Statistical and systematic errors are added in quadrature to obtain
$\Delta_m^{\MR{c}}$:
\begin{equation}
   \chi^2 = D^T \, V^{-1} \, D  
          + \sum_m \left( \frac{\Omega_m^{\MR{c}} - \Omega_m^{\MR{th}}}
                               {\Delta_m^{\MR{c}}}  \right)^2 \, .
\label{eq:def_chisq2}
\end{equation}
The small correlation of 8\% between $A_{\MR{e}}$ and $A_\tau$ measured 
from tau-polarisation is included by means of a covariance 
matrix.

%
%
\newpage
\typeout{   }     
\typeout{Using author list for the Z LINESHAPE PAPER ONLY}
\typeout{Using author list for the Z LINESHAPE PAPER ONLY}
\typeout{Using author list for the Z LINESHAPE PAPER ONLY}
\typeout{Using author list for the Z LINESHAPE PAPER ONLY}
\typeout{Using author list for the Z LINESHAPE PAPER ONLY}
\typeout{Using author list for the Z LINESHAPE PAPER ONLY}
\typeout{Using author list for the Z LINESHAPE PAPER ONLY}
\typeout{Using author list for the Z LINESHAPE PAPER ONLY}
\typeout{Using author list for the Z LINESHAPE PAPER ONLY}
\typeout{$Modified: Tue Feb  1 11:09:22 2000 by clare $}
\typeout{!!!!  This should only be used with document option a4p!!!!}
\typeout{   }
%
%
%
%
%
%

\newcount\tutecount  \tutecount=0
\def\tutenum#1{\global\advance\tutecount by 1 \xdef#1{\the\tutecount}}
\def\tute#1{$^{#1}$}
\tutenum\aachen            
\tutenum\nikhef            
\tutenum\mich              
\tutenum\lapp              
\tutenum\basel             
\tutenum\lsu               
\tutenum\beijing           
\tutenum\berlin            
\tutenum\bologna           
\tutenum\tata              
\tutenum\ne                
\tutenum\bucharest         
\tutenum\budapest          
\tutenum\mit               
\tutenum\debrecen          
\tutenum\florence          
\tutenum\cern              
\tutenum\wl                
\tutenum\geneva            
\tutenum\hefei             
\tutenum\seft              
\tutenum\lausanne          
\tutenum\lecce             
\tutenum\lyon              
\tutenum\madrid            
\tutenum\milan             
\tutenum\moscow            
\tutenum\naples            
\tutenum\cyprus            
\tutenum\nymegen           
\tutenum\caltech           
\tutenum\perugia           
\tutenum\cmu               
\tutenum\prince            
\tutenum\rome              
\tutenum\peters            
\tutenum\potenza           
\tutenum\salerno           
\tutenum\ucsd              
\tutenum\santiago          
\tutenum\sofia             
\tutenum\korea             
\tutenum\alabama           
\tutenum\utrecht           
\tutenum\purdue            
\tutenum\psinst            
\tutenum\zeuthen           
\tutenum\eth               
\tutenum\hamburg           
\tutenum\taiwan            
\tutenum\tsinghua          
{
\parskip=0pt
\noindent
{\bf The L3 Collaboration:}
\ifx\selectfont\undefined
 \baselineskip=10.8pt
 \baselineskip\baselinestretch\baselineskip
 \normalbaselineskip\baselineskip
 \ixpt
\else
 \fontsize{9}{10.8pt}\selectfont
\fi
\medskip
\tolerance=10000
\hbadness=5000
\raggedright
\hsize=162truemm\hoffset=0mm
\def\r{\rlap,}
\noindent

M.Acciarri\r\tute\milan\
P.Achard\r\tute\geneva\ 
O.Adriani\r\tute{\florence}\ 
M.Aguilar-Benitez\r\tute\madrid\ 
J.Alcaraz\r\tute\madrid\ 
G.Alemanni\r\tute\lausanne\
J.Allaby\r\tute\cern\
A.Aloisio\r\tute\naples\ 
M.G.Alviggi\r\tute\naples\
G.Ambrosi\r\tute\geneva\
% ...
\newpage
\begin{list}{A}{\itemsep=0pt plus 0pt minus 0pt\parsep=0pt plus 0pt minus 0pt
                \topsep=0pt plus 0pt minus 0pt}
\item[\aachen]
 I. Physikalisches Institut, RWTH, D-52056 Aachen, FRG$^{\S}$\\
 III. Physikalisches Institut, RWTH, D-52056 Aachen, FRG$^{\S}$
\item[\nikhef] National Institute for High Energy Physics, NIKHEF, 
     and University of Amsterdam, NL-1009 DB Amsterdam, The Netherlands
\item[\mich] University of Michigan, Ann Arbor, MI 48109, USA
\item[\lapp] Laboratoire d'Annecy-le-Vieux de Physique des Particules, 
     LAPP,IN2P3-CNRS, BP 110, F-74941 Annecy-le-Vieux CEDEX, France
\item[\basel] Institute of Physics, University of Basel, CH-4056 Basel,
     Switzerland
\item[\lsu] Louisiana State University, Baton Rouge, LA 70803, USA
\item[\beijing] Institute of High Energy Physics, IHEP, 
  100039 Beijing, China$^{\triangle}$ 
\item[\berlin] Humboldt University, D-10099 Berlin, FRG$^{\S}$
\item[\bologna] University of Bologna and INFN-Sezione di Bologna, 
     I-40126 Bologna, Italy
\item[\tata] Tata Institute of Fundamental Research, Bombay 400 005, India
\item[\ne] Northeastern University, Boston, MA 02115, USA
\item[\bucharest] Institute of Atomic Physics and University of Bucharest,
     R-76900 Bucharest, Romania
\item[\budapest] Central Research Institute for Physics of the 
     Hungarian Academy of Sciences, H-1525 Budapest 114, Hungary$^{\ddag}$
\item[\mit] Massachusetts Institute of Technology, Cambridge, MA 02139, USA
\item[\debrecen] KLTE-ATOMKI, H-4010 Debrecen, Hungary$^\P$
\item[\florence] INFN Sezione di Firenze and University of Florence, 
     I-50125 Florence, Italy
\item[\cern] European Laboratory for Particle Physics, CERN, 
     CH-1211 Geneva 23, Switzerland
\item[\wl] World Laboratory, FBLJA  Project, CH-1211 Geneva 23, Switzerland
\item[\geneva] University of Geneva, CH-1211 Geneva 4, Switzerland
\item[\hefei] Chinese University of Science and Technology, USTC,
      Hefei, Anhui 230 029, China$^{\triangle}$
\item[\seft] SEFT, Research Institute for High Energy Physics, P.O. Box 9,
      SF-00014 Helsinki, Finland
\item[\lausanne] University of Lausanne, CH-1015 Lausanne, Switzerland
\item[\lecce] INFN-Sezione di Lecce and Universit\'a Degli Studi di Lecce,
     I-73100 Lecce, Italy
\item[\lyon] Institut de Physique Nucl\'eaire de Lyon, 
     IN2P3-CNRS,Universit\'e Claude Bernard, 
     F-69622 Villeurbanne, France
\item[\madrid] Centro de Investigaciones Energ{\'e}ticas, 
     Medioambientales y Tecnolog{\'\i}cas, CIEMAT, E-28040 Madrid,
     Spain${\flat}$ 
\item[\milan] INFN-Sezione di Milano, I-20133 Milan, Italy
\item[\moscow] Institute of Theoretical and Experimental Physics, ITEP, 
     Moscow, Russia
\item[\naples] INFN-Sezione di Napoli and University of Naples, 
     I-80125 Naples, Italy
\item[\cyprus] Department of Natural Sciences, University of Cyprus,
     Nicosia, Cyprus
\item[\nymegen] University of Nijmegen and NIKHEF, 
     NL-6525 ED Nijmegen, The Netherlands
\item[\caltech] California Institute of Technology, Pasadena, CA 91125, USA
\item[\perugia] INFN-Sezione di Perugia and Universit\'a Degli 
     Studi di Perugia, I-06100 Perugia, Italy   
\item[\cmu] Carnegie Mellon University, Pittsburgh, PA 15213, USA
\item[\prince] Princeton University, Princeton, NJ 08544, USA
\item[\rome] INFN-Sezione di Roma and University of Rome, ``La Sapienza",
     I-00185 Rome, Italy
\item[\peters] Nuclear Physics Institute, St. Petersburg, Russia
\item[\potenza] INFN-Sezione di Napoli and University of Potenza, 
     I-85100 Potenza, Italy
\item[\salerno] University and INFN, Salerno, I-84100 Salerno, Italy
\item[\ucsd] University of California, San Diego, CA 92093, USA
\item[\santiago] Dept. de Fisica de Particulas Elementales, Univ. de Santiago,
     E-15706 Santiago de Compostela, Spain
\item[\sofia] Bulgarian Academy of Sciences, Central Lab.~of 
     Mechatronics and Instrumentation, BU-1113 Sofia, Bulgaria
\item[\korea]  Laboratory of High Energy Physics, 
     Kyungpook National University, 702-701 Taegu, Republic of Korea
\item[\alabama] University of Alabama, Tuscaloosa, AL 35486, USA
\item[\utrecht] Utrecht University and NIKHEF, NL-3584 CB Utrecht, 
     The Netherlands
\item[\purdue] Purdue University, West Lafayette, IN 47907, USA
\item[\psinst] Paul Scherrer Institut, PSI, CH-5232 Villigen, Switzerland
\item[\zeuthen] DESY, D-15738 Zeuthen, 
     FRG
\item[\eth] Eidgen\"ossische Technische Hochschule, ETH Z\"urich,
     CH-8093 Z\"urich, Switzerland
\item[\hamburg] University of Hamburg, D-22761 Hamburg, FRG
\item[\taiwan] National Central University, Chung-Li, Taiwan, China
\item[\tsinghua] Department of Physics, National Tsing Hua University,
      Taiwan, China
\item[\S]  Supported by the German Bundesministerium 
        f\"ur Bildung, Wissenschaft, Forschung und Technologie
\item[\ddag] Supported by the Hungarian OTKA fund under contract
numbers T019181, F023259 and T024011.
\item[\P] Also supported by the Hungarian OTKA fund under contract
  numbers T22238 and T026178.
\item[$\flat$] Supported also by the Comisi\'on Interministerial de Ciencia y 
        Tecnolog{\'\i}a.
\item[$\sharp$] Also supported by CONICET and Universidad Nacional de La Plata,
        CC 67, 1900 La Plata, Argentina.
\item[$\diamondsuit$] Also supported by Panjab University, Chandigarh-160014, 
        India.
\item[$\triangle$] Supported by the National Natural Science
  Foundation of China.
\item[\dag] Deceased.
\end{list}
}
\vfill






%

%
%

%

\clearpage
%
%
%
%

\begin{table}
 \begin{center}
  \renewcommand{\arraystretch}{1.1}
   \begin{tabular}{|ll|rrrrrrr|}
    \hline
      &            & \MC{3}{c|}{1993}   & \MC{1}{c|}{1994}  & \MC{3}{c|}{1995}   \\
      &              & peak$-2$&  peak   &  \MC{1}{r|}{peak$+2$} & \MC{1}{r|}{peak}  
                                                    &  peak$-2$ & peak  & peak$+2$   \\
    \hline
          & peak$-2$  &$ 12.59 $&$  8.32 $&$  7.45 $&$  5.59 $&$  2.05 $&$  1.80 $&$  1.84 $ \\               
     1993 & peak      &$       $&$ 45.69 $&$  7.68 $&$  6.20 $&$  1.69 $&$  1.82 $&$  1.72 $ \\               
          & peak$+2$  &$       $&$       $&$  9.57 $&$  5.20 $&$  1.90 $&$  1.96 $&$  2.15 $ \\ \cline{1-2}   
     1994 & peak      &$       $&$       $&$       $&$ 14.30 $&$  1.90 $&$  2.07 $&$  1.92 $ \\ \cline{1-2}   
          & peak$-2$  &$       $&$       $&$       $&$       $&$  4.49 $&$  2.34 $&$  2.30 $ \\               
     1995 & peak      &$       $&$       $&$       $&$       $&$       $&$ 30.40 $&$  2.60 $ \\               
          & peak$+2$  &$       $&$       $&$       $&$       $&$       $&$       $&$  4.15 $ \\               
     \hline
  \end{tabular}
  \renewcommand{\arraystretch}{1.0}
  \parbox{\capwidth}{
  \caption[]{The covariance matrix, $v_{kl}^{\MR{LEP}}$, of the 
             LEP \protect\cms\ energy 
             uncertainty at the L3 interaction point 
             for the 1994 data set and the two scans of the Z resonance 
             performed in 1993 and 1995~\protect\cite{lep-09}.
             All values are given in units of $\MeV^2$.
   \label{tab:leperror1}
            }
   }
 \end{center}
\end{table}

\begin{table}
 \begin{center}
  \renewcommand{\arraystretch}{1.1}
  \begin{tabular}{|l|rrr|}
   \hline
                 & \multicolumn{3}{c|}{$\epsilon_{\MR{cms}}$\ [\MeV]} \\
                 &     1993  &     1994  &      1995  \\
   \hline
   pre-scan      &     56.8  &           &      56.9  \\
   peak$-2$      &     56.6  &           &      55.9  \\
   peak          &     57.0  &     56.5  &      56.4  \\
   peak$+2$      &     57.1  &           &      56.9  \\
   \hline
   $\Delta \epsilon_{\MR{cms}}$\ [\MeV] 
                 & $\pm 1.1$ & $\pm 1.1$ & $\pm 1.3$  \\
   \hline
  \end{tabular}
  \renewcommand{\arraystretch}{1.0}
  \parbox{\capwidth}{
  \caption[]{The spread on the average \protect\cms\ energy, 
             $\epsilon_{\MR{cms}}$, 
             and its error, $\Delta \epsilon_{\MR{cms}}$,
             at the L3 interaction point~\protect\cite{lep-09} 
             for the nine data sets.
             The additional scatter from the time variation
             of the mean \protect\cms\ energies is included.
             The uncertainty does not depend on the energy.
   \label{tab:cms_spread}
            }
   }
 \end{center}
\end{table}

\clearpage
\thispagestyle{empty}

%
\begin{table}
 \begin{sideways}
 \begin{minipage}[b]{\textheight}
 \ \vspace*{10mm} \\
 \small
 \begin{center}
  \renewcommand{\arraystretch}{1.1}
  \hspace*{-20mm}%
  \begin{tabular}{|l|rrrrrrr|rrrrrrrr|r|}
   \hline
           & \MC{7}{c|}{1990}                               &  \MC{8}{c|}{1991}                                     &  1992\\  \hline
$\RS\ [\GeV]$
           &88.231&89.236&90.238&91.230&92.226&93.228&94.223&91.254&88.480&89.470&90.228&91.222&91.967&92.966&93.716&91.294\\ \hline
  88.231   & 696  & 675  & 676  & 677  & 678  & 678  & 679  &   0  & 159  & 154  & 150  & 145  & 141  & 136  & 132  &   0  \\
  89.236   &      & 696  & 682  & 685  & 688  & 691  & 694  &   0  & 151  & 148  & 145  & 142  & 140  & 137  & 135  &   0  \\
  90.238   &      &      & 706  & 692  & 698  & 703  & 709  &   0  & 142  & 141  & 141  & 140  & 139  & 139  & 138  &   0  \\
  91.230   &      &      &      & 702  & 708  & 715  & 723  &   0  & 133  & 135  & 136  & 137  & 139  & 140  & 141  &   0  \\
  92.226   &      &      &      &      & 743  & 728  & 738  &   0  & 125  & 128  & 131  & 135  & 138  & 142  & 145  &   0  \\
  93.228   &      &      &      &      &      & 764  & 753  &   0  & 116  & 122  & 126  & 133  & 137  & 143  & 148  &   0  \\
  94.223   &      &      &      &      &      &      & 788  &   0  & 107  & 115  & 122  & 130  & 136  & 145  & 151  &   0  \\ \hline 
  91.254   &      &      &      &      &      &      &      & 333  &   0  &   0  &   0  &   0  &   0  &   0  &   0  &   0  \\
  88.480   &      &      &      &      &      &      &      &      &  93.5&  61.6&  54.1&  44.3&  36.9&  27.1&  19.7&   0  \\
  89.470   &      &      &      &      &      &      &      &      &      &  74.8&  48.2&  40.6&  34.9&  27.3&  21.6&   0  \\
  90.228   &      &      &      &      &      &      &      &      &      &      &  66.7&  37.8&  33.4&  27.5&  23.0&   0  \\
  91.222   &      &      &      &      &      &      &      &      &      &      &      &  45.3&  31.4&  27.7&  24.9&   0  \\
  91.967   &      &      &      &      &      &      &      &      &      &      &      &      &  53.2&  27.9&  26.3&   0  \\
  92.966   &      &      &      &      &      &      &      &      &      &      &      &      &      &  45.7&  28.3&   0  \\
  93.716   &      &      &      &      &      &      &      &      &      &      &      &      &      &      &  57.6&   0  \\ \hline  
  91.294   &      &      &      &      &      &      &      &      &      &      &      &      &      &      &      & 324  \\
   \hline  
  \end{tabular}
  \renewcommand{\arraystretch}{1.0}
  \parbox{\capwidth}{
  \caption[]{The covariance matrix of the LEP \protect\cms\ energy 
             uncertainty for the $1990 - 92$ data 
             sets obtained following 
             references~\protect\cite{lep-02}.
             All values are given in units of $\MeV^2$.
             The \protect\cms\ energies listed correspond to our measurements
             in Reference~\protect\cite{l3-69}.
            }
   \label{tab:leperror9092}
   }
 \end{center}
 \normalsize
 \end{minipage}
 \end{sideways}
\end{table}

%
%
%
%
\begin{table}
 \begin{center}
  \renewcommand{\arraystretch}{1.1}
  \begin{tabular}{|l|r|r|r|}
  \hline
   Source                                    &   1993 &  1994   & 1995  \\
  \hline                                                         
   Selection Criteria     \hfill [\permille] & $0.48$ &  $0.42$ & $0.47$ \\
   Detector Geometry      \hfill [\permille] & $0.63$ &  $0.34$ & $0.34$ \\
   Monte Carlo Statistics \hfill [\permille] & $0.35$ &  $0.35$ & $0.35$ \\
  \hline                                                                 
   Total Experimental ($\delta_k^{\MR{lum,exp}}$)                        
                          \hfill [\permille] & $0.86$ &  $0.64$ & $0.68$ \\
  \hline                                                                 
  \hline                                                                 
   Theory  ($\delta_k^{\MR{lum,th}}$)                                    
                          \hfill [\permille] & $0.61$ &  $0.61$ & $0.61$ \\
  \hline                                                                 
   Total Uncertainty      \hfill [\permille] & $1.05$ &  $0.88$ & $0.91$ \\
  \hline
  \end{tabular}
  \renewcommand{\arraystretch}{1.0}
  \parbox{\capwidth}{
  \caption[]{Experimental and theoretical contributions to the systematic 
             error on the luminosity measurement for different years.
             Additional contributions to the error from statistics and 
             from the uncertainty on the \protect\cms\ energy are also
             taken into account in the fitting procedure.
  \label{tab:lumi_sys}
            }
  }
 \end{center}
\end{table}

\begin{table}
 \begin{center}
  \renewcommand{\arraystretch}{1.1}
  \begin{tabular}{|l|rrr|}
  \hline
           &    1993 &  1994  &    1995  \\
  \hline             
    1993   &    1.00 &  0.59  &   0.59   \\
    1994   &         &  1.00  &   0.93   \\
    1995   &         &        &   1.00   \\
  \hline
  \end{tabular}
  \renewcommand{\arraystretch}{1.0}
  \parbox{\capwidth}{
  \caption[]{Correlation coefficients, $\rho_{kl}^{\MR{lum,exp}}$,
             between the data sets of different years of the total 
             experimental systematic error on the luminosity
             measurement, $\delta_k^{\MR{lum,exp}}$, 
             as given in Table~\protect\ref{tab:lumi_sys}.
  \label{tab:lumi_correlations}
            }
  }
 \end{center}
\end{table}

\begin{table}
 \begin{center}
  \renewcommand{\arraystretch}{1.1}
   \begin{tabular}{|ll|r|r|}
    \hline
          &           & $N_{\mathrm{events}}$ & $\delta^{\MR{lum,stat}}$\ [\permille]    \\
    \hline
     1993 & pre-scan & $   362\,500$ & $1.66$ \\
          & peak$-2$ & $   604\,535$ & $1.29$ \\               
          & peak     & $   651\,931$ & $1.24$ \\               
          & peak$+2$ & $   588\,962$ & $1.30$ \\ \cline{1-2}   
     1994 & peak     & $3\,129\,424$ & $0.57$ \\ \cline{1-2}   
     1995 & pre-scan & $   480\,342$ & $1.44$ \\
          & peak$-2$ & $   541\,580$ & $1.36$ \\               
          & peak     & $   283\,887$ & $1.88$ \\               
          & peak$+2$ & $   554\,371$ & $1.34$ \\               
     \hline
     \MC{2}{|l|}{peak combined}     
                     & $4\,908\,084$ & $0.45$ \\
     \hline
  \end{tabular}
  \renewcommand{\arraystretch}{1.0}
  \parbox{\capwidth}{
  \caption[]{Number of events used for the measurement of the total luminosity
             in the nine data taking periods, $N_{\mathrm{events}}$, 
             and the corresponding contributions to the error of 
             cross section measurements, $\delta^{\MR{lum,stat}}$.
             The numbers correspond to the luminosity used for the hadron
             cross section measurements and are used for leptonic channels
             too.
             The last line is the sum of the five data sets taken at the peak,
             indicating this error contribution to the measurements of the pole
             cross sections. 
  \label{tab:lumi_statistics}
            }
   }
 \end{center}
\end{table}

\begin{table}
 \begin{center}
  \renewcommand{\arraystretch}{1.1}
  \begin{tabular}{|l|r|}
   \hline
    Source                                     & $1993 - 1995$ \\ 
   \hline
    Monte Carlo statistics  \hfill [\permille] & $0.04 - 0.10$ \\
    Acceptance              \hfill [\permille] &        $0.21$ \\
    Selection cuts          \hfill [\permille] &        $0.30$ \\
    Trigger                 \hfill [\permille] &        $0.12$ \\
   \hline                                             
    Total scale             \hfill [\permille] & $0.39 - 0.40$ \\
   \hline                                             
   \hline                                             
    Non-resonant background \hfill [\pb]       &  $ 3$         \\
    Detector noise          \hfill [\pb]       &  $ 1$         \\
   \hline                                             
    Total  absolute         \hfill [\pb]       &  $3.2$        \\
   \hline
  \end{tabular}
  \parbox{\capwidth}{
  \caption[]{Contributions to the systematic uncertainty on the cross section
             \protect\EEHADG.
             Except for the contribution from Monte Carlo statistics all
             errors are fully correlated among the data sets.
             The resulting correlated scale error is 
             $\delta^{\MR{cor}} = 0.39$\protect\permille.
  \label{tab:had_sys}
            }
   }
 \end{center}
\end{table}

\begin{table}
 \begin{center}
  \renewcommand{\arraystretch}{1.1}
  \begin{tabular}{|l|r|r|r|}
   \hline
    Source                                   & 1993      &   1994 & 1995       \\
   \hline
    Monte Carlo statistics\hfill [\permille] & $0.9-1.5$ &  $0.4$ & $1.7-2.4$  \\
    Acceptance            \hfill [\permille] & $2.7$     &  $2.7$ & $3.2$      \\
    Selection cuts        \hfill [\permille] & $1.3$     &  $1.3$ & $1.4-2.2$  \\
    Trigger               \hfill [\permille] & $0.6$     &  $0.6$ & $0.5-0.7$  \\
    Resonant background   \hfill [\permille] & $0.3$     &  $0.3$ & $0.3$      \\
   \hline
     Total scale          \hfill [\permille] & $3.2-3.4$ &  $3.1$ & $3.9-4.6$  \\
   \hline
   \hline
    $\mathrm{\ee\rightarrow \ee\,\mm}$ \hfill [\pb]
                                             & $-$       &  $-$   & $0.1$      \\
    Cosmic rays          \hfill [\pb]        & $0.3$     & $0.3$  & $0.3$      \\
   \hline
    Total  absolute      \hfill [\pb]        & $0.3$     & $0.3$  & $0.3$      \\
   \hline
  \end{tabular}
  \renewcommand{\arraystretch}{1.0}
  \parbox{\capwidth}{
  \caption[]{Contributions to the systematic uncertainty on the cross section
             \protect\EEMMG.
             Except for the contribution from Monte Carlo statistics, all
             errors are fully correlated among the data sets yielding
             a correlated scale error of 
             $\delta^{\MR{cor}} = 3.1$\protect\permille\ for $1993-94$ data.
             For the 1995 data this error is estimated to be 
             $3.6$\protect\permille\ and it is taken to be fully correlated 
             with the other years.
             }
  \label{tab:muon_sys}
        }
 \end{center}
\end{table}

\begin{table}
 \begin{center}
  \renewcommand{\arraystretch}{1.1}
  \begin{tabular}{|l|r|r|}
   \hline
    Source                  &        $1993-94$   & 1995        \\
   \hline                                                              
    Fit procedure           &       $<0.0003$  & $<0.0003$   \\
    Detector asymmetry      &        $0.0006$  &  $0.0006$   \\
    Charge confusion        &        $0.0004$  &  $0.0010$   \\
    Momentum reconstruction &        $0.0004$  &  $0.0009$   \\
    Background              & $0.0001-0.0005$  & $<0.0003$   \\
   \hline                                                   
     Total uncertainty      & $0.0008-0.0009$  &  $0.0015$   \\
   \hline
  \end{tabular}
  \renewcommand{\arraystretch}{1.0}
  \parbox{\capwidth}{
  \caption[]{Contributions to the systematic uncertainty on the
             forward-backward asymmetry in \protect\EEMMG.
             The total uncertainty is assumed to be fully correlated
             among the data sets.
             }
  \label{tab:muon_sys_afb}
         }
 \end{center}
\end{table}

\begin{table}
 \begin{center}
  \renewcommand{\arraystretch}{1.1}
  \begin{tabular}{|l|r|r|r|}
   \hline
    Source                                     &     1993  &    1994  &      1995 \\
   \hline
    Monte Carlo statistics  \hfill [\permille] & $2.3-4.2$ &   $1.1$  & $1.5-1.7$ \\
    Tau branching fractions \hfill [\permille] & $2.0$     &   $2.0$  &     $2.0$ \\
    Selection cuts          \hfill [\permille] & $5.3-8.0$ &   $6.0$  & $6.4-7.5$ \\
    Trigger                 \hfill [\permille] & $0.2$     &   $0.2$  &     $0.2$ \\
    Resonant background     \hfill [\permille] & $1.4-3.3$ &   $1.0$  & $1.3-3.0$ \\
   \hline
    Total scale             \hfill [\permille] & $6.8-9.5$ &   $6.5$  & $7.5-8.0$ \\
   \hline
   \hline
    $\ee\rightarrow \ee\,
     \tautau,\,\ee\,\MR{hadrons}$
                            \hfill [\pb]       &  \multicolumn{3}{c|}{$0.6$}      \\
    Cosmic rays             \hfill [\pb]       &  \multicolumn{3}{c|}{$1.0$}      \\
   \hline
    Total  absolute         \hfill [\pb]       &  \multicolumn{3}{c|}{$1.2$}      \\
   \hline
  \end{tabular}
  \renewcommand{\arraystretch}{1.0}
  \parbox{\capwidth}{
  \caption[]{Contributions to the systematic uncertainty on the cross section
             \protect\EETTG.
             The total absolute and $5.7$\protect\permille\ of the total scale
             error are assumed to be fully correlated among all data sets.
            }
  \label{tab:tau_sys}
      }
 \end{center}
\end{table}

\begin{table}
 \begin{center}
  \renewcommand{\arraystretch}{1.1}
  \begin{tabular}{|l|r|}
   \hline
    Source                  & $1993-1995$  \\ 
   \hline
    Detector asymmetry      & $ 0.0030$    \\
    Charge confusion        & $<0.0001$    \\
    Background              & $ 0.0010$    \\
    Helicity bias           & $ 0.0004$    \\
   \hline                                
    Total uncertainty       & $ 0.0032$    \\
   \hline
  \end{tabular}
  \renewcommand{\arraystretch}{1.0}
  \parbox{\capwidth}{
  \caption[]{Contributions to the systematic uncertainty on the
             forward-backward asymmetry in \protect\EETTG.
             The total uncertainty is assumed to be fully correlated
             among the data sets.
            }
  \label{tab:tau_sys_afb}
        }
 \end{center}
\end{table}

\begin{table}
 \begin{center}
  \renewcommand{\arraystretch}{1.1}
  \begin{tabular}{|l|r|r|r|}
   \hline
    Source                                         &    1993   &   1994  & 1995   \\ 
   \hline                                                                         
    Monte Carlo statistics     \hfill [\permille]  & $0.4-1.0$ &  $0.4$  & $0.4$  \\
    $^\ast$ Generator          \hfill [\permille]  &     $1.0$ &  $1.0$  & $1.0$  \\  
    $^\ast$ $\MR{e}$-\Photon\ discrimination
                               \hfill [\permille]  &     $0.6$ &  $0.6$  & $1.5$  \\
    $^\ast$ z-vertex position  \hfill [\permille]  &     $0.5$ &  $0.5$  & $ - $  \\
    $^\ast$ BGO geometry       \hfill [\permille]  &     $0.5$ &  $0.5$  & $ - $  \\  
    $^\ast$ fiducial volume    \hfill [\permille]  &     $ - $ &  $ - $  & $1.5$  \\
    Background                 \hfill [\permille]  & $0.7-1.0$ &  $0.4$  & $0.4$  \\
    Selection cuts             \hfill [\permille]  & $1.6-2.7$ &  $0.8$  & $1.5$  \\
   \hline                                                                         
    Total scale                \hfill [\permille]  & $2.3-3.2$ &  $1.7$  & $2.8$  \\
   \hline
  \end{tabular}
  \renewcommand{\arraystretch}{1.0}
  \parbox{\capwidth}{
  \caption[]{Contributions to the systematic uncertainty on the cross section
             \protect\EEEEG.
             Error sources indicated as $^\ast$ are correlated yielding a
             total correlated scale error of  
             $\delta^{\MR{cor}} = 1.4$\protect\permille\ for the $1993-94$ data
             and $2.3$\protect\permille\ for the 1995 data.
           }
  \label{tab:electron_sys}
        }
 \end{center}
\end{table}

\begin{table}
 \begin{center}
  \renewcommand{\arraystretch}{1.1}
  \begin{tabular}{|l|r|}
   \hline
     Source                 & $1993 - 1995$\\ 
   \hline                                      
    Charge confusion        & $0.0020$     \\
    $z$-vertex              & $0.0015$     \\
    Background              & $0.0005$     \\
   \hline                                      
    Total uncertainty       & $0.0025$     \\
   \hline
  \end{tabular}
  \parbox{\capwidth}{
  \caption[]{Contributions to the systematic uncertainty on the
             forward-backward asymmetry in \protect\EEEEG.
             }
  \label{tab:electron_sys_afb}
        }
 \end{center}
\end{table}

\clearpage

%
%
%
%
\begin{table}
\begin{center}
\renewcommand{\arraystretch}{1.1}
\begin{tabular}{|r|r|r|r@{$\,\pm\,$}l|r|}
\hline
    $\RS\ [\GeV]$ & $N_{\mathrm{events}}$ & $\IL\ [\pbinv]$ &
    \MC{2}{c|}{$\sigma\ [\nb]$} &
    $\Delta_i^{\MR{unc}}\ [\nb]$  \\
   \hline
      $91.3217$ & $     158\,736 $ & $   5.21$ & $30.665$ & $ 0.077$ & $ 0.003$ \\   
      $89.4498$ & $      83\,681 $ & $   8.32$ & $10.087$ & $ 0.035$ & $ 0.001$ \\   
      $91.2057$ & $     281\,359 $ & $   9.34$ & $30.309$ & $ 0.057$ & $ 0.003$ \\   
      $93.0352$ & $     121\,926 $ & $   8.79$ & $13.909$ & $ 0.040$ & $ 0.001$ \\   
   \hline
    1993 Totals & $     645\,702 $ & $  31.66$ & \multicolumn{3}{c|}{ } \\
   \hline
   \hline
      $91.2202$ & $  1\,359\,490 $ & $  44.84$ & $30.513$ & $ 0.026$ & $ 0.001$ \\   
   \hline
   \hline
      $91.3093$ & $     209\,195 $ & $   6.90$ & $30.512$ & $ 0.066$ & $ 0.003$ \\   
      $89.4517$ & $      75\,102 $ & $   7.46$ & $10.081$ & $ 0.037$ & $ 0.001$ \\   
      $91.2958$ & $     123\,791 $ & $   4.08$ & $30.493$ & $ 0.086$ & $ 0.003$ \\   
      $92.9827$ & $     117\,555 $ & $   8.28$ & $14.232$ & $ 0.041$ & $ 0.001$ \\   
   \hline
    1995 Totals & $     525\,643 $ & $  26.72$ & \multicolumn{3}{c|}{ } \\
   \hline
   \hline
      Total sum & $  2\,530\,835 $ & $ 103.21$ & \multicolumn{3}{c|}{ } \\
   \hline
\end{tabular}
\renewcommand{\arraystretch}{1.0}
\parbox{\capwidth}{
\caption[]{Average \protect\cms\ energies, number of selected events, 
           integrated luminosities and measured cross sections with statistical
           errors for \protect\EEHADG.
           The cross sections are quoted for 
           $\protect\RSprime > 0.1\protect\RS$.
           Apart from the uncorrelated part listed,
           $\Delta_i^{\MR{unc}}$, systematic errors consist
           in addition of a fully correlated multiplicative contribution,
           $\delta_i^{\MR{cor}} = 0.39$\protect\permille\ 
           and an absolute uncertainty, $\Delta_i^{\MR{abs}} = 3.2\ \pb$.
           Systematic errors from the luminosity measurement 
           (Tables~\protect\ref{tab:lumi_sys} 
               and~\protect\ref{tab:lumi_statistics}) 
           have to be added.
          } 
\label{tab:had_xs}
      }
\end{center}
\end{table}

\begin{table}
\begin{center}
\renewcommand{\arraystretch}{1.1}
\begin{tabular}{|r|r|r|r@{$\,\pm\,$}l|r|}
\hline
    $\RS\ [\GeV]$ & $N_{\mathrm{events}}$ & $\IL\ [\pbinv]$ &
    \MC{2}{c|}{$\sigma\ [\nb]$} &
    $\Delta_i^{\MR{unc}}\ [\nb]$ \\
   \hline
      $91.3216$ & $       5\,134 $ & $   5.00$ & $ 1.504$ & $ 0.021$ & $ 0.002$ \\   
      $89.4499$ & $       2\,739 $ & $   8.03$ & $ 0.504$ & $ 0.010$ & $ 0.001$ \\   
      $91.2056$ & $       9\,237 $ & $   9.15$ & $ 1.482$ & $ 0.016$ & $ 0.001$ \\   
      $93.0357$ & $       3\,679 $ & $   7.97$ & $ 0.684$ & $ 0.011$ & $ 0.001$ \\   
   \hline
    1993 Totals & $      20\,789 $ & $  30.15$ & \multicolumn{3}{c|}{ } \\
   \hline
   \hline
      $91.2197$ & $      41\,768 $ & $  41.38$ & $ 1.484$ & $ 0.007$ & $ 0.001$ \\   
   \hline
   \hline
      $91.3090$ & $       5\,772 $ & $   5.09$ & $ 1.467$ & $ 0.020$ & $ 0.003$ \\   
      $89.4517$ & $       2\,789 $ & $   7.36$ & $ 0.490$ & $ 0.009$ & $ 0.001$ \\   
      $91.2949$ & $       3\,967 $ & $   3.47$ & $ 1.483$ & $ 0.024$ & $ 0.003$ \\   
      $92.9825$ & $       4\,305 $ & $   7.83$ & $ 0.703$ & $ 0.011$ & $ 0.002$ \\   
   \hline
    1995 Totals & $      16\,833 $ & $  23.75$ & \multicolumn{3}{c|}{ } \\
   \hline
   \hline
      Total sum & $      79\,390 $ & $  95.29$ & \multicolumn{3}{c|}{ } \\
   \hline
\end{tabular}
\renewcommand{\arraystretch}{1.0}
\parbox{\capwidth}{
\caption[]{Same as Table~\protect\ref{tab:had_xs} for \protect\EEMMG.
           The cross sections are extrapolated to the full phase space.
           Apart from the uncorrelated part listed, systematic errors consist
           in addition of a fully correlated multiplicative contribution,
           $\delta_i^{\MR{cor}} = 3.1$\protect\permille\ 
            (3.6\protect\permille\ for 1995 data),
           and an absolute uncertainty, $\Delta_i^{\MR{abs}} = 0.3\ \pb$.
          } 
\label{tab:muon_xs}
      }
\end{center}
\end{table}

\begin{table}
\begin{center}
\renewcommand{\arraystretch}{1.1}
\begin{tabular}{|r|r|r|r@{$\,\pm\,$}l|r|}
\hline
    $\RS\ [\GeV]$ & $N_{\mathrm{events}}$ & $\IL\ [\pbinv]$ &
    \MC{2}{c|}{$\sigma\ [\nb]$} &
    $\Delta_i^{\MR{unc}}\ [\nb]$ \\
   \hline
      $91.3221$ & $       4\,805 $ & $   5.08$ & $ 1.492$ & $ 0.021$ & $ 0.006$ \\   
      $89.4500$ & $       2\,706 $ & $   8.08$ & $ 0.509$ & $ 0.010$ & $ 0.004$ \\   
      $91.2058$ & $       8\,506 $ & $   9.00$ & $ 1.474$ & $ 0.017$ & $ 0.007$ \\   
      $93.0358$ & $       3\,637 $ & $   7.93$ & $ 0.718$ & $ 0.012$ & $ 0.003$ \\   
   \hline
    1993 Totals & $      19\,654 $ & $  30.09$ & \multicolumn{3}{c|}{ } \\
   \hline
   \hline
      $91.2197$ & $      41\,439 $ & $  43.53$ & $ 1.472$ & $ 0.007$ & $ 0.005$ \\   
   \hline
   \hline
      $91.3096$ & $       7\,314 $ & $   7.75$ & $ 1.474$ & $ 0.017$ & $ 0.007$ \\   
      $89.4518$ & $       2\,352 $ & $   7.48$ & $ 0.483$ & $ 0.010$ & $ 0.002$ \\   
      $91.2951$ & $       3\,509 $ & $   3.67$ & $ 1.503$ & $ 0.025$ & $ 0.009$ \\   
      $92.9828$ & $       3\,723 $ & $   8.25$ & $ 0.707$ & $ 0.012$ & $ 0.004$ \\   
   \hline
    1995 Totals & $      16\,898 $ & $  27.15$ & \multicolumn{3}{c|}{ } \\
   \hline
   \hline
      Total sum & $      77\,991 $ & $ 100.77$ & \multicolumn{3}{c|}{ } \\
   \hline
\end{tabular}
\renewcommand{\arraystretch}{1.0}
\parbox{\capwidth}{
\caption[]{Same as Table~\protect\ref{tab:had_xs} for \protect\EETTG.
           The cross sections are extrapolated to the full phase space.
           Apart from the uncorrelated part listed, systematic errors consist
           in addition of a fully correlated multiplicative contribution,
           $\delta_i^{\MR{cor}} = 5.7$\protect\permille\ and an absolute 
           uncertainty, $\Delta_i^{\MR{abs}} = 1.2\ \pb$.
          } 
\label{tab:tau_xs}
      }
\end{center}
\end{table}

\begin{table}
\begin{center}
\renewcommand{\arraystretch}{1.1}
\begin{tabular}{|r|r|r|r@{$\,\pm\,$}l|r|r@{$\,\pm\,$}l|}
\hline
    $\RS\ [\GeV]$ & $N_{\mathrm{events}}$ & $\IL\ [\pbinv]$ &
    \MC{2}{c|}{$\sigma\ [\nb]$} &
    $\Delta_i^{\MR{unc}}\ [\nb]$ &
    \MC{2}{c|}{$\sigma^s\ [\nb]$} \\
   \hline
      $91.3213$ & $       5\,267 $ & $   5.19$ & $ 1.017$ & $ 0.014$ & $ 0.003$ & $ 1.438  $ & $ 0.023 $ \\   
      $89.4497$ & $       4\,610 $ & $   8.28$ & $ 0.566$ & $ 0.008$ & $ 0.001$ & $ 0.515  $ & $ 0.014 $ \\   
      $91.2057$ & $       9\,834 $ & $   9.23$ & $ 1.075$ & $ 0.011$ & $ 0.002$ & $ 1.497  $ & $ 0.018 $ \\   
      $93.0358$ & $       3\,610 $ & $   8.39$ & $ 0.431$ & $ 0.007$ & $ 0.001$ & $ 0.703  $ & $ 0.012 $ \\   
   \hline
    1993 Totals & $      23\,321 $ & $  31.09$ & \multicolumn{3}{c|}{ } &\multicolumn{2}{c|}{ } \\
   \hline
   \hline
      $91.2197$ & $      43\,300 $ & $  40.64$ & $ 1.075$ & $ 0.005$ & $ 0.001$ & $ 1.501  $ & $ 0.008 $ \\   
   \hline
   \hline
      $91.3106$ & $       8\,200 $ & $   7.88$ & $ 1.042$ & $ 0.012$ & $ 0.002$ & $ 1.476  $ & $ 0.019 $ \\   
      $89.4517$ & $       3\,891 $ & $   6.91$ & $ 0.564$ & $ 0.009$ & $ 0.001$ & $ 0.511  $ & $ 0.015 $ \\   
      $91.2960$ & $       4\,310 $ & $   4.02$ & $ 1.072$ & $ 0.017$ & $ 0.002$ & $ 1.520  $ & $ 0.027 $ \\   
      $92.9828$ & $       3\,405 $ & $   7.90$ & $ 0.427$ & $ 0.007$ & $ 0.001$ & $ 0.696  $ & $ 0.012 $ \\   
   \hline
    1995 Totals & $      19\,806 $ & $  26.72$ & \multicolumn{3}{c|}{ } &\multicolumn{2}{c|}{ } \\
   \hline
   \hline
      Total sum & $      86\,427 $ & $  98.45$ & \multicolumn{3}{c|}{ } &\multicolumn{2}{c|}{ } \\
   \hline
\end{tabular}
\renewcommand{\arraystretch}{1.0}
\parbox{\capwidth}{
\caption[]{Average \protect\cms\ energies, number of selected events,
           integrated luminosity and measured cross sections with their 
           statistical and uncorrelated systematic errors in the fiducial volume 
           $44^\circ < \theta < 136^\circ$ and for 
           $\xi < 25^\circ$ for the reaction \protect\EEEEG.
           The systematic error consists in addition of a fully correlated 
           multiplicative contribution,
           $\delta_i^{\MR{cor}} = 1.4$\protect\permille\ 
           (2.3\protect\permille\ in 1995).
           In the rightmost column the $s$-channel contribution to the total
           cross section in the full solid angle is listed
           with the statistical error
           (see Appendix~\protect\ref{app:tchannel}).
          } 
\label{tab:electron_xs}
      }
\end{center}
\end{table}

\begin{table}
\begin{center}
\renewcommand{\arraystretch}{1.1}
\begin{tabular}{|r|r|r@{$\,\pm\,$}l|}
\hline
    $\RS\ [\GeV]$ & $N_{\mathrm{events}}$ &
    \MC{2}{c|}{\AFB}      \\
   \hline
      $91.3217$ & $       5\,385 $ & $  0.009$ & $  0.015$ \\    
      $89.4497$ & $       2\,631 $ & $ -0.182$ & $  0.020$ \\    
      $91.2054$ & $       9\,150 $ & $  0.000$ & $  0.011$ \\    
      $93.0352$ & $       3\,635 $ & $  0.119$ & $  0.017$ \\    
   \hline
   \hline
      $91.2203$ & $      43\,416 $ & $ 0.0086$ & $ 0.0051$ \\    
   \hline
   \hline
      $91.3090$ & $       5\,643 $ & $  0.022$ & $  0.013$ \\    
      $89.4517$ & $       2\,686 $ & $ -0.175$ & $  0.019$ \\    
      $91.2949$ & $       3\,858 $ & $  0.030$ & $  0.016$ \\    
      $92.9825$ & $       4\,193 $ & $  0.104$ & $  0.015$ \\    
   \hline
\end{tabular}
\renewcommand{\arraystretch}{1.0}
\parbox{\capwidth}{
\caption[]{Results on the forward-backward asymmetry,
           including an acollinearity cut $\xi < 15^\circ$,
           for \protect\EEMMG\ together with the average 
           \protect\cms\ energies and the number of events.
           The errors are statistical only and a correlated absolute
           systematic error of $\Delta_i^{\MR{abs}} = 0.0008$
           ($0.0015$ for 1995 data) has to be added.
          } 
\label{tab:muon_afb}
      }
\end{center}
\end{table}

\begin{table}
\begin{center}
\renewcommand{\arraystretch}{1.1}
\begin{tabular}{|r|r|r@{$\,\pm\,$}l|}
\hline
    $\RS\ [\GeV]$ & $N_{\mathrm{events}}$ &
    \MC{2}{c|}{\AFB}      \\
   \hline
      $91.3221$ & $       3\,655 $ & $ -0.003$ & $  0.017$ \\    
      $89.4500$ & $       2\,090 $ & $ -0.138$ & $  0.022$ \\    
      $91.2058$ & $       6\,669 $ & $  0.020$ & $  0.013$ \\    
      $93.0358$ & $       2\,822 $ & $  0.133$ & $  0.019$ \\    
   \hline
   \hline
      $91.2202$ & $      36\,509 $ & $ 0.0062$ & $ 0.0053$ \\    
   \hline
   \hline
      $91.3096$ & $       6\,317 $ & $  0.044$ & $  0.013$ \\    
      $89.4518$ & $       2\,020 $ & $ -0.134$ & $  0.023$ \\    
      $91.2951$ & $       3\,017 $ & $  0.001$ & $  0.018$ \\    
      $92.9828$ & $       3\,263 $ & $  0.134$ & $  0.017$ \\    
   \hline
\end{tabular}
\renewcommand{\arraystretch}{1.0}
\parbox{\capwidth}{
\caption[]{Same as Table~\protect\ref{tab:muon_afb} for
           \protect\EETTG.
           The cut on the acollinearity angle is $\xi < 10^\circ$.
           The errors are statistical only and a correlated absolute
           systematic error of $\Delta_i^{\MR{abs}} = 0.0032$
           has to be added.
          }
\label{tab:tau_afb}
      }
\end{center}
\end{table}

\begin{table}
\begin{center}
\renewcommand{\arraystretch}{1.1}
\begin{tabular}{|r|r|r@{$\,\pm\,$}l|r@{$\,\pm\,$}l|}
\hline
    $\RS\ [\GeV]$ & $N_{\mathrm{events}}$ &
    \MC{2}{c|}{\AFB}      &
    \MC{2}{c|}{$\AFB^s$}    \\
   \hline
      $91.3213$ & $       4\,009 $ & $  0.083$ & $  0.016$ & $ -0.027 $ & $ 0.022 $ \\   
      $89.4497$ & $       3\,434 $ & $  0.311$ & $  0.016$ & $ -0.119 $ & $ 0.039 $ \\   
      $91.2057$ & $       7\,330 $ & $  0.111$ & $  0.012$ & $ -0.003 $ & $ 0.016 $ \\   
      $93.0358$ & $       2\,679 $ & $  0.101$ & $  0.019$ & $  0.110 $ & $ 0.024 $ \\   
   \hline                                                                           
   \hline                                                                           
      $91.2197$ & $      31\,636 $ & $ 0.1213$ & $ 0.0056$ & $  0.014 $ & $ 0.008 $ \\   
   \hline                                                                           
   \hline                                                                           
      $91.3106$ & $       7\,861 $ & $  0.101$ & $  0.011$ & $ -0.001 $ & $ 0.016 $ \\   
      $89.4517$ & $       3\,722 $ & $  0.257$ & $  0.016$ & $ -0.244 $ & $ 0.039 $ \\   
      $91.2960$ & $       4\,083 $ & $  0.080$ & $  0.016$ & $ -0.030 $ & $ 0.023 $ \\   
      $92.9828$ & $       3\,203 $ & $  0.055$ & $  0.018$ & $  0.054 $ & $ 0.022 $ \\   
   \hline
\end{tabular}
\renewcommand{\arraystretch}{1.0}
\parbox{\capwidth}{
\caption[]{Measured forward-backward asymmetry
           in the fiducial volume 
           $44^\circ < \theta < 136^\circ$ and for 
           $\xi < 25^\circ$ for the reaction \protect\EEEEG.
           The errors are statistical only and a correlated absolute
           systematic error of $\Delta_i^{\MR{abs}} = 0.0025$
           has to be added. 
           In the rightmost column the $s$-channel contribution to the
           forward-backward asymmetry in the full solid angle is listed for
           $\xi < 25^\circ$
           with the statistical error
           (see Appendix~\protect\ref{app:tchannel}).
          } 
\label{tab:electron_afb}
      }
\end{center}
\end{table}

\clearpage
%
%
%
%
\begin{table}
 \begin{center}
  \renewcommand{\arraystretch}{1.1}
  \begin{tabular}{|l|r@{$\,\pm\,$}l|r@{$\,\pm\,$}l|c|}
   \hline
    Parameter  & \multicolumn{4}{c|}{Treatment of Charged Leptons}  & Standard \\
               & \multicolumn{2}{c|}{non-universality}
                                  & \multicolumn{2}{c|}{universality}
                                                                    & Model  \\
   \hline
   \MZ \hfill [\MeV]   &  91\,189.7  &  3.1  & \mZqval & \mZqerr &    ---   \\
   \GZ \hfill [\MeV]   &   2\,502.4  &  4.2  & \GZqval & \GZqerr & $2\,492.7\,^{+3.8}_{-5.2}$  \\
   \hline
   \Gh   \hfill [\MeV] &   1\,750.9  &  4.7  & \Ghqval & \Ghqerr & $1\,739.8\,^{+3.2}_{-4.1}$  \\        
   \GE   \hfill [\MeV] &   84.16     &  0.22 & \MC{2}{c|}{---}   & $ 83.91\,^{+0.10}_{-0.14}$  \\        
   \GM   \hfill [\MeV] &   83.95     &  0.44 & \MC{2}{c|}{---}   & $ 83.91\,^{+0.10}_{-0.14}$  \\        
   \GT   \hfill [\MeV] &   84.23     &  0.58 & \MC{2}{c|}{---}   & $ 83.72\,^{+0.10}_{-0.14}$  \\        
   \GL   \hfill [\MeV] &    \MC{2}{c|}{---}  & \Glqval & \Glqerr & $ 83.91\,^{+0.10}_{-0.14}$  \\
   \hline
   $\chi^2$/dof        &\MC{2}{c|}{$91/94$}&\MC{2}{c|}{$91/96$}&\MC{1}{c|}{---} \\
   \hline
  \end{tabular}
  \caption{Results of the fits to the $1990-95$ total cross section data with
           and without assuming lepton universality.
          }
  \label{tab:fitwidth}
 \end{center}
\end{table}

\begin{table}
 \begin{center}
  \renewcommand{\arraystretch}{1.1}
  \begin{tabular}{|l|rrrrrr|}
   \hline
          &     \MZ &    \GZ  &    \Gh  &    \GE  &    \GM   &    \GT   \\
   \hline
   \MZ    &$\phantom{-}1.00 $
                    &$\phantom{-}0.06 $
                              &$\phantom{-}0.14 $
                                        &$ -0.03 $&$\phantom{-}0.07  $
                                                             &$\phantom{-}0.05  $\\
   \GZ    &$       $&$  1.00 $&$  0.54 $&$  0.55 $&$  0.28  $&$  0.21  $\\
   \Gh    &$       $&$       $&$  1.00 $&$ -0.29 $&$  0.48  $&$  0.36  $\\
   \GE    &$       $&$       $&$       $&$  1.00 $&$ -0.14  $&$ -0.11  $\\
   \GM    &$       $&$       $&$       $&$       $&$  1.00  $&$  0.19  $\\
   \GT    &$       $&$       $&$       $&$       $&$        $&$  1.00  $\\
   \hline  
  \end{tabular}
  \renewcommand{\arraystretch}{1.0}
  \parbox{\capwidth}{
   \caption[]{Correlation matrix for the six parameter fit without assuming 
              lepton universality in Table~\protect\ref{tab:fitwidth}.
             }  
   \label{tab:fitcor6}
   }
 \end{center}
\end{table}

\begin{table}
 \begin{center}
  \renewcommand{\arraystretch}{1.1}
  \begin{tabular}{|l|rrrr|}
   \hline
          &     \MZ &    \GZ  &    \Gh  &    \GL  \\
   \hline                                         
   \MZ    &$\phantom{-}1.00 $
                    &$\phantom{-}0.07 $
                              &$\phantom{-}0.12 $
                                        &$\phantom{-}0.03 $\\
   \GZ    &$       $&$  1.00 $&$  0.68 $&$  0.71 $\\
   \Gh    &$       $&$       $&$  1.00 $&$  0.12 $\\
   \GL    &$       $&$       $&$       $&$  1.00 $\\
   \hline  
  \end{tabular}
  \renewcommand{\arraystretch}{1.0}
  \parbox{\capwidth}{
   \caption[]{Correlation matrix for the four parameter fit assuming 
              lepton universality in Table~\protect\ref{tab:fitwidth}.
             }  
   \label{tab:fitcor4}
   }
 \end{center}
\end{table}


\begin{table}
 \begin{center}
  \renewcommand{\arraystretch}{1.1}  
  \begin{tabular}{|l|r@{$\,\pm\,$}l@{$\,\pm\,$}l|c|}
   \hline
 & \MC{3}{r|}{$\Gamma^{\MR{exp}} - \Gamma^{\MR{SM}}\ [\MeV]$}
                                 &   $\Gamma^{\MR{NP}}_{\MR{95}}\ [\MeV]$ \\
   \hline
    \GZ     &  $  14.6  $  & $4.2 $  & $1.7 $  & $ 22.0 $ \\
    \Gh     &  $  15.2  $  & $3.8 $  & $1.5 $  & $ 21.9 $ \\
    \Ginv   &  $  -1.7  $  & $2.9 $  & $0.23$  & $  4.8 $ \\
    \GL     &  $   0.37 $  & $0.17$  & $0.05$  & $  0.66$ \\
    \GE     &  $  0.38 $  & $0.22$  & $0.05$  & $ 0.75$ \\
    \GM     &  $  0.17 $  & $0.44$  & $0.05$  & $ 0.99$ \\
    \GT     &  $  0.64 $  & $0.58$  & $0.05$  & $ 1.64$ \\
   \hline
  \end{tabular}
  \renewcommand{\arraystretch}{1.0}
  \parbox{\capwidth}{
   \caption[]{The one-sided upper limits (95\% C.L.) on non-\SM\ contributions 
              to the \protect\Zo\ widths, $\Gamma^{\MR{NP}}_{\MR{95}}$,
              as derived from \SM\ calculations and our 
              measurements.
              Also given are the differences of our measurements and the
              \protect\SM\ expectations for $\protect\MH = 1\ \TeV$,
              $\Gamma^{\MR{exp}}-
               \Gamma^{\MR{SM}}$,
              with the experimental and theoretical errors.
              The results on the total and partial widths are correlated and
              hence the limits cannot be applied simultaneously.
             } 
   \label{tab:gamma_limits}
   }
 \end{center}
\end{table}

\clearpage

\begin{table}
 \begin{center}
  \renewcommand{\arraystretch}{1.1}  
  \begin{tabular}{|l|r@{$\,\pm\,$}l|r@{$\,\pm\,$}l|c|}
   \hline
    Parameter  & \multicolumn{4}{c|}{Treatment of Charged Leptons}  & Standard \\
               & \multicolumn{2}{c|}{non-universality}
                                  & \multicolumn{2}{c|}{universality}
                                                                    & Model    \\
   \hline
   \MZ \hfill [\MeV]   & 91\,189.8 &    3.1 & 91\,189.5 &    3.1 &    ---   \\
   \GZ \hfill [\MeV]   &  2\,502.5 &    4.2 &  2\,502.5 &    4.2 & $2\,492.7\,^{+3.8}_{-5.2}   $ \\
   \hline
   \s0had\hfill [\nb]  &    41.535 &  0.055 &    41.535 &  0.055 & $ 41.476 \pm 0.012          $ \\
   \RE                 &    20.816 &  0.089 &   \MC{2}{c|}{---}  & $ 20.733 \pm 0.018$ \\
   \RM                 &    20.861 &  0.097 &   \MC{2}{c|}{---}  & $ 20.733 \pm 0.018$ \\
   \RT                 &    20.792 &  0.133 &   \MC{2}{c|}{---}  & $ 20.780 \pm 0.018$ \\
   \RL                 & \MC{2}{c|}{---}    &   20.810  & 0.060  & $ 20.733 \pm 0.018$ \\
   \hline
   \AFBZe              &    0.0106 & 0.0058 &   \MC{2}{c|}{---}  & $ 0.0151 \pm 0.0012$\\
   \AFBZm              &    0.0188 & 0.0033 &   \MC{2}{c|}{---}  & $ 0.0151 \pm 0.0012$\\
   \AFBZt              &    0.0260 & 0.0047 &   \MC{2}{c|}{---}  & $ 0.0151 \pm 0.0012$\\
   \AFBZl              & \MC{2}{c|}{---}    &    0.0192 & 0.0024 & $ 0.0151 \pm 0.0012$\\
   \hline
   $\chi^2$/dof        &\MC{2}{c|}{$158/166$}&\MC{2}{c|}{$163/170$}&\MC{1}{c|}{---} \\
   \hline
  \end{tabular}
  \renewcommand{\arraystretch}{1.0}
  \parbox{\capwidth}{
   \caption[]{Results on the mass, total width,
             the hadronic pole cross section, \protect\s0had,
             the ratios of hadronic to leptonic widths, \protect\RL,
             and the leptonic pole asymmetries, \AFBZl, determined from
             cross section and forward-backward asymmetry data with and
             without assuming lepton universality.
             The \protect\SM\ expectations are calculated using the parameters
             listed in Equation~\protect\ref{eq:smpara}.
             } 
   \label{tab:fitpar59}
   }
 \end{center}
\end{table}

\begin{table}
 \begin{center}
  \renewcommand{\arraystretch}{1.1}
  \begin{tabular}{|l|rrrrrrrrr|}
   \hline
          &     \MZ &    \GZ  &  \s0had &     \RE &     \RM &    \RT  &  \AFBZe &  \AFBZm &  \AFBZt \\
   \hline
   \MZ    &$\phantom{-}1.00 $
                    &$\phantom{-}0.07 $
                              &$ -0.01 $&$\phantom{-}0.11 $
                                                  &$\phantom{-}0.00 $
                                                            &$\phantom{-}0.00 $
                                                                      &$ -0.05 $&$\phantom{-}0.05 $
                                                                                          &$\phantom{-}0.03 $\\
   \GZ    &$       $&$  1.00 $&$ -0.35 $&$  0.00 $&$  0.00 $&$  0.00 $&$  0.00 $&$  0.00 $&$   0.00 $\\
   \s0had &$       $&$       $&$  1.00 $&$  0.07 $&$  0.08 $&$  0.05 $&$  0.01 $&$  0.01 $&$   0.00 $\\
   \RE    &$       $&$       $&$       $&$  1.00 $&$  0.03 $&$  0.02 $&$ -0.15 $&$  0.02 $&$   0.01 $\\
   \RM    &$       $&$       $&$       $&$       $&$  1.00 $&$  0.02 $&$  0.00 $&$  0.00 $&$   0.00 $\\
   \RT    &$       $&$       $&$       $&$       $&$       $&$  1.00 $&$  0.00 $&$  0.00 $&$   0.01 $\\
   \AFBZe &$       $&$       $&$       $&$       $&$       $&$       $&$  1.00 $&$  0.01 $&$  -0.01 $\\
   \AFBZm &$       $&$       $&$       $&$       $&$       $&$       $&$       $&$  1.00 $&$   0.01 $\\
   \AFBZt &$       $&$       $&$       $&$       $&$       $&$       $&$       $&$       $&$   1.00 $\\
   \hline
  \end{tabular}
  \renewcommand{\arraystretch}{1.0}
  \parbox{\capwidth}{
   \caption[]{Correlation matrix for the nine parameter fit not assuming 
              lepton universality in Table~\protect\ref{tab:fitpar59}.
             } 
   \label{tab:fitcor9}
   }
 \end{center}
\end{table}

\begin{table}
 \begin{center}
  \renewcommand{\arraystretch}{1.1}
  \begin{tabular}{|l|rrrrr|}
   \hline
          &     \MZ &    \GZ  &  \s0had &      \RL & \AFBZl  \\
   \hline
   \MZ    &$\phantom{-}1.00 $
                    &$\phantom{-}0.06 $
                              &$\phantom{-}0.01 $
                                        &$\phantom{-}0.07 $
                                                  &$\phantom{-}0.04  $\\
   \GZ    &$       $&$  1.00 $&$ -0.35 $&$  0.00 $&$  0.02  $\\
   \s0had &$       $&$       $&$  1.00 $&$  0.12 $&$  0.01  $\\
   \RL    &$       $&$       $&$       $&$  1.00 $&$ -0.02  $\\
   \AFBZl &$       $&$       $&$       $&$       $&$  1.00  $\\
   \hline  
  \end{tabular}
  \renewcommand{\arraystretch}{1.0}
  \parbox{\capwidth}{
   \caption[]{Correlation matrix for the five parameter fit assuming 
              lepton universality in Table~\protect\ref{tab:fitpar59}.
             }  
   \label{tab:fitcor5}
   }
 \end{center}
\end{table}

\begin{table}
 \begin{center}
  \renewcommand{\arraystretch}{1.1}
  \begin{tabular}{|l|c|}
   \hline
    \multicolumn{2}{|c|}{Forward-backward asymmetry} \\
   \hline
    $A_{\MR{e}}$ &   $0.119\,^{+0.029}_{-0.039}$  \\
    $A_\mu$      &   $0.210\,^{+0.108}_{-0.054}$  \\
    $A_\tau$     &   $0.291\,^{+0.153}_{-0.076}$  \\
   \hline
     $A_\ell$    &  $ 0.160 \pm 0.010$            \\
   \hline
   \hline
    \multicolumn{2}{|c|}{Tau-polarisation}        \\
   \hline
    $A_{\MR{e}}$ & $0.1678 \pm 0.0130$            \\
    $A_\tau$     & $0.1476 \pm 0.0108$            \\
   \hline
   \hline
    \multicolumn{2}{|c|}{Average}                 \\
   \hline
    $A_{\MR{e}}$ &  $0.163 \pm  0.011$            \\  
    $A_\mu$      &  $0.153 \pm  0.029$            \\
    $A_\tau$     &  $0.152 \pm  0.010$            \\
   \hline
    $A_\ell$     & $\Alepval \pm \Aleperr$        \\
   \hline
  \end{tabular}
  \renewcommand{\arraystretch}{1.0}
  \parbox{\capwidth}{
  \caption[]{Measurements of the polarisation parameter $A_\ell$
             from forward-backward asymmetries for the three
             leptons and the combined value.
             Also listed are the results for $A_{\MR{e}}$ and
             $A_\tau$ obtained from our measurements of 
             tau-polarisation~\protect\cite{l3-140}.
             At the bottom the averages from the combined fit are 
             given.
             } 
   \label{tab:Alepton}
   }
 \end{center}
\end{table}

\begin{table}
 \begin{center}
  \renewcommand{\arraystretch}{1.1}
  \begin{tabular}{|l|r@{$\,\pm\,$}l|r@{$\,\pm\,$}l|c|}
   \hline
    Parameter  & \multicolumn{4}{c|}{Treatment of Charged Leptons}  & Standard \\
               & \multicolumn{2}{c|}{non-universality}
                                  & \multicolumn{2}{c|}{universality}
                                                                    & Model    \\
   \hline
   \gVbe       &$ -0.0412 $&$ 0.0027 $&  \MC{2}{c|}{---}       &           \\
   \gVbm       &$ -0.0386 $&$ 0.0073 $&  \MC{2}{c|}{---}       &           \\
   \gVbt       &$ -0.0384 $&$ 0.0026 $&  \MC{2}{c|}{---}       &           \\
   \gVbl       & \MC{2}{c|}{---}      & \gVqval    & \gVqerr   & $0.0358 \pm 0.0014$   \\
   \hline
   \gAbe       &$ -0.5015 $&$ 0.0007 $&  \MC{2}{c|}{---}       &            \\
   \gAbm       &$ -0.5009 $&$ 0.0014 $&  \MC{2}{c|}{---}       &            \\
   \gAbt       &$ -0.5023 $&$ 0.0017 $&  \MC{2}{c|}{---}       &            \\
   \gAbl       &  \MC{2}{c|}{---}     &  \gAqval    & \gAqerr  & $-0.50113\,^{+0.00034}_{-0.00022}$ \\
   \hline
  \end{tabular}
  \renewcommand{\arraystretch}{1.0}
  \parbox{\capwidth}{
  \caption[]{Results for the vector and axial-vector coupling 
             constants of charged leptons as obtained from fits
             to the total cross sections and forward-backward asymmetries
             and including our results from tau polarisation.
             The errors are correlated.
             } 
   \label{tab:gagv}
   }
 \end{center}
\end{table}

\begin{table}
 \begin{center}
  \renewcommand{\arraystretch}{1.1}
  \begin{tabular}{|l|r@{$\,\pm\,$}l|}
   \hline
   \gVbm/\gVbe  & 0.94   & 0.21   \\
   \gVbt/\gVbe  & 0.93   & 0.09   \\
   \hline
   \gAbm/\gAbe  & 0.9988 & 0.0033 \\
   \gAbt/\gAbe  & 1.0017 & 0.0038 \\
   \hline
  \end{tabular}
  \renewcommand{\arraystretch}{1.0}
  \parbox{\capwidth}{
  \caption[]{Ratios of vector and axial-vector coupling constants
             obtained from a fit to the total cross sections and 
             forward-backward asymmetries including our results from
             tau polarisation.
             } 
   \label{tab:gagvratio}
   }
 \end{center}
\end{table}

\begin{table}
 \begin{center}
  \renewcommand{\arraystretch}{1.1}
  \begin{tabular}{|l|r@{$\,\pm\,$}l|}
   \hline
   Input data       &  \MC{2}{c|}{\swsqb} \\
   \hline
   \AFBZl           & 0.2299  & 0.0013    \\ 
   Tau polarisation & 0.2304  & 0.0011    \\ 
   \AFBZb           & 0.2318  & 0.0013    \\
   \QFB             & 0.2327  & 0.0017    \\
   \hline
   Average          & \swsqval& \swsqerr  \\   
   \hline
  \end{tabular}
  \renewcommand{\arraystretch}{1.0}
  \parbox{\capwidth}{
  \caption[]{Determination of the effective weak mixing angle 
             \protect\swsqb\ from different reactions:
             the leptonic forward-backward pole asymmetry,
             \protect\AFBZl, tau polarisation (Table~\protect\ref{tab:Alepton}), 
             b-quark pole asymmetry~\protect\cite{l3-163}, 
             \protect\AFBZb, 
             and the quark charge asymmetry~\protect\cite{l3-157}, 
             \protect\QFB.
             The SM prediction is 
             $\protect\swsqb = 0.23215\,^{+0.00072}_{-0.00066}$.
            } 
   \label{tab:sweff}
   }
 \end{center}
\end{table}

\begin{table}
 \begin{center}
  \renewcommand{\arraystretch}{1.1}
  \begin{tabular}{|l|r@{$\,\pm\,$}l|r@{$\,\pm\,$}l|c|}
   \hline
    Parameter  & \multicolumn{4}{c|}{Treatment of Charged Leptons}  & Standard \\
               & \multicolumn{2}{c|}{non-universality}
                                  & \multicolumn{2}{c|}{universality}
                                                                    & Model    \\
   \hline
   \MZ \hfill [\MeV] & 91\,189.3 &    11.2  & \mZSmatrixqval &\mZSmatrixqerr &   ---     \\
   \GZ \hfill [\MeV] &  2\,502.8 &     4.6  &  2\,503.1 &  4.5     & $2\,492.7\,^{+3.8}_{-5.2}    $    \\
   \rtxsha           &   2.986   & 0.010    &    2.986  &  0.010   & $   2.958\,^{+0.009}_{-0.012}$    \\
   \rtxsel           &   0.14316 & 0.00089  &   \MC{2}{c|}{---}    &                                   \\
   \rtxsmu           &   0.14302 & 0.00082  &   \MC{2}{c|}{---}    &                                   \\
   \rtxsta           &   0.14386 & 0.00104  &   \MC{2}{c|}{---}    &                                   \\
   \rtxsle           &   \MC{2}{c|}{---}    &   0.14336 &  0.00066 & $0.14243\,^{+0.00035}_{-0.00049}$ \\ 
   \hline          
   \jtxsha           &   0.21    & 0.63     &   0.44    &   0.59   & $ 0.2133\,^{+0.0086}_{-0.0093}  $ \\
   \jtxsel           &  -0.029   & 0.054    &   \MC{2}{c|}{---}    &                                   \\
   \jtxsmu           &   0.035   & 0.046    &   \MC{2}{c|}{---}    &                                   \\
   \jtxsta           &   0.073   & 0.048    &   \MC{2}{c|}{---}    &                                   \\
   \jtxsle           &   \MC{2}{c|}{---}    &    0.045  &   0.035  & $0.00409 \pm 0.00032$             \\
   \hline          
   \rafbel           &   0.00174 & 0.00113  &   \MC{2}{c|}{---}    &                                   \\
   \rafbmu           &   0.00341 & 0.00066  &   \MC{2}{c|}{---}    &                                   \\
   \rafbta           &   0.00456 & 0.00093  &   \MC{2}{c|}{---}    &                                   \\
   \rafble           &   \MC{2}{c|}{---}    &   0.00333 &  0.00048 & $0.00255 \pm 0.00023$             \\
   \hline          
   \jafbel           &   0.698   & 0.080    &   \MC{2}{c|}{---}    &                                   \\
   \jafbmu           &   0.820   & 0.047    &   \MC{2}{c|}{---}    &                                   \\
   \jafbta           &   0.754   & 0.055    &   \MC{2}{c|}{---}    &                                   \\
   \jafble           &   \MC{2}{c|}{---}    &     0.777 &   0.033  & $0.7986\,^{+0.0009}_{-0.0012}   $ \\
   \hline
   $\chi^2$/dof      & \MC{2}{c|}{$153/159$}&\MC{2}{c|}{$161/167$}&\MC{1}{c|}{---} \\
  \hline
  \end{tabular}
  \renewcommand{\arraystretch}{1.0}
  \parbox{\capwidth}{
  \caption[]{Results of the fits to the data taken at the \protect\Zo\ resonance
             within the S-Matrix framework with and without the assumption
             of lepton universality.
             } 
   \label{tab:smatrix}
   }
 \end{center}
\end{table}

\begin{table}
 \begin{sideways}
 \begin{minipage}[b]{\textheight}
 \ \vspace*{10mm} \\
 \small
 \begin{center}
 \renewcommand{\arraystretch}{1.1}
 \begin{tabular}{|l|rrrrrrrrrrrrrrrr|}
  \hline
        & \MZ  &    \GZ & \rtxsha 
                               & \rtxsel 
                                       & \rtxsmu 
                                               & \rtxsta  
                                                       & \jtxsha 
                                                               & \jtxsel 
                                                                       & \jtxsmu        
                                                                               & \jtxsta 
                                                                                       & \rafbel        
                                                                                               &\rafbmu        
                                                                                                       & \rafbta        
                                                                                                               & \jafbel
                                                                                                                       & \jafbmu        
                                                                                                                               &\jafbta \\
 \hline
\MZ    &$\phantom{-}1.00$
               &$-0.38$&$-0.39$&$-0.53$&$-0.23$&$-0.18$&$-0.96$&$-0.53$&$-0.63$&$-0.60$&$-0.20$&$\phantom{-}0.22$
                                                                                                       &$\phantom{-}0.14$
                                                                                                               &$\phantom{-}0.01$
                                                                                                                       &$-0.01$&$\phantom{-}0.01$\\
\GZ    &$     $&$ 1.00$&$ 0.93$&$ 0.64$&$ 0.56$&$ 0.44$&$ 0.41$&$ 0.20$&$ 0.27$&$ 0.26$&$ 0.08$&$-0.07$&$-0.04$&$-0.01$&$ 0.04$&$ 0.03$\\
\rtxsha&$     $&$     $&$ 1.00$&$ 0.66$&$ 0.57$&$ 0.45$&$ 0.43$&$ 0.20$&$ 0.28$&$ 0.27$&$ 0.09$&$-0.07$&$-0.04$&$-0.01$&$ 0.04$&$ 0.03$\\
\rtxsel&$     $&$     $&$     $&$ 1.00$&$ 0.40$&$ 0.31$&$ 0.54$&$ 0.32$&$ 0.36$&$ 0.34$&$ 0.21$&$-0.12$&$-0.07$&$-0.03$&$ 0.03$&$ 0.01$\\
\rtxsmu&$     $&$     $&$     $&$     $&$ 1.00$&$ 0.27$&$ 0.24$&$ 0.11$&$ 0.25$&$ 0.16$&$ 0.05$&$-0.02$&$-0.02$&$-0.01$&$ 0.08$&$ 0.02$\\
\rtxsta&$     $&$     $&$     $&$     $&$     $&$ 1.00$&$ 0.19$&$ 0.09$&$ 0.13$&$ 0.20$&$ 0.04$&$-0.03$&$ 0.01$&$-0.01$&$ 0.02$&$ 0.09$\\
\jtxsha&$     $&$     $&$     $&$     $&$     $&$     $&$ 1.00$&$ 0.53$&$ 0.63$&$ 0.60$&$ 0.20$&$-0.21$&$-0.14$&$-0.01$&$ 0.01$&$-0.01$\\
\jtxsel&$     $&$     $&$     $&$     $&$     $&$     $&$     $&$ 1.00$&$ 0.35$&$ 0.33$&$ 0.10$&$-0.12$&$-0.08$&$ 0.17$&$ 0.00$&$ 0.00$\\
\jtxsmu&$     $&$     $&$     $&$     $&$     $&$     $&$     $&$     $&$ 1.00$&$ 0.39$&$ 0.13$&$-0.11$&$-0.09$&$-0.01$&$-0.04$&$ 0.00$\\
\jtxsta&$     $&$     $&$     $&$     $&$     $&$     $&$     $&$     $&$     $&$ 1.00$&$ 0.12$&$-0.13$&$-0.07$&$-0.01$&$ 0.00$&$-0.04$\\
\rafbel&$     $&$     $&$     $&$     $&$     $&$     $&$     $&$     $&$     $&$     $&$ 1.00$&$-0.05$&$-0.03$&$ 0.03$&$ 0.00$&$ 0.00$\\
\rafbmu&$     $&$     $&$     $&$     $&$     $&$     $&$     $&$     $&$     $&$     $&$     $&$ 1.00$&$ 0.04$&$ 0.00$&$ 0.17$&$ 0.00$\\
\rafbta&$     $&$     $&$     $&$     $&$     $&$     $&$     $&$     $&$     $&$     $&$     $&$     $&$ 1.00$&$ 0.00$&$ 0.00$&$ 0.15$\\
\jafbel&$     $&$     $&$     $&$     $&$     $&$     $&$     $&$     $&$     $&$     $&$     $&$     $&$     $&$ 1.00$&$ 0.00$&$ 0.00$\\
\jafbmu&$     $&$     $&$     $&$     $&$     $&$     $&$     $&$     $&$     $&$     $&$     $&$     $&$     $&$     $&$ 1.00$&$ 0.00$\\
\jafbta&$     $&$     $&$     $&$     $&$     $&$     $&$     $&$     $&$     $&$     $&$     $&$     $&$     $&$     $&$     $&$ 1.00$\\
 \hline
 \end{tabular}
 \renewcommand{\arraystretch}{1.0}
 \parbox{\capwidth}{
 \caption[]{Correlation of the S-Matrix parameters listed in 
            Table~\protect\ref{tab:smatrix} 
            not assuming lepton universality.
            } 
   \label{tab:cor16}
   }
 \end{center}
 \normalsize
 \end{minipage}
 \end{sideways}
\end{table}

\begin{table}
 \begin{center}
  \renewcommand{\arraystretch}{1.1}
  \begin{tabular}{|l|rrrrrrrr|}
   \hline
              & \MZ
                     & \GZ 
                             & \rtxsha 
                                      & \rtxsle 
                                             & \jtxsha 
                                                     & \jtxsle 
                                                             & \rafble 
                                                                     & \jafble \\
    \hline
    \MZ       &$\phantom{-}1.00$
                     &$-0.34$&$-0.35$&$-0.41$&\mZjhadcor&$-0.76$&$\phantom{-}0.14$
                                                                     &$-0.03$ \\
    \GZ       &      &$ 1.00$&$ 0.93$&$ 0.74$&$ 0.38$&$ 0.29$&$-0.03$&$ 0.06$ \\
    \rtxsha   &      &       &$ 1.00$&$ 0.76$&$ 0.39$&$ 0.30$&$-0.03$&$ 0.06$ \\
    \rtxsle   &      &       &       &$ 1.00$&$ 0.43$&$ 0.38$&$-0.01$&$ 0.08$ \\
    \jtxsha   &      &       &       &       &$ 1.00$&$ 0.76$&$-0.14$&$ 0.03$ \\
    \jtxsle   &      &       &       &       &       &$ 1.00$&$-0.09$&$ 0.02$ \\
    \rafble   &      &       &       &       &       &       &$ 1.00$&$ 0.14$ \\
    \jafble   &      &       &       &       &       &       &       &$ 1.00$ \\
   \hline
  \end{tabular}
  \renewcommand{\arraystretch}{1.0}
  \parbox{\capwidth}{
  \caption[]{Correlation of the S-Matrix parameters listed in 
             Table~\protect\ref{tab:smatrix} 
             assuming lepton universality.
             } 
   \label{tab:cor8}
   }
 \end{center}
\end{table}

\begin{table}
 \begin{center}
  \renewcommand{\arraystretch}{1.1}
  \begin{tabular}{|l|r@{$\,\pm\,$}l|}
   \hline
   \SM\ parameter &  \MC{2}{c|}{Fit result} \\
   \hline
   \MZ\ [\MeV]         & $91\,189.0$  & $    3.1$ \\
   \Mt\ [\GeV]         & $177.4    $  & $    4.8$ \\
   $\log_{10}\MH/\GeV$ & $1.56     $  & $  0.33 $ \\
   \as                 & $0.1226   $  & $ 0.0040$ \\
   \Da5had             & $0.02787  $  & $0.00062$ \\
   \hline
  \end{tabular}
  \renewcommand{\arraystretch}{1.0}
  \parbox{\capwidth}{
  \caption[]{Results of the fit for \protect\SM\ parameters.
             In addition to our measurements, constraints
             on \protect\Mt\ and \protect\Da5had\ as given
             in Equation~\protect\ref{eq:smpara} are used.
             } 
   \label{tab:Hfit}
   }
 \end{center}
\end{table}

\begin{table}
 \begin{center}
  \renewcommand{\arraystretch}{1.1}
  \begin{tabular}{|l|rrcrr|}
   \hline 
                        & \MC{1}{c}{\ \MZ} 
                                  & \MC{1}{c}{\ \Mt}  
                                          &  $\log_{10}\MH/\GeV$  
                                                               & \MC{1}{c}{\as} 
                                                                         & \MC{1}{c|}{\Da5had} \\
   \hline
                 \MZ    &$\phantom{-}1.00 $
                                  &$ -0.01 $&$          -0.02 $&$\phantom{-}0.01 $
                                                                &$\phantom{-}0.02  $\\
                 \Mt    &$       $&$  1.00 $&$\phantom{-}0.08 $&$ -0.07 $&$  0.11  $\\
  $\log_{10}\MH/\GeV$   &$       $&$       $&$\phantom{-}1.00 $&$ -0.03 $&$ -0.14  $\\
                 \as    &$       $&$       $&$                $&$  1.00 $&$  0.08  $\\
             \Da5had    &$       $&$       $&$                $&$       $&$  1.00  $\\
   \hline  
  \end{tabular}
  \renewcommand{\arraystretch}{1.0}
  \parbox{\capwidth}{
   \caption[]{Correlation matrix for the fit for \protect\SM\ parameters
              in Table~\protect\ref{tab:Hfit}.
             }  
   \label{tab:Hfitcor}
   }
 \end{center}
\end{table}


\begin{table}
\begin{center}
\renewcommand{\arraystretch}{1.1}
\begin{tabular}{|l|r|r|r|r|r|r|}
\hline
       & $\sqrt{s}\ [\GeV]$ 
                  & $\sigma^{t+s/t, \MR{AL}}\ [\pb]$
                            & \MC{1}{c|}{$R$}   & $\sigma_{\MR{FB}}^{t+s/t, \MR{AL}}\ [\pb]$        
                                                & \MC{1}{c|}{$R_{\MR{F}}$} 
                                                          &  \MC{1}{c|}{$R_{\MR{B}}$}  \\
\hline
  1990 & 88.232   &  226.5 & 0.9935 &  183.2  & 0.9922 & 0.9943 \\
       & 89.236   &  250.7 & 0.9946 &  202.7  & 0.9934 & 0.9956 \\
       & 90.237   &  261.6 & 0.9955 &  211.7  & 0.9943 & 0.9965 \\
       & 91.230   &  146.8 & 0.9957 &  118.5  & 0.9949 & 0.9965 \\
       & 92.226   &   17.7 & 0.9943 &   14.5  & 0.9944 & 0.9943 \\
       & 93.228   &    6.3 & 0.9911 &    5.4  & 0.9922 & 0.9898 \\
       & 94.223   &   17.2 & 0.9866 &   14.5  & 0.9887 & 0.9838 \\
\hline                                                            
  1991 & 91.253   &  142.2 & 0.9957 &  114.8  & 0.9949 & 0.9965 \\
       & 88.480   &  232.1 & 0.9938 &  187.5  & 0.9926 & 0.9946 \\
       & 89.470   &  256.4 & 0.9949 &  207.3  & 0.9936 & 0.9958 \\
       & 90.228   &  261.9 & 0.9955 &  211.9  & 0.9943 & 0.9965 \\
       & 91.222   &  148.3 & 0.9957 &  119.7  & 0.9949 & 0.9965 \\
       & 91.967   &   36.1 & 0.9949 &   29.4  & 0.9947 & 0.9951 \\
       & 92.966   &    5.0 & 0.9921 &    4.4  & 0.9929 & 0.9911 \\
       & 93.716   &   11.6 & 0.9890 &    9.7  & 0.9905 & 0.9870 \\
\hline                                                            
  1992 & 91.294   &  134.7 & 0.9957 &  108.8  & 0.9949 & 0.9964 \\
\hline                                                            
  1993 & 91.3213  &  129.5 & 0.9957 &  104.7  & 0.9950 & 0.9964 \\
       & 89.4497  &  255.9 & 0.9949 &  206.9  & 0.9936 & 0.9958 \\
       & 91.2057  &  151.3 & 0.9957 &  122.2  & 0.9949 & 0.9965 \\
       & 93.0358  &    5.0 & 0.9918 &    4.5  & 0.9927 & 0.9908 \\
\hline                                                            
  1994 & 91.2197  &  148.7 & 0.9957 &  120.1  & 0.9949 & 0.9965 \\
\hline                                                            
  1995 & 91.3105  &  131.5 & 0.9957 &  106.3  & 0.9950 & 0.9964 \\
       & 89.4517  &  256.0 & 0.9949 &  206.9  & 0.9936 & 0.9958 \\
       & 91.2960  &  134.3 & 0.9957 &  108.5  & 0.9950 & 0.9964 \\
       & 92.9828  &    5.0 & 0.9921 &    4.4  & 0.9929 & 0.9911 \\
\hline
\end{tabular}
  \renewcommand{\arraystretch}{1.0}
\parbox{\capwidth}{
\caption[]{Contribution from the $t$-channel and \protect\stint\
           and corrections to the process \protect\EEEEG\ as calculated
           with the ALIBABA and TOPAZ0 programs for all \protect\cms\ energies:
           $\sigma^{t+s/t, \MR{AL}}$ is the contribution to the 
           total cross section and $R$ the correction factor for the
           different polar angle cuts 
           (see Appendix~\protect\ref{app:tchannel}).
           The corresponding differences of forward and backward
           cross section, $\sigma_{\MR{FB}}^{t+s/t, \MR{AL}}$, and the
           correction factors for the forward and backward parts of the
           cross section, $R_{\MR{F}}$ and $R_{\MR{B}}$, are also given.
           }
\label{tab:tchann_numbers}
      }
\end{center}
\end{table}

\begin{table}
\begin{center}
\renewcommand{\arraystretch}{1.1}
\begin{tabular}{|l|rrrrrrrr|}
\hline
  $\RS$ [\GeV]        & 88.45 & 89.45 & 90.20 & 91.19 & 91.30 & 91.95 & 93.00 & 93.70 \\
\hline
  Cross section [\pb] &  1.1  &  1.2  &  1.2  &  1.0  &  1.0  &  1.2  &  1.0  &  1.0  \\  
  Asymmetry           &0.0012 &0.0014 &0.0007 &0.0007 &0.0006 &0.0001 &0.0002 &0.0002 \\
\hline
\end{tabular}
\parbox{\capwidth}{
\caption[]{Uncertainties of the ALIBABA program on the calculation of the 
           $t$-channels and \protect\stint\ contribution to 
           \protect\EEEEG~\protect\cite{Beenakker}.
           To account for the smaller fiducial volume used in our analysis the
           errors are scaled by a factor of 0.8.
           }
\label{tab:tchann_errors}
      }
\end{center}
\end{table}

\clearpage
%
%
%
%

%
%
%
%

\clearpage

%
%
%
%

\clearpage

%
%
%
%

\clearpage

%
%
%
%

\clearpage

%
%
%
%

\clearpage
%
%
%
%



\clearpage
%
%
%
%

\end{document}